\documentclass{article}%
\usepackage{graphicx}
\usepackage{amsmath}
\usepackage{amsfonts}
\usepackage{amssymb}%
\newtheorem{theorem}{Theorem}[section]
{}

\newtheorem{definition}{Definition}[section]

\newtheorem{lemma}{Lemma}[section]
{}
\newtheorem{notation}{Notation}[section]

\newtheorem{remark}{Remark}[section]

\newenvironment{proof}[1][Proof]{\textbf{#1.} }{\ \rule{0.5em}{0.5em}}

\begin{document}

\author{O. A. Veliev\\{\small \ Dept. of Math., Fac. of Arts and Sci., Dogus University,}\\{\small Acibadem, Kadikoy, Istanbul, Turkey,}\\{\small \ e-mail: oveliev@dogus.edu.tr}}
\title{\textbf{Perturbation Theory for the Periodic Multidimensional Schr\"{o}dinger
Operator and the Bethe-Sommerfeld Conjecture}}
\date{}
\maketitle

\begin{abstract}
In this paper we obtain asymptotic formulas of arbitrary order for the Bloch
eigenvalue and the Bloch function of the periodic Schr\"{o}dinger operator
$-\Delta+q(x),$ of arbitrary dimension, when corresponding quasimomentum lies
near a diffraction hyperplane. Besides, writing the asymptotic formulas for
the Bloch eigenvalue and the Bloch function, when corresponding quasimomentum
lies far from the diffraction hyperplanes, obtained in my previous papers in
improved and enlarged form, we obtain the complete perturbation theory for the
multidimensional Schr\"{o}dinger operator with a periodic potential. Moreover,
we estimate the measure of the isoenergetic surfaces in the high energy region
which implies the validity of the Bethe-Sommerfeld conjecture for arbitrary
dimension and arbitrary lattice.



\end{abstract}

\section{Introduction}

\bigskip In this paper we consider the Schr\"{o}dinger operator%
\begin{equation}
L(q)=-\Delta+q(x),\ x\in\mathbb{R}^{d},\ d\geq2 \tag{1.1}%
\end{equation}
with a periodic (relative to a lattice $\Omega$) potential $q(x),$ where%
\begin{equation}
q(x)\in W_{2}^{s}(F),\text{ }s\geq s_{0}\equiv\frac{3d-1}{2}(3^{d}%
+d+2)+\frac{d3^{d}}{4}+d+6, \tag{1.2}%
\end{equation}
$F\equiv\mathbb{R}^{d}/\Omega$ is a fundamental domain of $\Omega.$ Without
loss of generality it can be assumed that the measure $\mu(F)$ of $F$ is $1$
and $\int_{F}q(x)dx=0.$ Let $L_{t}(q)$ be the operator generated in $L_{2}(F)$
by (1.1) and the quasiperiodic conditions:%
\begin{equation}
u(x+\omega)=e^{i(t,\omega)}u(x),\ \forall\omega\in\Omega, \tag{1.3}%
\end{equation}
where $t\in F^{\star}\equiv\mathbb{R}^{d}/\Gamma$ and $\Gamma$ is the lattice
dual to $\Omega$, that is, $\Gamma$ is the set of all vectors $\gamma
\in\mathbb{R}^{d}$ satisfying $(\gamma,\omega)\in2\pi\mathbb{Z}$ for all
$\omega\in\Omega.$ It is well-known that ( see [2]) the spectrum of $L_{t}(q)$
consists of the eigenvalues $\Lambda_{1}(t)\leq\Lambda_{2}(t)\leq....$The
$n$th band function $\Lambda_{n}(t)$ is continuous with respect to $t$ and its
range $\left\{  \Lambda_{n}(t):t\in F^{\ast}\right\}  $ is $n$th band of the
spectrum $Spec(L)$ of $L$:
\[
Spec(L)=\cup_{n=1}^{\infty}\left\{  \Lambda_{n}(t):t\in F^{\ast}\right\}  .
\]
The normalized eigenfunction $\Psi_{n,t}(x)$ of $L_{t}(q)$ corresponding to
the eigenvalue $\Lambda_{n}(t)$ is known as Bloch functions:%
\begin{equation}
L_{t}(q)\Psi_{n,t}(x)=\Lambda_{n}(t)\Psi_{n,t}(x). \tag{1.4}%
\end{equation}
In the case $q(x)=0$ the eigenvalues and eigenfunctions of $L_{t}(q)$ are
$\mid\gamma+t\mid^{2}$ and $e^{i(\gamma+t,x)}$ for $\gamma\in\Gamma$:%
\begin{equation}
\text{ }L_{t}(0)e^{i(\gamma+t,x)}=\mid\gamma+t\mid^{2}e^{i(\gamma+t,x)}.
\tag{1.5}%
\end{equation}

This paper consists of 6 section. First section is the introduction, where we
describe briefly the scheme of this paper and discuss the related papers.

In the papers [13-17] for the first time the eigenvalues $\left\vert
\gamma+t\right\vert ^{2}$, for large $\ \gamma\in\Gamma,$ were divided into
two groups: non-resonance ( roughly speaking, if $\gamma+t$ far from the
diffraction planes) ones and resonance ( if $\gamma+t$ near a diffraction
plane) ones and for the perturbations of each group various asymptotic
formulae were obtained. To give the precise definition of the non-resonance
and resonance eigenvalue $\left\vert \gamma+t\right\vert ^{2}$ of order
$\rho^{2}$ ( written as $\left\vert \gamma+t\right\vert ^{2}\sim\rho^{2},$ for
definiteness suppose $\gamma+t\in R(\frac{3}{2}\rho)\backslash R(\frac{1}%
{2}\rho)),$ where $R(\rho)=\{x\in\mathbb{R}^{d}:\mid x\mid<\rho\}$) for large
parameter $\rho$ we write the potential $q(x)\in W_{2}^{s}(F)$ in the form
\begin{equation}
q(x)=P(x)+O(\rho^{-p\alpha}),\text{ }P(x)=\sum_{\gamma\in\Gamma(\rho^{\alpha
})}q_{\gamma}e^{i(\gamma,x)}, \tag{1.6}%
\end{equation}
where $p=s-d,$ $\alpha=\frac{1}{\varkappa},$ $\varkappa=3^{d}+d+2,$
$q_{\gamma}=(q(x),e^{i(\gamma,x)})=\int_{F}q(x)e^{-i(\gamma,x)}dx,$

$\Gamma(\rho^{\alpha})=\{\gamma\in\Gamma:0<$ $\mid\gamma\mid<\rho^{\alpha
})\}.$ The relation $\left\vert \gamma+t\right\vert ^{2}\sim\rho^{2}$ means
that there exist a constants $c_{1}$ and $c_{2}$ such that $c_{1}%
\rho<\left\vert \gamma+t\right\vert <c_{2}\rho.$ Here and in subsequent
relations we denote by $c_{i}$ ($i=1,2,...)$ the positive, independent of
$\rho$ constants. Note that the relation $q(x)\in W_{2}^{s}(F)$ ( see (1.2))
means that
\[
\sum_{\gamma\in\Gamma}\mid q_{\gamma}\mid^{2}(1+\mid\gamma\mid^{2s})<\infty.
\]
This implies that if $s\geq d,$ then
\begin{equation}
\sum_{\gamma\in\Gamma}\mid q_{\gamma}\mid<c_{3},\text{ }\sup\mid\sum
_{\gamma\notin\Gamma(\rho^{\alpha})}q_{\gamma}e^{i(\gamma,x)}\mid\leq
\sum_{\mid\gamma\mid\geq\rho^{\alpha}}\mid q_{\gamma}\mid=O(\rho^{-p\alpha}),
\tag{1.7}%
\end{equation}
i.e., (1.6) holds. It follows from (1.6) and (1.7) that the influence of
$q(x)-P(x)$\ to the eigenvalue $\left\vert \gamma+t\right\vert ^{2}$ is
$O(\rho^{-p\alpha}).$ To observe the influence of the trigonometric polynomial
$P(x)$ to the eigenvalue $\left\vert \gamma+t\right\vert ^{2},$ we use the
formula
\begin{equation}
(\Lambda_{N}-\mid\gamma+t\mid^{2})b(N,\gamma)=(\Psi_{N,t}(x)q(x),e^{i(\gamma
+t,x)}), \tag{1.8}%
\end{equation}
where $b(N,\gamma)=(\Psi_{N,t}(x),e^{i(\gamma+t,x)}),$ which is obtained
from\ (1.4) by multiplying by $e^{i(\gamma+t,x)}$ and using (1.5). We say that
(1.8) is the binding formula for $L_{t}(q)$ and $L_{t}(0),$ since it connects
the eigenvalues and eigenfunctions of $L_{t}(q)$ and $L_{t}(0)$. Introducing
into (1.8) the expansion (1.6) of $q(x)$, we get
\begin{equation}
(\Lambda_{N}(t)-\mid\gamma+t\mid^{2})b(N,\gamma)=\sum_{\gamma_{1}\in
\Gamma(\rho^{\alpha})}q_{\gamma_{1}}b(N,\gamma-\gamma_{1})+O(\rho^{-p\alpha}).
\tag{1.9}%
\end{equation}
If $\Lambda_{N}$ is close to \ $\mid\gamma+t\mid^{2}$ and $\gamma+t$ does not
belong to any of the sets%
\begin{equation}
V_{\gamma_{1}}(\rho^{\alpha_{1}})\equiv\{x\in\mathbb{R}^{d}:\mid\mid x\mid
^{2}-\mid x+\gamma_{1}\mid^{2}\mid\leq\rho^{\alpha_{1}}\}\cap(R(\frac{3\rho
}{2})\backslash R(\frac{\rho}{2})) \tag{1.10}%
\end{equation}
for $\gamma_{1}\in\Gamma(\rho^{\alpha}),$ where $\alpha_{1}=3\alpha,$ that is,
$\gamma+t$ far from the diffraction planes

$\{x\in\mathbb{R}^{d}:\mid x\mid^{2}-\mid x+\gamma_{1}\mid^{2}=0\}$ for
$\gamma_{1}\in\Gamma(\rho^{\alpha}),$ then
\begin{equation}
\mid\mid\gamma+t\mid^{2}-\mid\gamma-\gamma_{1}+t\mid^{2}\mid>\rho^{\alpha_{1}%
},\text{ }\mid\Lambda_{N}(t)-\mid\gamma-\gamma_{1}+t\mid^{2}\mid>\frac{1}%
{2}\rho^{\alpha_{1}}\text{ } \tag{1.11}%
\end{equation}
for all $\gamma_{1}\in\Gamma(\rho^{\alpha}).$ Therefore, it follows from (1.8)
that
\begin{equation}
b(N,\gamma-\gamma_{1})=\dfrac{(\Psi_{N,t}(x)q(x),e^{i(\gamma-\gamma_{1}%
+t,x)})}{\Lambda_{N}(t)-\mid\gamma-\gamma_{1}+t\mid^{2}}=O(\rho^{-\alpha_{1}%
}). \tag{1.12}%
\end{equation}
This with the first inequality of (1.7) implies that the right-hand side of
(1.9) is $O(\rho^{-\alpha_{1}}).$ Moreover we prove that there exist an index
$N$ such that $\frac{1}{b(N,\gamma)}$ times the right-hand side of (1.9) is
$O(\rho^{-\alpha_{1}}),$ i.e.,
\begin{equation}
\Lambda_{N}(t)=\mid\gamma+t\mid^{2}+O(\rho^{-\alpha_{1}}). \tag{1.13}%
\end{equation}
Thus we see that if $\gamma+t$ does not belong to any of the sets
$V_{\gamma_{1}}(\rho^{\alpha_{1}})$ ( see (1.10)) for $\gamma_{1}\in
\Gamma(\rho^{\alpha}),$ then the influence of the trigonometric polynomial
$P(x)$ and hence the influence of the potential $q(x)$ ( see (1.6)) to the
eigenvalue $\left\vert \gamma+t\right\vert ^{2}$ is not significant and there
exists an eigenvalue of the operator $L_{t}(q)$ satisfying (1.13). This case
is called the non-resonance case. More precisely, we give the following definitions:

\begin{definition}
Let $\rho$ be a large parameter, $\alpha_{k}=3^{k}\alpha$ for $k=1,2,...,$ and%
\[
V_{\gamma_{1}}(c_{4}\rho^{\alpha_{1}})\equiv\{x\in\mathbb{R}^{d}:\mid\mid
x\mid^{2}-\mid x+\gamma_{1}\mid^{2}\mid\leq c_{4}\rho^{\alpha_{1}}%
\}\cap(R(\frac{3}{2}\rho)\backslash R(\frac{1}{2}\rho),
\]%
\[
E_{1}(c_{4}\rho^{\alpha_{1}},p)\equiv\bigcup_{\gamma_{1}\in\Gamma
(p\rho^{\alpha})}V_{\gamma_{1}}(c_{4}\rho^{\alpha_{1}}),\text{ }U(c_{4}%
\rho^{\alpha_{1}},p)\equiv(R(\frac{3}{2}\rho)\backslash R(\frac{1}{2}%
\rho))\backslash E_{1}(c_{4}\rho^{\alpha_{1}},p),
\]%
\[
E_{k}(c_{4}\rho^{\alpha_{k}},p)\equiv\bigcup_{\gamma_{1},\gamma_{2}%
,...,\gamma_{k}\in\Gamma(p\rho^{\alpha})}(\cap_{i=1}^{k}V_{\gamma_{i}}%
(c_{4}\rho^{\alpha_{k}})),
\]
where $p$ is defined in (1.6), the intersection $\cap_{i=1}^{k}V_{\gamma_{i}}$
in the definition of $E_{k}$ is taken over $\gamma_{1},\gamma_{2}%
,...,\gamma_{k}$ that are linearly independent. The set $U(\rho^{\alpha_{1}%
},p)$ is said to be a non-resonance domain and $\left\vert \gamma+t\right\vert
^{2}$ is called a non-resonance eigenvalue if $\gamma+t\in U(\rho^{\alpha_{1}%
},p).$ The domains $V_{\gamma_{1}}(\rho^{\alpha_{1}})$ for $\gamma_{1}%
\in\Gamma(p\rho^{\alpha})$ are called resonance domains and $\mid\gamma
+t\mid^{2}$ is called a resonance eigenvalue if $\gamma+t\in V_{\gamma_{1}%
}(\rho^{\alpha_{1}}).$ The domain $V_{\gamma_{1}}^{^{\prime}}(\rho^{\alpha
_{1}})\equiv V_{\gamma_{1}}(\rho^{\alpha_{1}})\backslash E_{2}$, i.e., the
part of the resonance domains $V_{\gamma_{1}}(\rho^{\alpha_{1}}),$ which does
not contain the intersection of two resonance domains is called a single
resonance domain.
\end{definition}

It is clear that the asymptotic formula (1.13) hold true if we replace
$V_{\gamma_{1}}(\rho^{\alpha_{1}})$ by $V_{\gamma_{1}}(c_{4}\rho^{\alpha_{1}%
}).$ Note that changing the value of $c_{4}$\ in the definition of
$V_{\gamma_{1}}(c_{4}\rho^{\alpha_{1}}),$ we obtain the different definitions
of the non-resonance eigenvalues ( for simplicity of notation we take
$c_{4}=1)$. However, in any case we obtain the same asymptotic formulas and
the same perturbation theory, that is, this changing does not change anything
for asymptotic formulas. Therefore we can define the non-resonance eigenvalue
in different way. In papers [15-17] instead of the resonance domain
$V_{\gamma_{1}}(c_{4}\rho^{\alpha_{1}})$ the set

$W_{\gamma_{1},\alpha_{1}}=\{x\in\mathbb{R}^{d}:\mid\mid x\mid^{2}-\mid
x+\gamma_{1}\mid^{2}\mid<\mid x\mid^{\alpha_{1}}\}$ is considered. Since
\[
V_{\gamma_{1}}(\frac{1}{2}\rho^{\alpha_{1}})\subset(R(\frac{3}{2}%
\rho)\backslash R(\frac{1}{2}\rho))\cap W_{\gamma_{1},\alpha_{1}}\subset
V_{\gamma_{1}}(\frac{3}{2}\rho^{\alpha_{1}}),
\]
\ in all considerations the domain $V_{\gamma_{1}}(\rho^{\alpha_{1}})$ can be
replaced by $W_{\gamma_{1},\alpha_{1}}\cap(R(\frac{3}{2}\rho)\backslash
R(\frac{1}{2}\rho)).$ In my first papers [13,14] instead of the domain
$V_{\gamma_{1}}(\rho^{\alpha_{1}})$ the cone $\{x\in\mathbb{R}^{d}%
:\mid(x,\gamma_{1})\mid<\varepsilon\mid x\mid\mid\gamma_{1}\mid\},$ where
$\varepsilon\ll1,$ is considered. In any case we use the same idea: roughly
speaking, the eigenvalues $\left\vert \gamma+t\right\vert ^{2}$, for large
$\ \gamma\in\Gamma,$ is non-resonance if $\gamma+t$ far from the diffraction
planes. Nevertheless it is suitable to define the non-resonance eigenvalue in
different way depending on the form of the potential. Namely, the domain
$W_{\gamma_{1},\alpha_{1}}$ is suitable, when the potential is the
trigonometric polynomial. In case of smooth potential we need to introduce a
large parameter $\rho$ and consider $V_{\gamma_{1}}(\rho^{\alpha_{1}}).$ Note
that all considered eigenvalues $\left\vert \gamma+t\right\vert ^{2}$ of
$L_{t}(0)$ satisfy the relations $\frac{1}{2}\rho<\left\vert \gamma
+t\right\vert <\frac{3}{2}\rho.$ Therefore in the asymptotic formulas instead
of $O(\rho^{a})$ one can take \ $O(\left\vert \gamma+t\right\vert ^{a}).$

In section 2 to investigate the perturbation of the non-resonance eigenvalues
$\mid\gamma+t\mid^{2}$ we take the operator $L_{t}(0)$ for an unperturbed
operator and $q(x)$ for a perturbation. Iterating the binding formula (1.8)
for $L_{t}(q)$ and $L_{t}(0),$ namely, using (1.12) in (1.9) and then using
the decomposition (1.6) and continuing this process, we prove that (1.13) and
an asymptotic formulas of arbitrary order hold. More precisely, we obtain the
following results. For each $\gamma+t\in U(\rho^{\alpha_{1}},p)$ there exists
an eigenvalue $\Lambda_{N}(t)$ of the operator $L_{t}(q)$ satisfying the
formulae
\begin{equation}
\Lambda_{N}(t)=\mid\gamma+t\mid^{2}+F_{k-1}(\gamma+t)+O(\mid\gamma
+t\mid^{-k\alpha_{1}}) \tag{1.14}%
\end{equation}
for $k=1,2,...,[\frac{1}{3}(p-\frac{1}{2}\varkappa(d-1))],$ where $[a]$
denotes the integer part of $a,$ $F_{0}=0,$ and $F_{k-1}$ for $k>1$ is
expressed by the potential $q(x)$ and the eigenvalues of $L_{t}(0).$ Besides,
we prove that if the conditions
\begin{align}
&  \mid\Lambda_{N}(t)-\mid\gamma+t\mid^{2}\mid<\frac{1}{2}\rho^{\alpha_{1}%
},\tag{1.15}\\
&  \mid b(N,\gamma)\mid>c_{5}\rho^{-c\alpha}, \tag{1.16}%
\end{align}
where $0\leq c<p-\frac{1}{4}d3^{d},$ hold, then $\ $the following statements
are valid:

(a) if $\gamma+t\in U(\rho^{\alpha_{1}},p)$, then $\Lambda_{N}(t)$ satisfies
(1.14) for $k=1,2,...,[\frac{1}{3}(p-c)]$ ;

(b) if $\gamma+t\in E_{s}\backslash E_{s+1},$ where $s=1,2,...,d-1,$ then%
\begin{equation}
\Lambda_{N}(t)=\lambda_{j}(\gamma+t)+O(\mid\gamma+t\mid^{-(p-c-\frac{1}%
{4}d3^{d})\alpha}), \tag{1.17}%
\end{equation}
where $\lambda_{j}$ is an eigenvalue of a matrix $C(\gamma+t)$ ( see below for
the explanation of $C$ \ in the three-dimensional case). Moreover, we prove
that every large eigenvalue of the operator $L_{t}(q)$ for all values of $t$
satisfies one of these formulae ( see Theorem 2.1 and Theorem 2.2).

The results of section 2 is considered in [15,17]. However, in those paper
these results are written only briefly. Here we write the non-resonance case
in an improved and enlarged form and so that it can easily be used in the next
sections. The non-resonance eigenvalues for the three-dimensional
Schr\"{o}dinger operator $L_{t}(q)$ was considered in [16]. Moreover, in [16]
we observed that if $\gamma+t\in V_{\delta}(\rho^{\alpha_{1}})\backslash
E_{2}$ and $\gamma_{1}\in\Gamma(\rho^{\alpha})\backslash\{n\delta
:n\in\mathbb{Z}\},$ where $\delta$ is the element of \ $\Gamma$ of minimal
norm in its direction, then it follows from the definition of $E_{2}$ that the
inequalities obtained from (1.11) by replacing $\alpha_{1}$ with $\alpha_{2}$
hold. Hence

$b(N,\gamma-\gamma_{1})=$ $O(\rho^{-\alpha_{2}})$ (see (1.12)) and (1.9) has
the form
\begin{equation}
(\Lambda_{N}(t)-\mid\gamma+t\mid^{2})b(N,\gamma)=\sum_{n\in\mathbb{Z}%
,n\delta\in\Gamma(\rho^{\alpha})}q_{n\delta}b(N,\gamma-n\delta)+O(\frac
{1}{\rho^{\alpha_{2}}}). \tag{1.18}%
\end{equation}
This gives an idea that the influence of $q(x)-q^{\delta}(x),$ where%
\begin{equation}
q^{\delta}(x)=\sum_{n\in\mathbb{Z}}q_{n\delta}e^{in(\delta,x)} \tag{1.19}%
\end{equation}
is the directional potential, is not significant and\ there exist eigenvalues
of $L_{t}(q)$ which are close to the eigenvalues of $L_{t}(q^{\delta}).$ Note
that in [16] ( see Theorem 2 of [16]) writing the equations obtained from
(1.18) by replacing $\mid\gamma+t\mid^{2}$ with $\mid\gamma+t+n\delta\mid^{2}$
for $n\in\mathbb{Z},$ $n\delta\in\Gamma(\rho^{\alpha}),$ we got the system
from which we conclude that the probable approximations, besides $\mid
\gamma+t\mid^{2}$, of the eigenvalues of the three-dimensional Schr\"{o}dinger
operator $L_{t}(q)$ are the eigenvalue of the matrix $C,$ where $C$ is a
finite submatrix of the matrix corresponding to the operator $L_{t}(q^{\delta
}).$ However, in the $d$-dimensional case, to investigate the perturbation of
the eigenvalue $\mid\gamma+t\mid^{2}$ when corresponding quasimomentum
$\gamma+t$ lies in intersection of $k$ resonance domains we have to consider
more complicated system and matrix (see (2.15) and [15,17]). In [13,14] to
investigate the non-resonance and resonance eigenvalues we used the
approximation of the Green functions of $L_{t}(q)$\ by the Green functions of
$L_{t}(0)$\ and $L_{t}(q^{\delta})$ respectively.

Thus, in section 2 we write the asymptotic formulas obtained in [15,17] an
improved and enlarged form. Moreover it helps to read section 3, where we
consider in detail the single resonance domains $V_{\delta}(\rho^{\alpha_{1}%
})\backslash E_{2}$, since there are similarities between investigations \ of
the non-resonance and the single resonance case. To see the similarities and
differences between the non-resonance case and the single resonance case, that
is, between the section 2 and section 3, let us give the following comparison.
As we noted above in the non-resonance case the influence of the potential
$q(x)$ is not significant, while in the single resonance case the influence of
$q(x)-q^{\delta}(x)$ is not significant. Therefore, in the section 2 for the
investigation of the non-resonance case we take the operator$L_{t}(0)$ for an
unperturbed operator and $q(x)$ for a perturbation, while in the section 3 for
investigation of the single resonance case we take the operator $L_{t}%
(q^{\delta})$ for an unperturbed operator and $q(x)-q^{\delta}(x)$ for a
perturbation. In section 2 to obtain the asymptotic formula for the
non-resonance case we iterate the formula (1.8) (called binding formula for
$L_{t}(q)$ and $L_{t}(0)$) connecting the eigenvalues and eigenfunctions of
$L_{t}(q)$ and $L_{t}(0).$ Similarly, in section 3 for investigation of the
eigenvalues corresponding to the quasimomentum lying in the single resonance
domain $V_{\delta}(\rho^{\alpha_{1}})\backslash E_{2}$ ( see Definition 1.1),
we iterate a formula (called binding formula for $L_{t}(q)$ and $L_{t}%
(q^{\delta})$) connecting the eigenvalues and eigenfunctions of $L_{t}(q)$ and
$L_{t}(q^{\delta})$. The binding formula for $L_{t}(q)$ and $L_{t}(q^{\delta
})$ can be obtained from the binding formula (1.8) for $L_{t}(q)$ and
$L_{t}(0)$ by replacing the perturbation $q(x)$ and the eigenvalues
$\mid\gamma+t\mid^{2}$, the eigenfunctions $e^{i(\gamma+t,x)}$ of the
unperturbed ( for the non-resonance case) operator $L_{t}(0)$ with the
perturbation $q(x)-q^{\delta}(x)$ and the eigenvalues, the eigenfunctions of
the \ unperturbed ( for the single resonance case) operator $L_{t}(q^{\delta
})$ respectively. To write this formula first we consider the eigenvalues and
eigenfunctions of $L_{t}(q^{\delta})$. The eigenvalues of $L_{t}(q^{\delta})$
can be indexed by pair $($ $j,$ $\beta)$ of the Cartesian product
$\mathbb{Z}\times\Gamma_{\delta}:$
\begin{equation}
L_{t}(q^{\delta})\Phi_{j,\beta}(x)=\lambda_{j,\beta}\Phi_{j,\beta}(x),
\tag{1.20}%
\end{equation}
where $\Gamma_{\delta}$ is the dual lattice of $\Omega_{\delta}$ and
$\Omega_{\delta\text{ }}$ is the sublattice $\{h\in\Omega:(h,\delta)=0\}$ of
$\Omega$ in the hyperplane $H_{\delta}=$ $\{x\in\mathbb{R}^{d}:(x,\delta)=0\}$
( see Lemma 3.1 ). Thus the binding formula for $L_{t}(q)$ and $L_{t}%
(q^{\delta})$ is
\begin{equation}
(\Lambda_{N}(t)-\lambda_{j,\beta})b(N,j,\beta)=(\Psi_{N,t}(x),(q(x)-q^{\delta
}(x))\Phi_{j,\beta}(x)), \tag{1.21}%
\end{equation}
where $b(N,j,\beta)=(\Psi_{N,t}(x),\Phi_{j,\beta}(x)),$ which can be obtained
from (1.4) by multiplying by $\Phi_{j,\beta}(x)$ and using (1.20). To prove
the asymptotic formulas in the single resonance case we\ iterate the formula
(1.21). The iterations of the formulas (1.8) and (1.21) are similar. Therefore
the simple iterations of (1.8) in section 2 helps to read the complicated
iteration of (1.21) in section 3.\ The brief and rough scheme of the iteration
of (1.21) is following. Using (1.6), decomposing $(q(x)-q^{\delta}%
(x))\Phi_{j,\beta}(x)$ by eigenfunction of $L_{t}(q^{\delta})$ and putting
this decomposition into (1.21), we get
\[
(\Lambda_{N}(t)-\lambda_{j,\beta})b(N,j,\beta)=O(\rho^{-p\alpha})
\]

\begin{equation}
+\sum\limits_{(j_{1},\beta_{1})\in Q}A(j,\beta,j+j_{1,}\beta+\beta
_{1})b(N,j+j_{1},\beta+\beta_{1}), \tag{1.22}%
\end{equation}
where $Q$ is a subset of the Cartesian product $\mathbb{Z}\times\Gamma
_{\delta}.$ Now using%

\[
b(N,j+j_{1},\beta+\beta_{1})=\frac{(\Psi_{N,t}(x),(q(x)-q^{\delta}%
(x))\Phi_{j,\beta}(x))}{(\Lambda_{N}(t)-\lambda_{j+j_{1},\beta+\beta_{1}})},
\]
which is obtained from (1.21) by replacing $j,\beta$ with $j+j_{1},\beta
+\beta_{1},$ in (1.22), we get the one times iteration of (1.21):
\begin{equation}
(\Lambda_{N}(t)-\lambda_{j,\beta})b(N,j,\beta)=O(\rho^{-p\alpha})+ \tag{1.23}%
\end{equation}

\[
\sum\limits_{(j_{1},\beta_{1})\in Q}A(j,\beta,j+j_{1,}\beta+\beta_{1}%
)\frac{(\Psi_{N,t}(x),(q(x)-q^{\delta}(x))\Phi_{j,\beta}(x))}{(\Lambda
_{N}(t)-\lambda_{j+j_{1},\beta+\beta_{1}})}.
\]
Continuing this process we get the iterations of (1.21). Then we prove the
asymptotic formulas, by using the iterations of (1.21), as follows. First we
investigate, in detail, the multiplicand $A(j,\beta,j+j_{1,}\beta+\beta_{1})$
of (1.23) and prove the estimation%
\begin{equation}
\sum\limits_{(j_{1},\beta_{1})\in Q}\mid A(j,\beta,j+j_{1,}\beta+\beta
_{1})\mid<c_{6} \tag{1.24}%
\end{equation}
( see Lemma 3.2, Lemma 3.3, see Lemma 3.4). Then we investigate the distance
between eigenvalues $\lambda_{j,\beta}$ and $\lambda_{j+j_{1},\beta+\beta_{1}%
}$ ( see Lemma 3.5) and hence estimate the denominator of the fractions in
(1.23), since $\Lambda_{N}(t)$ is close to $\lambda_{j,\beta}$. Using this and
the estimation (1.24) we prove that there exists an index $N$ such that
$\frac{1}{b(N,j,\beta)}$ times the right-hand side of (1.23) is $O(\rho
^{-\alpha_{2}})$, from which we get
\begin{equation}
\Lambda_{N}(t)=\lambda_{j,\beta}+O(\rho^{-\alpha_{2}}) \tag{1.25}%
\end{equation}
( see Lemma 3.6, Theorem 3.1). At last using this formula in the arbitrary
times iterations of (1.21), we obtain the asymptotic formulas of arbitrary
order ( Theorem 3.2).

In Section 4 we investigate the Bloch function in the non-resonance domain. To
investigate the Bloch function we need to find the values of quasimomenta
$\gamma+t$ for which the corresponding eigenvalues of $L_{t}(q)$ are simple.
In the interval $(\rho^{2},\rho^{2}+1)$ of length $1$ there are , in average,
$\rho^{d-2}$ eigenvalues $\mid\gamma+t\mid^{2}$ of the unperturbed operator
$L_{t}(0).$ Under perturbation, all these eigenvalues move and some of them
move or order $1.$ Therefore, it seems it is impossible to find the values of
quasimomenta $\gamma+t$ for which the corresponding eigenvalues of $L_{t}(q)$
are simple. For the first time in papers [15-17] (in [16] for $d=3$ and in
[15,17] for the cases: $d=2,$ $q(x)\in L_{2}(F)$ and $d>2,$ $q(x)$ is a smooth
potential) we found the required values of quasimomenta, namely we constructed
the subset $B$ of $U(\rho^{\alpha_{1}},p)$ with the following property:

\textbf{Property 1 (Simplicity)}. If $\gamma+t\in B,$ then there exists a
unique eigenvalue $\Lambda_{N}(t)$, denoted by $\Lambda(\gamma+t),$ of the
operator $L_{t}(q)$ satisfying (1.13), (1.14). This is a simple eigenvalue of
$L_{t}(q)$. Therefore we call the set $B$ the simple set.

Construction of the set $B$ consists of two steps.

\textbf{Step 1.} We prove that all eigenvalues $\Lambda_{N}(t)\sim\rho^{2}$ of
the operator $L_{t}(q)$ lie in the $\varepsilon_{1}$ neighborhood of the
numbers $F(\gamma+t)$ and $\lambda_{j}(\gamma+t),$ where%
\begin{equation}
F(\gamma+t)=\mid\gamma+t\mid^{2}+F_{k_{1}-1}(\gamma+t),\text{ }\varepsilon
_{1}=\rho^{-d-2\alpha},\text{ }k_{1}=[\frac{d}{3\alpha}]+2 \tag{1.26}%
\end{equation}
( see (1.14), (1.17)). We call these numbers as the known parts of the
eigenvalues of $L_{t}(q)$. Moreover, for $\gamma+t\in U(\rho^{\alpha_{1}},p)$
there exists $\Lambda_{N}(t)$ satisfying
\begin{equation}
\Lambda_{N}(t)=F(\gamma+t)+o(\rho^{-d-2\alpha})=F(\gamma+t)+o(\varepsilon
_{1}). \tag{1.27}%
\end{equation}

\textbf{Step 2.} By eliminating the set of quasimomenta $\gamma+t$, for which
the known parts $F(\gamma+t)$ of $\Lambda_{N}(t)$ are situated from the known
parts $F(\gamma^{^{\prime}}+t),$ $\lambda_{j}(\gamma^{^{\prime}}+t)$
($\gamma^{^{\prime}}\neq\gamma)$ of other eigenvalues at a distance less than
$2\varepsilon_{1},$ we construct the set $B$ with the following properties: if
$\gamma+t\in B,$ then the following conditions (called simplicity conditions
for the eigenvalue $\Lambda_{N}(t)$ satisfying (1.27))
\begin{equation}
\mid F(\gamma+t)-F(\gamma^{^{\prime}}+t)\mid\geq2\varepsilon_{1}\text{ }
\tag{1.28}%
\end{equation}
for $\gamma^{^{\prime}}\in K\backslash\{\gamma\},$ $\gamma^{^{\prime}}+t\in
U(\rho^{\alpha_{1}},p)$ and%
\begin{equation}
\mid F(\gamma+t)-\lambda_{j}(\gamma^{^{\prime}}+t)\mid\geq2\varepsilon_{1}
\tag{1.29}%
\end{equation}
for $\gamma^{^{\prime}}\in K,\gamma^{^{\prime}}+t\in E_{k}\backslash E_{k+1},$
$j=1,2,...,$ where $K$ is the set of $\gamma^{^{\prime}}\in\Gamma$ satisfying
\begin{equation}
\mid F(\gamma+t)-\mid\gamma^{^{\prime}}+t\mid^{2}\mid<\frac{1}{3}\rho
^{\alpha_{1}}, \tag{1.30}%
\end{equation}
hold. Thus the the simple set $B$ is defined as follows:

\begin{definition}
The simple set $B$ is the set of $x\in U(\rho^{\alpha_{1}},p)\cap(R(\frac
{3}{2}\rho-\rho^{\alpha_{1}-1})\backslash R(\frac{1}{2}\rho+\rho^{\alpha
_{1}-1}))$ such that $x=\gamma+t,$ where $\gamma\in\Gamma,t\in F^{\star},$ and
the simplicity conditions (1.28), (1.29) hold.
\end{definition}

As a consequence of the conditions (1.28), (1.29) the eigenvalue $\Lambda
_{N}(t)$ satisfying (1.27) does not coincide with other eigenvalues.

To check the simplicity of $\Lambda_{N}(t)\equiv\Lambda(\gamma+t)$ ( see
Property 1) we prove that for any normalized eigenfunction $\Psi_{N,t}(x)$
corresponding to $\Lambda_{N}(t)$ the equality
\begin{equation}
\sum_{\gamma^{^{\prime}}\in\Gamma\backslash\gamma}\mid b(N,\gamma^{^{\prime}%
})\mid^{2}=O(\rho^{-2\alpha_{1}}), \tag{1.31}%
\end{equation}
which equivalent to
\begin{equation}
\mid b(N,\gamma)\mid^{2}=1+O(\rho^{-2\alpha_{1}}), \tag{1.31a}%
\end{equation}
holds. The equality (1.31a) implies the simplicity of \ $\Lambda_{N}(t).$
Indeed, if \ $\Lambda_{N}(t)$ is multiple eigenvalue, then there exist two
orthogonal normalized eigenfunctions satisfying (1.31a), which is impossible.
In fact to prove the simplicity of $\Lambda_{N}(t)$ it is enough to show that
for any normalized eigenfunction $\Psi_{N,t}(x)$ corresponding to $\Lambda
_{N}(t)$ the inequality%
\begin{equation}
\mid b(N,\gamma)\mid^{2}>\frac{1}{2} \tag{1.31b}%
\end{equation}
holds. We proved this inequality in [15-17] and as noted in Theorem 3 of [16]
and in [18] the proof of this inequality does not differ from the proof of
(1.31a) which equivalent to the following property:

\textbf{Property 2 (Asymptotic formulas for the Bloch function).} If
$\gamma+t\in B,$ then the eigenfunction $\Psi_{N,t}(x),$ denoted by
$\Psi_{\gamma+t}(x),$ corresponding to the eigenvalue $\Lambda_{N}%
(t)\equiv\Lambda(\gamma+t)$ ( see property 1) is close to $e^{i(\gamma+t,x)}$,
namely
\begin{equation}
\Psi_{N,t}(x)\equiv\Psi_{\gamma+t}(x)=e^{i(\gamma+t,x)}+O(\mid\gamma
+t\mid^{-\alpha_{1}}). \tag{1.32}%
\end{equation}

From (1.32), by iteration, we get
\begin{equation}
\Psi_{\gamma+t}(x)=F_{k-1}^{\ast}(\gamma+t)+O(\mid\gamma+t\mid^{-k\alpha_{1}})
\tag{1.33}%
\end{equation}
for $k=1,2,...$ , where $F_{k-1}^{\ast}(\gamma+t)$ is expressed by $q(x)$ and
by the eigenvalues and eigenfunctions of $L_{t}(0)$ ( see Theorem 4.2, formula
(4.20), and [18])).

Note that the main difficulty and the crucial point of the investigation of
the Bloch functions and hence the main difficulty of the perturbation theory
of $L(q)$\ is the construction of the simple set $B$. This difficulty of the
perturbation theory of $L(q)$ is of a physical nature and it is connected with
the complicated picture of the crystal diffraction. In the multidimensional
case this becomes extremely difficult since in the $1$ neighborhood of
$\rho^{2}$ there are , in average, $\rho^{d-2}$ eigenvalues and hence the
eigenvalues can be highly degenerate. To see that the main part of the
perturbation theory is the construction of the set $B$ let us briefly prove
that ( the precise proof is given in Theorem 4.1) from the construction of $B$
it easily follows the simplicity of the eigenvalues and the asymptotic formula
(1.32) for Bloch function. \ As we noted above to prove the simplicity of
$\Lambda_{N}(t)$ and the asymptotic formula (1.32) it is enough to prove that
(1.31) holds, that is, we need to prove that the terms $b(N,\gamma^{^{\prime}%
})$ in (1.31) is very small. If $\mid b(N,\gamma^{^{\prime}})\mid>c_{5}%
\rho^{-c\alpha}$ , then in (1.15), (1.16), (1.14), (1.17), (1.27) replacing
$\gamma$ by $\gamma^{^{\prime}},$ we see that $\Lambda_{N}(t)$ lies in
$\varepsilon_{1}$ neighborhood of one of the numbers $F(\gamma^{^{\prime}}+t)$
and $\lambda_{j}(\gamma^{^{\prime}}+t),$ which contradicts to the simplicity
conditions (1.28), (1.29), since (1.27) holds.

Since the main part of the perturbation theory is the construction of the set
$B$ let us discuss the construction and the history of the construction of the
simple set. For the first time in [15-17] we constructed the simple set $B$.
In [16] we constructed the simple set for the three dimensional
Schr\"{o}dinger operator $L(q).$ If $d=2,3,$ then the simplicity conditions
(1.28), (1.29) are relatively simple, namely in this case $F(\gamma
+t)=\mid\gamma+t\mid^{2}$ and the matrix $C(\gamma^{^{\prime}}+t),$ when
$\gamma^{^{\prime}}+t$ lies in the single resonance domain, corresponds to the
Schr\"{o}dinger operator with directional potential (1.19) ( see Theorem 1 and
2 in [16]). Therefore the simple set is constructed in such way that if
$\gamma+t\in B,$ then the inequality%
\begin{equation}
\mid\mid\gamma+t\mid^{2}-\mid\gamma^{^{\prime}}+t\mid^{2}\mid\geq\rho
^{-a}\text{ } \tag{1.34}%
\end{equation}
for $\gamma^{^{\prime}}+t\in U(\rho^{\alpha_{1}},p)$, the inequality%
\begin{equation}
\mid\mid\gamma+t\mid^{2}-\lambda_{j}(\gamma^{^{\prime}}+t)\mid\geq\rho^{-a}
\tag{1.35}%
\end{equation}
\ for $\gamma^{^{\prime}}+t$ lying in single resonance domain, and the
inequality
\begin{equation}
\mid\mid\gamma+t\mid^{2}-\mid\gamma^{^{\prime}}+t\mid^{2}\mid\geq c_{3}
\tag{1.36}%
\end{equation}
for $\gamma^{^{\prime}}+t$ lying in the intersection of two resonance domains
hold, where $a>0.$ Thus for construction of the simple set $B$ of quasimomenta
in case $d=3$ we eliminated the vicinities of the diffraction planes ( see
(1.34)), the sets connected with directional potential ( see (1.35)), and the
intersection of two resonance domains ( see (1.36)).

As dimension $d$ increases, the geometrical structure of $B$ becomes more
complicated for the following reason. Since the denseness of the eigenvalues
of the free operator increases as $d$ increases we need use the asymptotic
formulas of high accuracy and investigate the intersections of high order of
the resonance domains. Then the functions $F(\gamma+t),$ $\lambda_{j}%
(\gamma+t)$ ( see (1.28), (1.29)) taking part in the construction of $B$ ( see
definition 1.2) becomes more complicated. Therefore surfaces and sets defined
by these functions becomes more intricate. Besides of this construction in
[15] we gave the additional idea for nonsmooth potential, namely\ for
construction of the simple set $B$ for nonsmooth potentials $q(x)\in
L_{2}(\mathbb{R}^{2}/\Omega),$ we eliminated additionally a set, which is
described in the terms of the number of states ( see [15] page 47 and
[19,20]). More precisely, we eliminated the translations $A_{k}^{^{\prime}}$
of the set $A_{k}$ by vectors $\gamma\in\Gamma,$ where
\[
A_{1}=\{x:N_{x}(K_{\rho}(\frac{M_{0}}{\rho}))>b_{1},\text{ }A_{k}%
=\{x:N_{x}(K_{\rho}(\frac{2^{k-1}M_{0}}{\rho})\backslash K_{\rho}%
(\frac{2^{k-2}M_{0}}{\rho}))>b_{k}\},
\]
$M_{0}\gg1,b_{1}=(M_{0})^{\frac{3}{2}},$ $b_{k}=(2^{k}M_{0})^{\frac{3}{2}%
},k\geq2,$ $K_{\rho}(a)=\{x:\mid\mid x\mid-\rho\mid<a\}$ and $N_{x}(A)$ is the
number of the vectors $\gamma+x$ lying in $A.$ These eliminations imply that
if $\gamma+t$ is in the simple set then the number of vectors $\gamma
^{^{\prime}}$ in $A_{k}$ less than or equal to $b_{k}.$ On the other hand
using the formula (1.8) it can be proved that $\mid b(N,\gamma^{^{\prime}%
})\mid^{2}=O((2^{k}M_{0})^{-2}).$ As a result the left-hand side of (1.31)
becomes $o(1),$ which implies the simplicity of $\Lambda(\gamma+t)$ and the
closest of the functions $\Psi_{\gamma+t}(x),$ $e^{i(\gamma+t,x)}$. The simple
sets $B$ of quasimomenta for the first time is constructed and investigated (
hence the main difficulty and the crucial point of perturbation theory of
$L(q)$ is investigated) in [16] for $d=3$ and in [15] for the cases: 1. $d=2,$
$q(x)\in L_{2}(F);$ \ \ \ 2. $d>2,$ $q(x)$ is a smooth potential.

Then, Yu. E. Karpeshina proved ( see [6,7]) the convergence of the
perturbation series of two and three dimensional Schr\"{o}dinger operator
$L(q)$ with a wide class of nonsmooth potential $q(x)$ for a set, that is
similar to $B$ (see the section of geometric construction in [6] and footnote
in the page 110 in [7]). In [3] the asymptotic formulas for the eigenvalues
and Bloch function of the two and three dimensional operator $L_{t}(q)$ were
obtained by investigation of the corresponding infinity matrix.

In section 5 we consider the geometrical aspects of the simple set of the
Schr\"{o}dinger operator of arbitrary dimension. We prove that the simple sets
$B$ has asymptotically full measure on $\mathbb{R}^{d}$. Moreover, we
construct a part of isoenergetic surfaces $\{t\in F^{\ast}:\exists
N,\Lambda_{N}(t)=\rho^{2}\}$ corresponding to $\rho^{2},$ which is smooth
surfaces and has the measure asymptotically close to the measure of the
isoenergetic surfaces $\{t\in F^{\ast}:\exists\gamma\in\Gamma,\mid\gamma
+t\mid^{2}=\rho^{2}\}$ of the operator $L(0).$ For this we prove that the set
$B$ has the following third property:

\textbf{Property 3 (Intercept with the isoenergetic surface)}. The set $B$
contains the intervals $\{a+sb:s\in\lbrack-1,1]\}$ such that $\Lambda
(a-b)<\rho^{2},$ $\Lambda(a+b)>\rho^{2}$ and hence there exists $\gamma+t$
such that $\Lambda(\gamma+t)=\rho^{2},$ since in the intervals $\{a+sb:s\in
\lbrack-1,1]\}\subset B$\ the eigenvalue $\Lambda(\gamma+t)$ \ is simple ( see
Property 1) and the function $\Lambda(x)$ is continuous on these intervals.

Using this idea we construct the part of the isoenergetic surfaces. The
nonemptyness of the isoenergetic surfaces \ for $\rho\gg1$ implies that there
exist only a finite number of gaps in the spectrum of $L,$ that is, it implies
the validity of the Bethe-Sommerfeld conjecture for arbitrary dimension and
for arbitrary lattice.

For the first time M. M. Skriganov [11,12] proved the validity of the
Bethe-Sommerfeld conjecture for the Schr\"{o}dinger operator for dimension
$d=2,3$ for arbitrary lattice, for dimension $d>3$ for rational lattice. The
Skriganov's method is based on the detail investigation of the arithmetic and
geometric properties of the lattice. B. E. J. Dahlberg \ and E. Trubowits [1]
using an asymptotic of Bessel function, gave the simple proof of this
conjecture for the two dimensional Schr\"{o}dinger operator. Then in papers
[15-17] we proved the validity of the Bethe-Sommerfeld conjecture for
arbitrary lattice and for arbitrary dimension by using the asymptotic formulas
and by construction of the simple set $B,$ that is, by\ the method of
perturbation theory. Yu. E. Karpeshina ( see [6-9]) proved this conjecture for
two and three dimensional Schr\"{o}dinger operator $L(q)$ for a wide class of
singular potentials $q(x),$ including Coulomb potential, by \ the method of
perturbation theory. B. Helffer and A. Mohamed [5], by investigations the
integrated density of states, and recently L. Parnovski and A. V. Sobolev [10]
proved the validity of the Bethe-Sommerfeld conjecture for the Schr\"{o}dinger
operator for $d\leq4$ and for arbitrary lattice. The method of this paper and
papers [15-17] is a first and unique, for the present, by which the validity
of the Bethe-Sommerfeld conjecture for arbitrary lattice and for arbitrary
dimension is proved.

The results of the sections 4,5 is obtained in [15-18]. But in those papers
these results are written briefly. The enlarged variant is written in [19]
which can not be used as reference. In the sections 4,5 we write these results
in improved and enlarged form, namely we construct the simple set $B$ with the
properties 1, 2, 3. In the papers [15-17] we emphasized the Bethe-Sommerfeld
conjecture and for this conjecture it is enough to prove the properties 1, 3
and the inequality (1.31b). Therefore in [15-17] we constructed a simple set
satisfying the properties 1, 3 and the inequality (1.31b) and noted in Theorem
3 of [16] and in [18] that the proof of this inequality does not differ from
the proof of (1.31a) which equivalent to the property 2, that is, to the
asymptotic formula (1.32) for Bloch functions. From (1.32) we got \ (1.33) by
iteration (see [18]). Note that one can read Section 4 and Section 5 without
reading Section 3.

In section 6 we construct simple set in the resonance domain and obtain the
asymptotic formulas of arbitrary order for the Bloch functions of the $d$
dimensional Schr\"{o}dinger operator $L(q),$ where $q(x)\in W_{2}^{s}(F),$
$s\geq6(3^{d}(d+1)^{2})+d,$ when corresponding quasimomentum lies in this
simple set, by using the ideas of the sections 4, 5. For the first time the
asymptotic formulas for the Bloch function in the resonance case is obtained
in [4] for $d=2$ and then in [8,9] for $d=2,3.$ In this paper we obtain the
asymptotic formulas in the resonance domain for arbitrary dimension $d.$ Note
that we construct the simple sets in the non-resonance domain so that it
contains a big part of the isoenergetic surfaces of $L(q).$ However in the
case of resonance domain we construct the simple set so that it can be easily
used for the constructive determination ( in next papers) a family of the
spectral invariants by given Floquet spectrum and then to give an algorithm
for finding the potential $q(x)$ by these spectral invariants.

In this paper for the different types of the measures of the subset $A$ of
$\mathbb{R}^{d}$ we use the same notation $\mu(A).$ By $\mid A\mid$ we denote
the number of elements of the set $A$ and use the following obvious fact. If
$a\sim\rho,$ then
\begin{equation}
\mid\{\gamma+t:\gamma\in\Gamma,\mid\mid\gamma+t\mid-a\mid<1\}\mid=O(\rho
^{d-1}). \tag{1.37}%
\end{equation}
Therefore for the number of the eigenvalues $\Lambda_{N}(t)$ of $L_{t}(q)$
lying in $(a^{2}-\rho,a^{2}+\rho)$ the equality%
\begin{equation}
\mid\{N:\Lambda_{N}(t)\in(a^{2}-\rho,a^{2}+\rho)\}\mid=O(\rho^{d-1})
\tag{1.37a}%
\end{equation}
holds. Besides, we use the inequalities:%
\begin{align}
\alpha_{1}+d\alpha &  <1-\alpha\,,\ \ \ \ d\alpha<\frac{1}{2}\alpha
_{d},\tag{1.38}\\
\alpha_{k}+(k-1)\alpha &  <1,\text{ }\alpha_{k+1}>2(\alpha_{k}+(k-1))\alpha
\tag{1.39}\\
\ k_{1}  &  \leq\frac{1}{3}(p-\frac{1}{2}(\varkappa(d-1)),\text{ }3k_{1}%
\alpha>d+2\alpha,\ \ \ \ \ \ \tag{1.40}%
\end{align}
for $k=1,2,...,d,$ which follow from the definitions of the numbers
$p,\varkappa,\alpha,\alpha_{k},k_{1}$ ( see (1.6), (1.2), (1.26), and the
Definition 1.1).

\section{Asymptotic Formulae for the Eigenvalues}

First we obtain the asymptotic formulas for the non-resonance eigenvalues by
iteration of the formula (1.9). If (1.15) holds and $\gamma+t\in
U(\rho^{\alpha_{1}},p),$ then (1.11) holds. Therefore using the decomposition
(1.6) in (1.12), we obtain
\begin{equation}
b(N,\gamma-\gamma_{1})=%
{\displaystyle\sum_{\gamma_{2}\in\Gamma(\rho^{\alpha})}}
\dfrac{q_{\gamma_{2}}b(N,\gamma-\gamma_{1}-\gamma_{2})}{\Lambda_{N}%
(t)-\mid\gamma-\gamma_{1}+t\mid^{2}}+O(\rho^{-p\alpha}). \tag{2.1}%
\end{equation}
Substituting this for $b(N,\gamma-\gamma_{1})$ into the right-hand side of
(1.9) and isolating the terms containing the multiplicand $b(N,\gamma)$, we
get
\begin{equation}
(\Lambda_{N}(t)-\mid\gamma+t\mid^{2})b(N,\gamma)=%
{\displaystyle\sum_{\gamma_{1},\gamma_{2}\in\Gamma(\rho^{\alpha})}}
\dfrac{q_{\gamma_{1}}q_{\gamma_{2}}b(N,\gamma-\gamma_{1}-\gamma_{2})}%
{\Lambda_{N}(t)-\mid\gamma-\gamma_{1}+t\mid^{2}}+O(\rho^{-p\alpha})= \tag{2.2}%
\end{equation}%
\[%
{\displaystyle\sum_{\gamma_{1}\in\Gamma(\rho^{\alpha})}}
\dfrac{\mid q_{\gamma_{1}}\mid^{2}b(N,\gamma)}{\Lambda_{N}(t)-\mid
\gamma-\gamma_{1}+t\mid^{2}}+%
{\displaystyle\sum_{\substack{\gamma_{1},\gamma_{2}\in\Gamma(\rho^{\alpha
}),\\\gamma_{1}+\gamma_{2}\neq0}}}
\dfrac{q_{\gamma_{1}}q_{\gamma_{2}}b(N,\gamma-\gamma_{1}-\gamma_{2})}%
{\Lambda_{N}(t)-\mid\gamma-\gamma_{1}+t\mid^{2}}+O(\rho^{-p\alpha}),
\]
since $q_{\gamma_{1}}q_{\gamma_{2}}=\mid q_{\gamma_{1}}\mid^{2}$ for
$\gamma_{1}+\gamma_{2}=0$ and the last summation is taken under the condition
$\gamma_{1}+\gamma_{2}\neq0.$ The formula (2.2) is the one time iteration of
(1.9). Let us iterate it several times. It follows from the definition of
$U(\rho^{\alpha_{1}},p)$ that ( see Definition 1.1) if $\gamma+t\in
U(\rho^{\alpha_{1}},p)$, $\gamma_{1}\in\Gamma(\rho^{\alpha}),$ $\gamma_{2}%
\in\Gamma(\rho^{\alpha}),...,\gamma_{k}\in\Gamma(\rho^{\alpha}),$ $\gamma
_{1}+\gamma_{2}+...+\gamma_{k}\neq0,$ and (1.15) holds, then%
\begin{align}
&  \mid\mid\gamma+t\mid^{2}-\mid\gamma-\gamma_{1}-\gamma_{2}-...-\gamma
_{k}+t\mid^{2}\mid>\rho^{\alpha_{1}},\text{ }\nonumber\\
&  \mid\Lambda_{N}(t)-\mid\gamma-\gamma_{1}-\gamma_{2}-...-\gamma_{k}%
+t\mid^{2}\mid>\frac{1}{2}\rho^{\alpha_{1}}\text{ , }\forall k\leq p.
\tag{2.3}%
\end{align}
Therefore arguing as in the proof of (2.1), we get
\begin{equation}
b(N,\gamma-\sum_{j=1}^{k}\gamma_{j})=%
{\displaystyle\sum_{\gamma_{k+1}\in\Gamma(\rho^{\alpha})}}
\dfrac{q_{\gamma_{k+1}}b(N,\gamma-\gamma_{1}-\gamma_{2}-...-\gamma_{k+1}%
)}{\Lambda_{N}(t)-\mid\gamma-\gamma_{1}-\gamma_{2}-...-\gamma_{k}+t\mid^{2}%
}+O(\frac{1}{\rho^{p\alpha}}) \tag{2.4}%
\end{equation}
for $k\leq p,$ $\gamma_{1}+\gamma_{2}+...+\gamma_{k}\neq0.$ Now we iterate
(1.9), by using (2.4), as follows. In (2.2) replace $b(N,\gamma-\gamma
_{1}-\gamma_{2})$ by its expression from (2.4) ( in (2.4) replace $k$ by $2$)
and isolate the terms containing $b(N,\gamma),$ then replace $b(N,\gamma
-\gamma_{1}-\gamma_{2}-\gamma_{3})$ for $\gamma_{1}+\gamma_{2}+\gamma_{3}%
\neq0$ by its expression from (2.4) and isolate the terms containing
$b(N,\gamma).$ Repeating this $p_{1}$ times, we obtain
\begin{equation}
(\Lambda_{N}(t)-\mid\gamma+t\mid^{2})b(N,\gamma)=A_{p_{1}-1}(\Lambda
_{N},\gamma+t)b(N,\gamma)+C_{p_{1}}+O(\rho^{-p\alpha}), \tag{2.5}%
\end{equation}
where $p_{1}\equiv\lbrack\frac{p}{3}]+1,$ $A_{p_{1}-1}(\Lambda_{N}%
,\gamma+t)=\sum_{k=1}^{p_{1}-1}S_{k}(\Lambda_{N},\gamma+t)$ ,
\[
S_{k}(\Lambda_{N},\gamma+t)=%
{\displaystyle\sum_{\gamma_{1},...,\gamma_{k}\in\Gamma(\rho^{\alpha})}}
\dfrac{q_{\gamma_{1}}q_{\gamma_{2}}...q_{\gamma_{k}}q_{-\gamma_{1}-\gamma
_{2}-...-\gamma_{k}}}{\prod_{j=1}^{k}(\Lambda_{N}(t)-\mid\gamma+t-\sum
_{i=1}^{j}\gamma_{i}\mid^{2})},
\]%
\[
C_{p_{1}}=\sum_{\gamma_{1},...,\gamma_{p_{1}+1}\in\Gamma(\rho^{\alpha})}%
\dfrac{q_{\gamma_{1}}q_{\gamma_{2}}...q_{\gamma_{p_{1}+1}}b(N,\gamma
-\gamma_{1}-\gamma_{2}-...-\gamma_{p_{1}+1})}{\prod_{j=1}^{p_{1}}(\Lambda
_{N}(t)-\mid\gamma+t-\sum_{i=1}^{j}\gamma_{i}\mid^{2})}.
\]
Here the summations for $S_{k}$ and $C_{p_{1}}$ are taken under the additional
conditions $\gamma_{1}+\gamma_{2}+...+\gamma_{s}\neq0$ for $s=1,2,...,k$ and
$s=1,2,...,p_{1}$ respectively. These conditions and (2.3) shows that the
absolute values of the denominators of the fractions in $S_{k}$ and $C_{p_{1}%
}$ are greater than $(\frac{1}{2}\rho^{\alpha_{1}})^{k}$ and $(\frac{1}{2}%
\rho^{\alpha_{1}})^{p_{1}}$ respectively. Now using the first inequality in
(1.7), we get
\begin{align}
\text{ }S_{k}(\Lambda_{N},\gamma+t)  &  =O(\rho^{-k\alpha_{1}}),\forall
k=1,2,...,p_{1}-1,\tag{2.6}\\
C_{p_{1}}  &  =O(\rho^{-p_{1}\alpha_{1}})=O(\rho^{-p\alpha}),\nonumber
\end{align}
since $p_{1}\geq3p$ ( see (2.5)), $\alpha_{1}=3\alpha$ ( see Definition 1.1),
and hence $p_{1}\alpha_{1}\geq p\alpha.$ In the proof of (2.6) we used only
the condition (1.15) for $\Lambda_{N}.$ Therefore
\begin{equation}
S_{k}(a,\gamma+t)=O(\rho^{-k\alpha_{1}}) \tag{2.7}%
\end{equation}
for all $a\in\mathbb{R}$ satisfying $\mid a-\mid\gamma+t\mid^{2}\mid<\frac
{1}{2}\rho^{\alpha_{1}}$.

\begin{theorem}
$(a)$ Suppose $\gamma+t\in U(\rho^{\alpha_{1}},p).$ If (1.15) and (1.16) hold,
then $\Lambda_{N}(t)$ satisfies formulas (1.14) for $k=1,2,...,[\frac{1}%
{3}(p-c)],$ where
\begin{equation}
F_{0}(\gamma+t)=0,\text{ }F_{k}(\gamma+t)=O(\rho^{-\alpha_{1}}),\forall
k=0,1,..., \tag{2.8}%
\end{equation}%
\begin{equation}
F_{1}(\gamma+t)=%
{\displaystyle\sum_{\gamma_{1}\in\Gamma(\rho^{\alpha})}}
\dfrac{\mid q_{\gamma_{1}}\mid^{2}}{\mid\gamma+t\mid^{2}-\mid\gamma-\gamma
_{1}+t\mid^{2}}, \tag{2.9}%
\end{equation}%
\begin{align}
F_{s}  &  =A_{s}(\mid\gamma+t\mid^{2}+F_{s-1},\gamma+t)=\sum_{k=1}^{s}%
S_{k}(\mid\gamma+t\mid^{2}+F_{s-1},\gamma+t)=\tag{2.10}\\
&  \sum_{k=1}^{s}(%
{\displaystyle\sum_{\gamma_{1},...,\gamma_{k}\in\Gamma(\rho^{\alpha})}}
\dfrac{q_{\gamma_{1}}q_{\gamma_{2}}...q_{\gamma_{k}}q_{-\gamma_{1}-\gamma
_{2}-...-\gamma_{k}}}{\prod_{j=1}^{k}(\mid\gamma+t\mid^{2}+F_{s-1}-\mid
\gamma+t-\sum_{i=1}^{j}\gamma_{i}\mid^{2})})\nonumber
\end{align}
for $s=1,2,....$ and the last summations in (2.10) are taken under the
additional conditions $\gamma_{1}+\gamma_{2}+...+\gamma_{j}\neq0$ for
$j=2,3,...,k$

$(b)$ For each vector $\gamma+t$ from $U(\rho^{\alpha_{1}},p)$ there exists an
eigenvalue $\Lambda_{N}(t)$ of $L_{t}(q)$ satisfying (1.14) for
$k=1,2,...,[\frac{1}{3}(p-\frac{1}{2}\varkappa(d-1))].$
\end{theorem}

\begin{proof}
$(a)$ Dividing both side of (2.5) by $b(N,\gamma)$ and using (1.16), (2.6), we
get the proof of (1.13). Thus the formula (1.14) for $k=1$ holds and
$F_{0}=0.$ Hence (2.8) for $k=0$ is also proved. Moreover, from (2.7), we
obtain
\begin{equation}
S_{k}(\mid\gamma+t\mid^{2}+O(\rho^{-\alpha_{1}}),\gamma+t)=O(\rho
^{-k\alpha_{1}}) \tag{2.11}%
\end{equation}
for $k=1,2,....$ Therefore (2.8) for arbitrary $k$ follows from the definition
of $F_{k}$ ( see (2.10)) by induction . Now we prove (1.14) by induction on
$k$. Suppose (1.14) holds for

$k=j<[\frac{1}{3}(p-c)]\leq p_{1}$, that is,%
\[
\Lambda_{N}(t)=\mid\gamma+t\mid^{2}+F_{j-1}(\gamma+t)+O(\rho^{-j\alpha_{1}}).
\]
Substituting this into $A_{p_{1}-1}(\Lambda_{N},\gamma+t)$ in (2.5), dividing
both sides of (2.5) by $b(N,\gamma),$ using (1.16), and taking into account
that
\[
A_{p_{1}-1}(\Lambda_{N},\gamma+t)=A_{j}(\Lambda_{N},\gamma+t)+O(\rho
^{-(j+1)\alpha_{1}})
\]
( see (2.6) and the definition of $A_{p_{1}-1}$ in (2.5)), we get
\[
\Lambda_{N}(t)=\mid\gamma+t\mid^{2}+A_{j}(\mid\gamma+t\mid^{2}+F_{j-1}%
+O(\rho^{-j\alpha_{1}}),\gamma+t)+O(\rho^{-(j+1)\alpha_{1}})+O(\rho
^{-(p-c)\alpha}).
\]
On the other hand $O(\rho^{-(p-c)\alpha})=O(\rho^{-(j+1)\alpha_{1}}),$ since
$j+1\leq\frac{1}{3}[p-c],$ and $\alpha_{1}=3\alpha.$ Therefore to prove (1.14)
for $k=j+1$ it remains to show that
\begin{equation}
A_{j}(\mid\gamma+t\mid^{2}+F_{j-1}+O(\rho^{-j\alpha_{1}}),\gamma+t)=A_{j}%
(\mid\gamma+t\mid^{2}+F_{j-1},\gamma+t))+O(\rho^{-(j+1)\alpha_{1}}) \tag{2.12}%
\end{equation}
( see the definition of $F_{j}$ in (2.10)). It can be checked by using (1.7),
(2.8), (2.11) and the obvious relation
\begin{align*}
&  \frac{1}{\prod_{j=1}^{s}(\mid\gamma+t\mid^{2}+F_{j-1}+O(\rho^{-j\alpha_{1}%
})-\mid\gamma+t-\sum_{i=1}^{s}\gamma_{i}\mid^{2})}-\\
&  \frac{1}{\prod_{j=1}^{s}(\mid\gamma+t\mid^{2}+F_{j-1}-\mid\gamma
+t-\sum_{i=1}^{s}\gamma_{i}\mid^{2})}\\
&  =\frac{1}{\prod_{j=1}^{s}(\mid\gamma+t\mid^{2}+F_{j-1}-\mid\gamma
+t-\sum_{i=1}^{s}\gamma_{i}\mid^{2})}(\frac{1}{1-O(\rho^{-(j+1)\alpha_{1}}%
)}-1)\\
&  =O(\rho^{-(j+1)\alpha_{1}}),\text{ }\forall s=1,2,....
\end{align*}
The formula (2.9) is also proved, since by (2.10) and (2.8) we have
\begin{equation}
F_{1}=A_{1}(\mid\gamma+t\mid^{2},\gamma+t)=S_{1}(\mid\gamma+t\mid^{2}%
,\gamma+t)=%
{\displaystyle\sum_{\gamma_{1}\in\Gamma(\rho^{\alpha})}}
\dfrac{q_{\gamma_{1}}q_{-\gamma_{1}}}{\mid\gamma+t\mid^{2}-\mid\gamma
+t-\gamma_{1}\mid^{2}} \tag{2.13}%
\end{equation}

$(b)$ Let $A$ be the set of indices $N$ satisfying (1.15). Using (1.8) and
Bessel inequality, we obtain%
\[
\sum_{N\notin A}\mid b(N,\gamma)\mid^{2}=\sum_{N\notin A}\mid\dfrac{(\Psi
_{N}(x),q(x)e^{i(\gamma+t,x)})}{\Lambda_{N}-\mid\gamma+t\mid^{2}}\mid
^{2}=O(\rho^{-2\alpha_{1}})
\]
Hence, by the Parseval equality, we have%
\[
\sum_{N\in A}\mid b(N,\gamma)\mid^{2}=1-O(\rho^{-2\alpha_{1}}).
\]
This and the inequality $\mid A\mid=O(\rho^{d-1})=O(\rho^{(d-1)\varkappa
\alpha})$ ( see (1.37a) and the definition of $\alpha$ in (1.6)) imply that
there exists a number $N$ satisfying (1.16) for $c=\frac{1}{2}\varkappa(d-1)$.
Thus $\Lambda_{N}(t)$ satisfies (1.14) due to $(a)$
\end{proof}

Theorem 2.1 shows that in the non-resonance case the eigenvalue of the
operator $L_{t}(q)$ is close to the eigenvalue of the unperturbed operator
$L_{t}(0).$ However, in Theorem 2.2 we prove that if $\gamma+t\in\cap
_{i=1}^{k}V_{\gamma_{i}}(\rho^{\alpha_{k}})\backslash E_{k+1}$ for $k\geq1,$
where $\gamma_{1},\gamma_{2},...,\gamma_{k}$ are linearly independent vectors
of $\Gamma(p\rho^{\alpha}),$ then the corresponding eigenvalue of $L_{t}(q)$
is close to the eigenvalue of the matrix constructed as follows. Introduce the sets:

$B_{k}\equiv B_{k}(\gamma_{1},\gamma_{2},...,\gamma_{k})=\{b:b=\sum_{i=1}%
^{k}n_{i}\gamma_{i},n_{i}\in\mathbb{Z},\mid b\mid<\frac{1}{2}\rho^{\frac{1}%
{2}\alpha_{k+1}}\},$%
\begin{equation}
B_{k}(\gamma+t)=\gamma+t+B_{k}=\{\gamma+t+b:b\in B_{k}\}, \tag{2.14}%
\end{equation}
$B_{k}(\gamma+t,p_{1})=\{\gamma+t+b+a:b\in B_{k},\mid a\mid<p_{1}\rho^{\alpha
},a\in\Gamma\}=\{h_{i}+t:$ $i=1,2,...,b_{k}\},$

where $p_{1}$ is defined in (2.5), $h_{1}+t,$ $h_{2}+t,...,h_{b_{k}}+t$ are
the vectors of $B_{k}(\gamma+t,p_{1}),$ and $b_{k}\equiv b_{k}(\gamma
_{1},\gamma_{2},...,\gamma_{k})$ is the number of the vectors of $B_{k}%
(\gamma+t,p_{1})$. Define the matrix $C(\gamma+t,\gamma_{1},\gamma
_{2},...,\gamma_{k})\equiv(c_{i,j})$ by
\begin{equation}
c_{i,i}=\mid h_{i}+t\mid^{2},\text{ }c_{i,j}=q_{h_{i}-h_{j}},\text{ }\forall
i\neq j, \tag{2.15}%
\end{equation}
where $i,j=1,2,...,b_{k}.$ Now we consider the resonance eigenvalue
\ $\mid\gamma+t\mid^{2}$ for
\[
\gamma+t\in(\cap_{i=1}^{k}V_{\gamma_{i}}(\rho^{\alpha_{k}}))
\]
by using the following Lemma.

\begin{lemma}
Suppose $\gamma+t\in(\cap_{i=1}^{k}V_{\gamma_{i}}(\rho^{\alpha_{k}%
}))\backslash E_{k+1}$ and $h+t\in B_{k}(\gamma+t,p_{1})$. If

$(h-\gamma^{^{\prime}}+t)\notin B_{k}(\gamma+t,p_{1})$, where $\gamma
^{^{\prime}}\in\Gamma(\rho^{\alpha})$, then
\begin{equation}
\mid\mid\gamma+t\mid^{2}-\mid h-\gamma^{^{\prime}}-\gamma_{1}^{^{\prime}%
}-\gamma_{2}^{^{\prime}}-...-\gamma_{s}^{^{\prime}}+t\mid^{2}\mid>\frac{1}%
{5}\rho^{\alpha_{k+1}} \tag{2.16}%
\end{equation}
for $s=0,1,...,p_{1}-1,$ where $\gamma_{1}^{^{\prime}}\in\Gamma(\rho^{\alpha
}),$ $\gamma_{2}^{^{\prime}}\in\Gamma(\rho^{\alpha}),...,\gamma_{s}^{^{\prime
}}\in\Gamma(\rho^{\alpha}).$
\end{lemma}

\begin{proof}
It follows from the definitions of $p_{1}$ ( see (2.5)) and $p$ ( see (1.6),
(1.2)) that $p>2p_{1}.$ Therefore the conditions of Lemma 2.1 imply that%
\[
h-\gamma^{^{\prime}}-\gamma_{1}^{^{\prime}}-\gamma_{2}^{^{\prime}}%
-...-\gamma_{s}^{^{\prime}}+t\in B_{k}(\gamma+t,p)\backslash B_{k}(\gamma+t)
\]
for $s=0,1,...,p_{1}-1.$ By the definitions of $B_{k}(\gamma+t,p)$ and $B_{k}$
( see (2.14)) we have $h-\gamma^{^{\prime}}-\gamma_{1}^{^{\prime}}-\gamma
_{2}^{^{\prime}}-...-\gamma_{s}^{^{\prime}}+t=\gamma+t+b+a,$ where
\begin{equation}
\mid b\mid<\frac{1}{2}\rho^{\frac{1}{2}\alpha_{k+1}},\mid a\mid<p\rho^{\alpha
},\text{ }\gamma+t+b+a\notin\gamma+t+B_{k},\text{ }b\in B_{k}\subset P,
\tag{2.17}%
\end{equation}
and $P=Span\{\gamma_{1,}\gamma_{2},...,\gamma_{k}\}.$ In this notation (2.16)
has the form
\begin{equation}
\mid\mid\gamma+t+a+b\mid^{2}-\mid\gamma+t\mid^{2}\mid>\frac{1}{5}\rho
^{\alpha_{k+1}}, \tag{2.18}%
\end{equation}
where (2.17) holds. To prove (2.18) we consider two cases:

Case 1. $a\in P$. Since $b\in B_{k}\subset P$ ( see (2.17)) we have $a+b\in
P.$ This with the third relation in (2.17) imply that $a+b\in P\backslash
B_{k}$ ,i.e.,
\begin{equation}
a+b\in P,\text{ }\mid a+b\mid\geq\frac{1}{2}\rho^{\frac{1}{2}\alpha_{k+1}}
\tag{2.19}%
\end{equation}
( see the definition of $B_{k}$ in (2.14)). Now to prove (2.18) we consider
the orthogonal decomposition $\gamma+t=y+v$ of $\gamma+t,$ where $v\in P$ and
$y\bot P.$ First we prove that the projection $v$ of any vector $x\in
\cap_{i=1}^{k}V_{\gamma_{i}}(\rho^{\alpha_{k}})$ on $P$ satisfies
\begin{equation}
\mid v\mid=O(\rho^{(k-1)\alpha+\alpha_{k}}). \tag{2.20}%
\end{equation}
For this we turn the coordinate axis so that $P$ coincides with the span of
the vectors

$e_{1}=(1,0,0,...,0)$, $e_{2}=(0,1,0,...,0),...,$ $e_{k}$. Since $\gamma
_{s}\in P$ we have%
\[
\gamma_{s}=\sum_{i=1}^{k}\gamma_{s,i}e_{i},\text{ }\forall s=1,2,...,k
\]
Therefore the relation $x\in\cap_{i=1}^{k}V_{\gamma_{i}}(\rho^{\alpha_{k}})$
and (1.10) imply
\[
\sum_{i=1}^{k}\gamma_{s,i}x_{i}=O(\rho^{\alpha_{k}}),\text{ }\forall
s=1,2,...,k;
\]
where $x=(x_{1},x_{2},...,x_{d}),\gamma_{j}=(\gamma_{j,1},\gamma
_{j,2},...,\gamma_{j,k},0,0,...,0).$ Solving this system of equations by
Cramer's rule, we obtain%
\begin{equation}
\text{ }x_{n}=\frac{\det(b_{j,i}^{n})}{\det(\gamma_{j,i})}\text{, }\forall
n=1,2,...,k, \tag{2.21}%
\end{equation}
where $b_{j,i}^{n}=\gamma_{j,i}$ for $n\neq j$ and $b_{j,i}^{n}=O(\rho
^{\alpha_{k}})$ for $n=j.$ Since $\det(\gamma_{j,i})$ is the volume of the
parallelepiped generated by the vectors $\gamma_{1},\gamma_{2},...,\gamma_{k}$
we have $\det(\gamma_{j,i})\geq\mu(F)=1.$ On the other hand the relation
$\gamma_{j}\in\Gamma(p\rho^{\alpha})$ and the definition of $b_{j,i}^{n}$
imply that%
\[
\mid\gamma_{j,i}\mid<p\rho^{\alpha},\text{ }\det(b_{j,i}^{n})=O(\rho
^{\alpha_{k}+(k-1)\alpha}).
\]
Therefore using (2.21), we get
\begin{equation}
x_{n}=O(\rho^{\alpha_{k}+(k-1)\alpha})\text{, }\forall n=1,2,...,k;\text{
}\forall x\in\cap_{i=1}^{k}V_{\gamma_{i}}(\rho^{\alpha_{k}}). \tag{2.22}%
\end{equation}
Hence (2.20) holds. The conditions $a\in P,$ $b\in P$ and the orthogonal decomposition

$\gamma+t=y+v$ of $\gamma+t,$ where $v\in P$ and $y\bot P$ imply that
$(y,v)=(y,a)=(y,b)=0,$%
\begin{equation}
\mid\gamma+t+a+b\mid^{2}-\mid\gamma+t\mid^{2}=\mid a+b+v\mid^{2}-\mid
v\mid^{2}. \tag{2.23}%
\end{equation}
Therefore using (2.20), (2.19), and the inequality $\alpha_{k+1}>2(\alpha
_{k}+(k-1)\alpha)$ ( see the \ second inequality in (1.39)), we obtain the
estimation (2.18).

Case 2. $a\notin P.$ First we show that
\begin{equation}
\mid\mid\gamma+t+a\mid^{2}-\mid\gamma+t\mid^{2}\mid\geq\rho^{\alpha_{k+1}}.
\tag{2.24}%
\end{equation}
Suppose that (2.24) does not hold. Then $\gamma+t\in V_{a}(\rho^{\alpha_{k+1}%
}).$ On the other hand
\[
\gamma+t\in\cap_{i=1}^{k}V_{\gamma_{i}}(\rho^{\alpha_{k+1}})
\]
( see the conditions of Lemma 2.1). Therefore we have $\gamma+t\in E_{k+1}$
which contradicts the conditions of the lemma. \ Thus (2.24) is proved. Now,
to prove (2.18) we write the difference $\mid\gamma+t+a+b\mid^{2}-\mid
\gamma+t\mid^{2}$ as the sum of
\[
d_{1}\equiv\mid\gamma+t+a+b\mid^{2}-\mid\gamma+t+b\mid^{2}\text{ and }%
d_{2}\equiv\mid\gamma+t+b\mid^{2}-\mid\gamma+t\mid^{2}.
\]
Since $d_{1}=\mid\gamma+t+a\mid^{2}-\mid\gamma+t\mid^{2}+2(a,b),$ it follows
from the inequalities (2.24), (2.17) that $\mid d_{1}\mid>\frac{2}{3}$
$\rho^{\alpha_{k+1}}$. On the other hand, taking $a=0$ in (2.23), we have
$d_{2}=\mid b+v\mid^{2}-\mid v\mid^{2}.$ Therefore (2.20), the first
inequality in (2.17) and the \ second inequality in (1.39) imply that
\[
\mid d_{2}\mid<\frac{1}{3}\rho^{\alpha_{k+1}},\mid d_{1}\mid-\mid d_{2}%
\mid>\frac{1}{3}\rho^{\alpha_{k+1}},
\]
that is, (2.18) holds
\end{proof}

\begin{theorem}
$(a)$ Suppose $\gamma+t\in(\cap_{i=1}^{k}V_{\gamma_{i}}(\rho^{\alpha_{k}%
}))\backslash E_{k+1},$ where $1\leq k\leq d-1.$ If (1.15) and (1.16) hold,
then there is an index $j$ such that (1.17) holds, where

$\lambda_{1}(\gamma+t)\leq\lambda_{2}(\gamma+t)\leq...\leq\lambda_{b_{k}%
}(\gamma+t)$ are the eigenvalues of the matrix $C(\gamma+t,\gamma_{1}%
,\gamma_{2},...,\gamma_{k})$ defined in (2.15).

$(b)$ Every eigenvalue $\Lambda_{N}(t)$ of the operator $L_{t}(q)$ satisfies
one of the formulas (1.14) and (1.17) for $k=[\frac{1}{3}(p-\frac{1}%
{2}\varkappa(d-1))]$ and $c=\frac{\varkappa(d-1)}{2}$ respectively.
\end{theorem}

\begin{proof}
$(a)$Writing the equation (1.9) for all $h_{i}+t\in B_{k}(\gamma+t,p_{1}),$ we
obtain%
\begin{equation}
(\Lambda_{N}-\mid h_{i}+t\mid^{2})b(N,h_{i})=\sum_{\gamma^{^{\prime}}\in
\Gamma(\rho^{\alpha})}q_{\gamma^{^{\prime}}}b(N,h_{i}-\gamma^{^{\prime}%
})+O(\rho^{-p\alpha}) \tag{2.25}%
\end{equation}
for $i=1,2,...,b_{k}$ ( see (2.14) for the definition of $B_{k}(\gamma
+t,p_{1})$). It follows from (1.15) and Lemma 2.1 that if $(h_{i}%
-\gamma^{^{\prime}}+t)\notin B_{k}(\gamma+t,p_{1}),$ then
\begin{equation}
\mid\Lambda_{N}(t)-\mid h_{i}-\gamma^{^{\prime}}-\gamma_{1}-\gamma
_{2}-...-\gamma_{s}+t\mid^{2}\mid>\frac{1}{6}\rho^{\alpha_{k+1}}, \tag{2.26}%
\end{equation}
where $\gamma^{^{\prime}}\in\Gamma(\rho^{\alpha}),\gamma_{j}\in\Gamma
(\rho^{\alpha}),$ $j=1,2,...,s$ and $s=0,1,...,p_{1}-1.$ Therefore, using the
$p_{1}$ times iterations of (2.1) taking into account (2.26), (1.7) and the
obvious inequality $p_{1}\alpha_{k+1}>p\alpha$ ( see (2.5) and Definition 1.1
for the definitions of\ $p_{1}$ and $\alpha_{k+1}$), we see that if
$(h_{i}-\gamma^{^{\prime}}+t)\notin B_{k}(\gamma+t,p_{1}),$ then
\begin{equation}
b(N,h_{i}-\gamma^{^{\prime}})=\sum_{\gamma_{1},...,\gamma_{p_{1}-1}\in
\Gamma(\rho^{\alpha})}\dfrac{q_{\gamma_{1}}q_{\gamma_{2}}...q_{\gamma_{p_{1}}%
}b(N,h_{i}-\gamma^{^{\prime}}-\sum_{i=1}^{p_{1}}\gamma_{i})}{\prod
_{j=0}^{p_{1}-1}(\Lambda_{N}-\mid h_{i}-\gamma^{^{\prime}}+t-\sum_{i=1}%
^{j}\gamma_{i}\mid^{2})}+ \tag{2.27}%
\end{equation}%
\[
+O(\rho^{-p\alpha})=O(\rho^{p_{1}\alpha_{k+1}})+O(\rho^{-p\alpha}%
)=O(\rho^{-p\alpha}).
\]
Hence (2.25) has the form%
\[
(\Lambda_{N}-\mid h_{i}+t\mid^{2})b(N,h_{i})=\sum_{\substack{\gamma^{^{\prime
}}:\gamma^{^{\prime}}\in\Gamma(\rho^{\alpha}),\\h_{i}-\gamma^{^{\prime}}+t\in
B_{k}(\gamma+t,p_{1})}}q_{\gamma^{^{\prime}}}b(N,h_{i}-\gamma^{^{\prime}%
})+O(\rho^{-p\alpha})
\]
for $i=1,2,...,b_{k}.$ This system can be written in the matrix form
\[
(C-\Lambda_{N}I)(b(N,h_{1}),b(N,h_{2}),...b(N,h_{b_{k}}))=O(\rho^{-p\alpha}),
\]
where the right-hand side of this system is a vector having the norm%
\[
\parallel O(\rho^{-p\alpha})\parallel=O(\sqrt{b_{k}}\rho^{-p\alpha}).
\]
Using the last two equalities, taking into account that one of the vectors
$h_{1}+t,h_{2}+t,...,h_{b_{k}}+t$ is $\gamma+t$\ ( see the definition of
$B_{k}(\gamma+t,p_{1})$ in (2.14)) and (1.16) holds, we obtain
\begin{equation}
c_{5}\rho^{-c\alpha}<(\sum_{i=1}^{b_{k}}\mid b(N,h_{i})\mid^{2})^{\frac{1}{2}%
}\leq\parallel(C-\Lambda_{N}I)^{-1}\parallel\sqrt{b_{k}}c_{7}\rho^{-p\alpha}.
\tag{2.28}%
\end{equation}
Since $(C-\Lambda_{N}I)^{-1}$ is the symmetric matrix having the eigenvalues
$(\Lambda_{N}-\lambda_{i})^{-1}$ for

$i=1,2,...,b_{k},$ we have%
\begin{equation}
\max_{i=1,2,...,b_{k}}\mid\Lambda_{N}-\lambda_{i}\mid^{-1}=\parallel
(C-\Lambda_{N}I)^{-1}\parallel>c_{5}c_{7}^{-1}b_{k}^{-\frac{1}{2}}%
\rho^{-c\alpha+p\alpha}, \tag{2.29}%
\end{equation}
where $b_{k}$ is the number of the vectors of $B_{k}(\gamma+t,p_{1}).$ It
follows from the definition of $B_{k}(\gamma+t,p_{1})$ ( see (2.14)) and the
obvious relations
\[
\mid B_{k}\mid=O(\rho^{\frac{k}{2}\alpha_{k+1}}),\text{ }\mid\Gamma(p_{1}%
\rho^{\alpha})\mid=O(\rho^{d\alpha}),\text{ }d\alpha<\frac{1}{2}3^{d}%
\alpha=\frac{1}{2}\alpha_{d}\text{ that}%
\]%
\begin{equation}
b_{k}=O(\rho^{d\alpha+\frac{k}{2}\alpha_{k+1}})=O(\rho^{\frac{d}{2}\alpha_{d}%
})=O(\rho^{\frac{d}{2}3^{d}\alpha}),\forall k=1,2,...,d-1. \tag{2.30}%
\end{equation}
Thus formula (1.17) follows from (2.29) and (2.30).

$(b)$ Let $\Lambda_{N\text{ }}(t)$ be any eigenvalue of $L_{t}(q)$ lying in
$(\frac{3}{4}\rho^{2},\frac{5}{4}\rho^{2}).$ Denote by $D$ the set of all
vectors $\gamma\in\Gamma$ satisfying (1.15). Using (1.8), (1.15), Bessel's
inequality, Parseval's equality, we obtain%
\[
\sum_{\gamma\notin D}\mid b(N,\gamma)\mid^{2}=\sum_{\gamma\notin D}\mid
\frac{(\Psi_{N,t}(x)q(x),e^{i(\gamma+t,x)})}{\Lambda_{N}-\mid\gamma+t\mid^{2}%
}\mid^{2}=O(\rho^{-2\alpha_{1}})\parallel\Psi_{N,t}(x)q(x)\parallel
=O(\rho^{-2\alpha_{1}}),
\]%
\[
\text{ }\sum_{\gamma\in D}\mid b(N,\gamma)\mid^{2}=1-O(\rho^{-2\alpha_{1}}).
\]
Since $\mid D\mid=O(\rho^{d-1})$ ( see (1.37)), there exists $\gamma\in D$
such that%
\[
\mid b(N,\gamma)\mid>c_{8}\rho^{-\frac{(d-1)}{2}}=c_{8}\rho^{-\frac
{(d-1)\varkappa}{2}\alpha},
\]
that is, condition (1.16) for $c=\frac{(d-1)\varkappa}{2}$ holds. Now the
proof of $(b)$ follows from Theorem 2.1$(a)$ and Theorem 2.2$(a),$ since
either $\gamma+t$ $\in U(\rho^{\alpha_{1}},p)$ or $\gamma+t\in$ $E_{k}%
\backslash E_{k+1}$ ( see (2.33))
\end{proof}

\begin{remark}
The obtained asymptotic formulas hold true, without any change in their proof,
if we replace $V_{\gamma_{1}}(\rho^{\alpha_{1}})$ by $V_{\gamma_{1}}(c_{4}%
\rho^{\alpha_{1}})$ and the multiplicand $\frac{1}{2}$ in (1.15) by
$\frac{c_{4}}{2}.$ Here we note that the non-resonance domain%
\[
U\equiv U(c_{4}\rho^{\alpha_{1}},p)\equiv(R(\frac{3}{2}\rho)\backslash
R(\frac{1}{2}\rho))\backslash\bigcup_{\gamma_{1}\in\Gamma(p\rho^{\alpha}%
)}V_{\gamma_{1}}(c_{4}\rho^{\alpha_{1}})
\]
( see Definition 1.1) has an asymptotically full measure on $\mathbb{R}^{d}$
in the sense that $\frac{\mu(U\cap B(\rho))}{\mu(B(\rho))}$ tends to $1$ as
$\rho$ tends to infinity, where $B(\rho)=\{x\in\mathbb{R}^{d}:\mid x\mid
=\rho\}.$ Clearly, $B(\rho)\cap V_{b}(c_{4}\rho^{\alpha_{1}})$ is the part of
sphere $B(\rho),$ which is contained between two parallel hyperplanes%
\[
\{x:\mid x\mid^{2}-\mid x+b\mid^{2}=-c_{4}\rho^{\alpha_{1}}\}\text{ and
}\{x:\mid x\mid^{2}-\mid x+b\mid^{2}=c_{4}\rho^{\alpha_{1}}\}.
\]
The distance of these hyperplanes from origin is $O(\frac{\rho^{\alpha_{1}}%
}{\mid b\mid}).$ Therefore, the relations

$\mid\Gamma(p\rho^{\alpha})\mid=O(\rho^{d\alpha})$ and $\alpha_{1}%
+d\alpha<1-\alpha$ ( see (1.38)) imply
\begin{align}
\mu(B(\rho)\cap V_{b}(c_{4}\rho^{\alpha_{1}}))  &  =O(\frac{\rho^{\alpha
_{1}+d-2}}{\mid b\mid}),\text{ }\mu(E_{1}\cap B(\rho))=O(\rho^{d-1-\alpha
}),\tag{2.31}\\
\mu(U(c_{4}\rho^{\alpha_{1}},p)\cap B(\rho))  &  =(1+O(\rho^{-\alpha}%
))\mu(B(\rho)). \tag{2.32}%
\end{align}
If $x\in\cap_{i=1}^{d}V_{\gamma_{i}}(\rho^{\alpha_{d}}),$ then (2.22) holds
for $k=d$ and $n=1,2,...,d.$ Hence we have

$\mid x\mid=O(\rho^{\alpha_{d}+(d-1)\alpha}).$ It is impossible, since
$\alpha_{d}+(d-1)\alpha<1$ ( see the first inequality in (1.39)) and $x\in
B(\rho).$ It means that
\[
(\cap_{i=1}^{d}V_{\gamma_{i}}(\rho^{\alpha_{k}}))\cap B(\rho)=\emptyset
\]
for $\rho\gg1$. Thus for $\rho\gg1$ we have
\begin{equation}
R(\frac{3}{2}\rho)\backslash R(\frac{1}{2}\rho)=(U(\rho^{\alpha_{1}}%
,p)\cup(\cup_{s=1}^{d-1}(E_{s}\backslash E_{s+1}))). \tag{2.33}%
\end{equation}

\end{remark}

\begin{remark}
Here we note some properties of the known parts $\mid\gamma+t\mid^{2}%
+F_{k}(\gamma+t)$ (see Theorem 2.1) and $\lambda_{j}(\gamma+t)$ ( see Theorem
2.2) of the eigenvalues of $L_{t}(q).$ Denoting $\gamma+t$ by $x$ we consider
the function $F(x)=\mid x\mid^{2}+F_{k}(x).$ It follows from the definition of
$F_{k}(x)$ that ( see 2.10) $F(x)$ is continuous on $U(c_{4}\rho^{\alpha_{1}%
},p).$ Let us prove the equalities
\begin{equation}
\frac{\partial F_{k}(x)}{\partial x_{i}}=O(\rho^{-2\alpha_{1}+\alpha}),\forall
i=1,2,...,d;\forall k=1,2,..., \tag{2.34}%
\end{equation}
for $x\in U(\rho^{\alpha_{1}},p),$ by induction on $k.$ If $k=1$ then (2.34)
follows from the first inequality in (1.7) and the the obvious relation%
\begin{equation}
\frac{\partial}{\partial x_{i}}(\dfrac{1}{\mid x\mid^{2}-\mid x-\gamma_{1}%
\mid^{2}})=\dfrac{-2\gamma_{1}(i)}{(\mid x\mid^{2}-\mid x-\gamma_{1}\mid
^{2})^{2}}=O(\rho^{-2\alpha_{1}+\alpha}), \tag{2.35}%
\end{equation}
where $\gamma_{1}(i)$ is the $i$th component of the vector $\gamma_{1}%
\in\Gamma(p\rho^{\alpha}).$ Now suppose that (2.34) holds for $k=s.$ Using
this and (2.8), replacing $\mid x\mid^{2}$ by $\mid x\mid^{2}+F_{s}(x)$ in
(2.35) and evaluating as above we obtain
\[
\frac{\partial}{\partial x_{i}}(\dfrac{1}{\mid x\mid^{2}+F_{s}-\mid
x-\gamma_{1}\mid^{2}})=\dfrac{-2\gamma_{1}(i)+\frac{\partial F_{s}%
(x)}{\partial x_{i}}}{(\mid x\mid^{2}+F_{s}-\mid x-\gamma_{1}\mid^{2})^{2}%
}=O(\rho^{-2\alpha_{1}+\alpha}).
\]
This formula together with the definition (2.10) of $F_{k}$ give (2.34) for
$k=s+1.$

Now denoting $\lambda_{i}(\gamma+t)-\mid\gamma+t\mid^{2}$ by $r_{i}(\gamma+t)$
we prove that
\begin{equation}
\mid r_{i}(x)-r_{i}(x^{^{\prime}})\mid\leq2\rho^{\frac{1}{2}\alpha_{d}}\mid
x-x^{^{\prime}}\mid,\text{ }\forall i. \tag{2.36}%
\end{equation}
Clearly $r_{1}(x)\leq r_{2}(x)\leq...\leq$ $r_{b_{k}}(x)$ are the eigenvalue
of the matrix

$C(x)-\mid x\mid^{2}I\equiv\widetilde{C}(x),$ where $C(x)$ is defined in
(2.15). By definition, only the diagonal elements of the matrix $\widetilde
{C}(x)=(\widetilde{c}_{i,j}(x))$ depend on $x$ and they are%
\begin{equation}
\widetilde{c}_{i,j}(x)=\mid x+a_{i}\mid^{2}-\mid x\mid^{2}=2(x,a_{i})+\mid
a_{i}\mid^{2}, \tag{2.37}%
\end{equation}
where $x=\gamma+t,$ $a_{i}=h_{i}+t-x$ and $h_{i}+t\in B_{k}(\gamma+t,p_{1}).$
Using the equality $\alpha_{d}=3^{d}\alpha$ ( see Definition 1.1) and
definition of $B_{k}(\gamma+t,p_{1})$ ( see (2.14)), we get
\[
\mid a_{i}\mid<\frac{1}{2}\rho^{\frac{1}{2}\alpha_{k}}+p_{1}\rho^{\alpha}%
<\rho^{\frac{1}{2}\alpha_{d}}%
\]
for $k<d$. Therefore taking into account that $\widetilde{C}(x)-\widetilde
{C}(x^{^{\prime}})$ is a diagonal matrix with diagonal entries $\widetilde
{c}_{i,j}(x)-\widetilde{c}_{i,j}(x^{^{\prime}})=2(x-x^{^{\prime}},a_{i})$ (
see (2.37)), we have
\[
\parallel\widetilde{C}(x)-\widetilde{C}(x^{^{\prime}})\parallel\leq
2\rho^{\frac{1}{2}\alpha_{d}}\mid x-x^{^{\prime}}\mid
\]
which yields (2.36).
\end{remark}

\section{ Bloch Eigenvalues near\ the Diffraction Planes}

In this section we obtain the asymptotic formulae for the eigenvalues
corresponding to the quasimomentum $\gamma+t$ lying near the diffraction
hyperplane%
\[
D_{\delta}=\{x\in\mathbb{R}^{d}:\mid x\mid^{2}=\mid x+\delta\mid^{2}\},
\]
namely lying in the single resonance domain $V_{\delta}^{^{\prime}}%
(\rho^{\alpha_{1}})\equiv V_{\delta}(\rho^{\alpha_{1}})\backslash E_{2}$
defined in Definition 1.1, where $\delta$ is the element of \ $\Gamma$ of
minimal norm in its direction, that is, $\delta$ is the element of $\Gamma$
such that $\{(\delta,\omega):\omega\in\Omega\}=2\pi\mathbb{Z}$. In section 2
to obtain the asymptotic formula for the eigenvalues corresponding to the
quasimomentum $\gamma+t$ lying far from the diffraction planes we considered
the operator $L_{t}(q)$ as perturbation of the operator $L_{t}(0)$ with
$q(x).$ As a result the asymptotic formulas for these eigenvalues of
$L_{t}(q)$ is expressed in the term of the eigenvalues of $L_{t}(0)$. To
obtain the asymptotic formulae for the eigenvalues corresponding to the
quasimomentum $\gamma+t$ lying near the diffraction plane $D_{\delta}$ we
consider the operator $L_{t}(q)$ as the perturbation of the operator
$L_{t}(q^{\delta}),$ where the directional potential $q^{\delta}(x)$ is
defined in (1.19), with $q(x)-q^{\delta}(x).$ Hence it is natural that the
asymptotic formulas, which will be obtained in this section, will be expressed
in the term of the eigenvalues of $L_{t}(q^{\delta}).$ Therefore first of all
we need to investigate the eigenvalues and eigenfunctions of $L_{t}(q^{\delta
}).$ Let $\Omega_{\delta\text{ }}$be the sublattice $\{h\in\Omega
:(h,\delta)=0\}$ of $\Omega$ in hyperplane $H_{\delta}=$ $\{x\in\mathbb{R}%
^{d}:(x,\delta)=0\}$, and
\[
\Gamma_{\delta}\equiv\{a\in H_{\delta}:(a,k)\in2\pi\mathbb{Z},\forall
k\in\Omega_{\delta}\}
\]
be the dual lattice of $\Omega_{\delta}.$ Denote by $F_{\delta}$ the
fundamental domain $H_{\delta}/\Gamma_{\delta}$ of $\Gamma_{\delta}.$ Then
$t\in F^{\ast}$ has a unique decomposition
\begin{equation}
t=a+\tau+\mid\delta\mid^{-2}(t,\delta)\delta, \tag{3.1}%
\end{equation}
where $a\in\Gamma_{\delta},$ $\tau\in F_{\delta}.$ Define the sets
$\Omega^{^{\prime}}$ and $\Gamma^{^{\prime}}$ by $\Omega^{^{\prime}%
}=\{h+l\delta^{\ast}:h\in\Omega_{\delta},l\in\mathbb{Z}\},$ and by
$\Gamma^{^{\prime}}=\{b+(p-(2\pi)^{-1}(b,\delta^{\ast}))\delta:b\in
\Gamma_{\delta},p\in\mathbb{Z}\},$ where $\delta^{\ast}$ is the element of
$\Omega$ satisfying $(\delta^{\ast},\delta)=2\pi.$

\begin{lemma}
$(a)$ The following relations hold: $\Omega=\Omega^{^{\prime}},$
$\Gamma=\Gamma^{^{\prime}}.$

$(b)$ The eigenvalues and eigenfunctions of the operator $L_{t}(q^{\delta})$
are
\[
\lambda_{j,\beta}(v,\tau)=\mid\beta+\tau\mid^{2}+\mu_{j}(v(\beta,t)),\text{
}\Phi_{j,\beta}(x)=e^{i(\beta+\tau,x)}\varphi_{j,v(\beta,t))}(\zeta)
\]
for $j\in\mathbb{Z},$ $\beta\in\Gamma_{\delta},$ where $v(\beta,t)$ is the
fractional part of $\mid\delta\mid^{-2}(t,\delta)-(2\pi)^{-1}(\beta
-a,\delta^{\ast}),$ $\tau$ and $a$ are uniquely determined from decomposition
(3.1) and $\mu_{j}(v(\beta,t)),\varphi_{j,v(\beta,t)}(\zeta)$ are eigenvalues
and corresponding normalized eigenfunctions of the operator $T_{v(\beta
,t)}(Q(\zeta))$ generated by the boundary value problem
\[
-\mid\delta\mid^{2}y^{\prime\prime}(\zeta)+Q(\zeta)y(\zeta)=\mu y(\zeta
),\text{ }y(\zeta+2\pi)=e^{i2\pi v}y(\zeta),
\]
where, $\zeta=(\delta,x),$ $Q(\zeta)=q^{\delta}(x)$ and for simplicity of the
notation, instead of $v(\beta,t)$ we write $v(\beta)$ (or $v)$ if $t$ (or $t$
and $\beta$), for which we consider $v(\beta,t),$ is unambiguous.
\end{lemma}

\begin{proof}
$(a)$ For each vector $\omega$ of the lattice $\Omega$ assign $h=\omega
-(2\pi)^{-1}(\omega,\delta)\delta^{\ast}.$ Using the relations $(\omega
,\delta)\equiv2\pi l\in2\pi\mathbb{Z},$ and $(\delta^{\ast},\delta)=2\pi$ we
see that $h\in\Omega$ and $(h,\delta)=0$ ,i.e., $h\in\Omega_{\delta}.$ Hence
$\Omega\subset\Omega^{^{\prime}}.$ Now for each vector $\gamma$ of the lattice
$\Gamma$ assign $b=\gamma-\mid\delta\mid^{-2}(\gamma,\delta)\delta.$ It is not
hard to verify that $b\in H_{\delta}$ and $(b,\omega)=(\gamma,\omega)\in
2\pi\mathbb{Z}$ for $\omega\in\Omega_{\delta}$ $\subset\Omega.$ Therefore
$b\in\Gamma_{\delta}.$ Moreover $(b,\delta^{\ast})=(\gamma,\delta^{\ast}%
)-2\pi(\gamma,\delta)\mid\delta\mid^{-2}.$ Since $(\gamma,\delta^{\ast}%
)\in2\pi\mathbb{Z},$ that is, $(\gamma,\delta^{\ast})=2\pi n$ where
$n\in\mathbb{Z}$, we have $(\gamma,\delta)\mid\delta\mid^{-2}=n-(2\pi
)^{-1}(b,\delta^{\ast}).$ Therefore we obtain an orthogonal decomposition
\begin{equation}
\gamma=b+\langle\gamma,\frac{\delta}{\mid\delta\mid}\rangle\frac{\delta}%
{\mid\delta\mid}=b+(n-(2\pi)^{-1}(b,\delta^{\ast}))\delta\tag{3.2}%
\end{equation}
of $\gamma\in\Gamma$, where $b\in\Gamma_{\delta},$ and $n\in\mathbb{Z}.$ Hence
$\Gamma\subset\Gamma^{^{\prime}}.$ On the other hand if $b\in\Gamma_{\delta},$
$h\in\Omega_{\delta}$ and $n,l\in\mathbb{Z}$, then $(h+l\delta^{\ast
},b+(n-(2\pi)^{-1}(b,\delta^{\ast}))\delta)=(h,b)+2\pi nl\in2\pi\mathbb{Z}.$
Thus we have the relations ( see definition of the sets $\Omega^{^{\prime}%
},\Gamma^{^{\prime}}$ )
\begin{equation}
\Omega\subset\Omega^{^{\prime}},\Gamma\subset\Gamma^{^{\prime}},(\omega
^{^{\prime}},\gamma^{^{\prime}})\in2\pi\mathbb{Z},\forall\omega^{^{\prime}}%
\in\Omega^{^{\prime}},\forall\gamma^{^{\prime}}\in\Gamma^{^{\prime}}.
\tag{3.3}%
\end{equation}
Since $\Omega$ is the set of all vectors $\omega\in\mathbb{R}^{d}$ satisfying
$(\omega,\gamma)\in2\pi\mathbb{Z}$ for all $\gamma\in\Gamma$ and $\Gamma$ is
the set of all vectors $\gamma\in\mathbb{R}^{d}$ satisfying $(\omega
,\gamma)\in2\pi\mathbb{Z}$ for all $\omega\in\Omega$ the relations in (3.3)
imply $\Omega^{^{\prime}}\subset\Omega,$ $\Gamma^{^{\prime}}\subset\Gamma$ and
hence $\Omega=\Omega^{^{\prime}},$ $\Gamma=\Gamma^{^{\prime}}$.

$(b)$ Since $\beta+\tau$ is orthogonal to $\delta,$ turning the coordinate
axis so that $\delta$ coincides with one of the coordinate axis and taking
into account that the Laplace operator is invariant under rotation, one can
easily verify that
\[
(-\Delta+q^{\delta}(x))\Phi_{j,\beta}(x)=\lambda_{j,\beta}\Phi_{j,\beta}(x)
\]
Now using the relation $(\delta,\omega)=2\pi l,$ where $\omega\in\Omega,$
$l\in\mathbb{Z},$ and the definitions of $\Phi_{j,\beta}(x),\varphi
_{j,v}(\delta,x)$ we obtain
\[
\Phi_{j,\beta}(x+\omega)=e^{i(\beta+\tau,x+\omega)}\varphi_{j,v}%
(\delta,x+\omega)=\Phi_{j,\beta}(x)e^{i(\beta+\tau,\omega)+i2\pi lv(\beta
,t)}.
\]
Replacing $\tau$ and $\omega$ by $t-a-\mid\delta\mid^{-2}(t,\delta)\delta$ and
$h+l\delta^{\ast}$, where $h\in\Omega_{\delta},l\in\mathbb{Z},$ (see (3.1) and
the first equality of $(a)$) respectively, and then using

$(h,\delta)=0,$ $(\delta^{\ast},\delta)=2\pi$ one can easily verify that%
\[
(\beta+\tau,\omega)=(t,\omega)+(\beta-a,h)-2\pi l[\mid\delta\mid^{-2}%
(t,\delta)-(2\pi)^{-1}(\beta-a,\delta^{\ast})].
\]
From this using that $(\beta-a,h)\in2\pi\mathbb{Z},$ ( since $\beta-a\in
\Gamma_{\delta},$ $h\in\Omega_{\delta}$), and $v(\beta,t)$ is a fractional
part of the expression in the last square bracket, we infer
\[
\Phi_{j,\beta}(x+\omega)=e^{i(t,\omega)}\Phi_{j,\beta}(x).
\]
Thus $\Phi_{j,\beta}(x)$ is an eigenfunction of $L_{t}(q^{\delta}(x)).$

Now we prove that the system $\{\Phi_{j,\beta}(x):j\in\mathbb{Z},\beta
\in\Gamma_{\delta}\}$ contains all eigenfunctions of $L_{t}(q^{\delta}(x))$.
Assume the converse. Then there exists a nonzero function $f(x)\in L_{2}(F),$
which is orthogonal to all elements of this system. Using (3.1), (3.2) of and
the definition of $v(\beta,t)$ ( see Lemma 3.1(b)), we get%
\begin{equation}
\gamma+t=\beta+\tau+(j+v)\delta, \tag{3.4}%
\end{equation}
where $\beta\in\Gamma_{\delta},\tau\in F_{\delta},$ $j\in\mathbb{Z},$ and
$v=v(\beta,t).$ Since $e^{i(j+v)\zeta}$ can be decomposed by basis
$\{\varphi_{j,v(\beta,t))}(\zeta):$ $j\in\mathbb{Z}\}$ the function
$e^{i(\gamma+t,x)}=e^{i(\beta+\tau,x)}e^{i(j+v)\zeta}$ (see (3.4)) can be
decomposed by system
\[
\{\Phi_{j,\beta}(x)=e^{i(\beta+\tau,x)}\varphi_{j,v(\beta,t))}(\zeta
):j\in\mathbb{Z}\}.
\]
Then the assumption $(\Phi_{j,\beta}(x),$ $f(x))=0$ for $j\in\mathbb{Z},$
$\beta\in\Gamma_{\delta}$ implies that $(f(x),e^{i(\gamma+t,x)})=0$ for all
$\gamma\in$ $\Gamma.$ This is impossible, since the system $\{e^{i(\gamma
+t,x)}:\gamma\in\Gamma$ $\}$ is a basis of $L_{2}(F)$
\end{proof}

\begin{remark}
\bigskip Clearly every vectors $x$ of $\mathbb{R}^{d}$ has decompositions:

$x=\gamma+t$ and $x=\beta+\tau+(j+v)\delta$, where $\gamma\in\Gamma,$ $t\in F$
and $\beta\in\Gamma_{\delta},$ $\tau\in F_{\delta},$ $j\in\mathbb{Z},$
$v\in\lbrack0,1).$ We say that the first and second decompositions are
$\Gamma$ and $\Gamma_{\delta}$ decompositions respectively. Writing
$\gamma+t\equiv\beta+\tau+(j+v(\beta,t))\delta$ (see (3.4)) we mean the
$\Gamma_{\delta}$ decomposition of $\gamma+t.$ As it is noted in lemma 3.1
instead of $v(\beta,t)$ we write $v(\beta)$ (or $v)$ if $t$ (or $t$ and
$\beta$), for which we consider $v(\beta,t),$ is unambiguous. The
decompositions (3.4) of $\gamma+t$ is orthogonal decompositions, since
$\beta\in\Gamma_{\delta},$ $\tau\in F_{\delta},$ and $\delta$ orthogonal to
$\Gamma_{\delta}$ and $F_{\delta}$. Hence
\[
\mid\gamma+t\mid^{2}=\mid\beta+\tau\mid^{2}+\mid(j+v)\delta\mid^{2}.
\]
Therefore, one can easily verify that, if $\gamma+t\in V_{\delta}(\rho
^{\alpha_{1}})$ ( see Definition 1.1), then

$\mid\mid(j+v+1)\delta\mid^{2}-\mid(j+v)\delta\mid^{2}\mid<\rho^{\alpha_{1}}.$
Using this and $\alpha_{1}=3\alpha,$ we get
\begin{equation}
\mid(j+v)\delta\mid<r_{1},\text{ }\mid j\delta\mid<r_{1},\text{ }r_{1}%
>2\rho^{\alpha}, \tag{3.5}%
\end{equation}
where $r_{1}=\frac{\rho^{\alpha_{1}}}{\mid2\delta\mid}+\mid2\delta\mid.$ To
the eigenvalue $\mid\gamma+t\mid^{2}=\mid\beta+\tau\mid^{2}+\mid
(j+v)\delta\mid^{2}$ of $L_{t}(0)$ assign the eigenvalue $\lambda_{j,\beta
}(v,\tau)=\mid\beta+\tau\mid^{2}+\mu_{j}(v)$ of $L_{t}(q^{\delta}),$ where
$\mid(j+v)\delta\mid^{2}$ is the eigenvalue of $T_{v}(0)$ and $\mu_{j}(v)$ is
the eigenvalue of $T_{v}(Q)$ ( see Lemma 3.1(b)) satisfying
\begin{equation}
\mid\mu_{j}(v)-\mid(j+v)\delta\mid^{2}\mid\leq\sup\mid Q(\zeta)\mid,\text{
}\forall j\in\mathbb{Z}. \tag{3.6}%
\end{equation}
The eigenvalue $\lambda_{j,\beta}(v,\tau)$ of $L_{t}(q^{\delta})$ can be
considered as the perturbation of the eigenvalue \ $\mid\gamma+t\mid^{2}%
=\mid\beta+\tau\mid^{2}+\mid(j+v)\delta\mid^{2}$ of $L_{t}(0)$ by $q^{\delta
}(x).$ Then we see that the influence of $q^{\delta}(x)$ is significant for
$\beta+\tau+(j+v)\delta\in V_{\delta}(\rho^{\alpha_{1}}),$ namely for small
value of $j.$
\end{remark}

Now we prove that if $\beta+\tau+(j+v)\delta\in V_{\delta}(\rho^{\alpha_{1}%
}),$ then there is an eigenvalue $\Lambda_{N}(t)$ of $L_{t}(q)$ which is close
to the eigenvalue $\lambda_{j,\beta}(v,\tau)$ of $L_{t}(q^{\delta}),$ that is,
we prove that the influence of $q(x)-q^{\delta}(x)$ is not significant if the
quasimomentum lies in $V_{\delta}(\rho^{\alpha_{1}})\backslash E_{2}.$ To
prove this we consider the operator $L_{t}(q)$ as perturbation of the operator
$L_{t}(q^{\delta})$ with $q(x)-q^{\delta}(x)$ and use the formula (1.21)
called binding formula for $L_{t}(q)$ and $L_{t}(q^{\delta}).$ Recall that we
have obtained the asymptotic formulas for the perturbation of the
non-resonance eigenvalue $\mid\gamma+t\mid^{2}$ by iteration the binding
formula (1.8) for the unperturbed operator $L_{t}(0)$ and the perturbed
operator $L_{t}(q)$ ( see section 2). Similarly, now to obtain the asymptotic
formulas for the perturbation of the resonance eigenvalue we iterate the
binding formula (1.21) for the unperturbed operator$\ L_{t}(q^{\delta})$ and
perturbed operator$\ L_{t}(q)$. For this ( as in the non-resonance case) we
decompose $(q(x)-q^{\delta}(x))\Phi_{j,\beta}(x)$ by the basis $\{\Phi
_{j^{^{\prime}},\beta^{^{\prime}}}(x):j^{^{\prime}}\in\mathbb{Z}%
,\beta^{^{\prime}}\in\Gamma_{\delta}\}$ and put this decomposition into
(1.21). Let us find this decomposition. Using (3.2) for $\gamma_{1}\in
\Gamma(\rho^{\alpha})$ and (1.6), we get

$\gamma_{1}=\beta_{1}+(n_{1}-(2\pi)^{-1}(\beta_{1},\delta^{\ast}))\delta,$
$e^{i(\gamma_{1},x)}=e^{i(\beta_{1},x)}e^{i(n_{1}-(2\pi)^{-1}(\beta_{1}%
,\delta^{\ast}))\zeta},$%
\[
q(x)-Q(\zeta)=\sum\limits_{(n_{1},\beta_{1})\in\Gamma^{^{\prime}}(\rho
^{\alpha})}c(n_{1},\beta_{1})e^{i(\beta_{1},x)}e^{i(n_{1}-(2\pi)^{-1}%
(\beta_{1},\delta^{\ast}))\zeta}+O(\rho^{-p\alpha}),
\]%
\begin{equation}
(q(x)-Q(\zeta))\Phi_{j,\beta}(x)= \tag{3.7}%
\end{equation}%
\[
\sum\limits_{(n_{1},\beta_{1})\in\Gamma^{^{\prime}}(\rho^{\alpha})}%
c(n_{1},\beta_{1})e^{i(\beta_{1}+\beta+\tau,x)}e^{i(n_{1}-(2\pi)^{-1}%
(\beta_{1},\delta^{\ast}))\zeta}\varphi_{j,v(\beta)}(\zeta)+O(\rho^{-p\alpha
}),
\]
where $c(n_{1},\beta_{1})=q_{\gamma_{1}},$%
\[
\Gamma^{^{\prime}}(\rho^{\alpha})=\{(n_{1},\beta_{1}):\beta_{1}\in
\Gamma_{\delta}\backslash\{0\},n_{1}\in\mathbb{Z},\beta_{1}+(n_{1}-(2\pi
)^{-1}(\beta_{1},\delta^{\ast}))\delta\in\Gamma(\rho^{\alpha})\}.
\]
Note that if $(n_{1},\beta_{1})\in\Gamma^{^{\prime}}(\rho^{\alpha}),$ then
$\mid\beta_{1}+(n_{1}-(2\pi)^{-1}(\beta_{1},\delta^{\ast}))\delta\mid
<\rho^{\alpha}$ and
\begin{equation}
\mid\beta_{1}\mid<\rho^{\alpha},\text{ }\mid(n_{1}-(2\pi)^{-1}(\beta
_{1},\delta^{\ast}))\delta\mid<\rho^{\alpha}<\frac{1}{2}r_{1}, \tag{3.8}%
\end{equation}
since $\beta_{1}$ is orthogonal to $\delta$ and $r_{1}>2\rho^{\alpha}$ ( see
(3.5)). To decompose the right-hand side of (3.7) by basis $\{\Phi
_{j^{^{\prime}},\beta^{^{\prime}}}(x)\}$ we use the following lemma

\begin{lemma}
$(a)$ If $j$, $m$ satisfy the inequalities $\mid m\mid>2\mid j\mid,$ $\mid
m\delta\mid\geq2r,$ then
\begin{align}
(\varphi_{j,v}(\zeta),e^{i(m+v)\zeta})  &  =O(\mid m\delta\mid^{-s-1}%
)=O(\rho^{-(s+1)\alpha}),\tag{3.9}\\
(\varphi_{m,v},e^{i(j+v)\zeta})  &  =O(\mid m\delta\mid^{-s-1}). \tag{3.10}%
\end{align}
where $r\geq r_{1}=\frac{\rho^{\alpha_{1}}}{\mid2\delta\mid}+\mid2\delta\mid$,
$\varphi_{j,v}(\zeta)$ is the eigenfunctions of the operator $T_{v}%
(Q(\zeta)),$ and $Q(\zeta)\in W_{2}^{s}[0,2\pi].$
\end{lemma}

\begin{proof}
$(a).$ To prove (3.9) we iterate the formula
\begin{equation}
(\mu_{j}(v)-\mid(m+v)\delta\mid^{2})(\varphi_{j,v}(\zeta),e^{i(m+v)\zeta
})=(\varphi_{j,v}(\zeta)Q(\zeta),e^{i(m+v)\zeta}), \tag{3.11}%
\end{equation}
by using the decomposition
\begin{equation}
Q(\zeta)=\sum_{\mid l_{1}\mid<\frac{\mid m\mid}{2s}}q_{l_{1}\delta}%
e^{il_{1}\zeta}+O(\mid m\delta\mid^{-(s-1)}) \tag{3.12}%
\end{equation}
Note that (3.11), (3.12) is one dimensional case of (1.8), (1.6) and the
iteration of (3.11) is simpler than the iteration of (1.8) ( see (1.9),
(2.5)). If $\mid j\mid<\frac{\mid m\mid}{2},$ and$\mid l_{i}\mid<\frac{\mid
m\mid}{2s}$ for $i=1,2,...k\equiv\lbrack\frac{s}{2}],$ then the inequalities
\begin{align*}
&  \mid m+v-l_{1}-l_{2}-...-l_{q}\mid-\mid j\mid>\frac{1}{5}\mid m\mid,\\
&  \mid m\mid-\mid j+v-l_{1}-l_{2}-...-l_{q}\mid>\frac{1}{5}\mid m\mid
\end{align*}
hold for $q=0,1,...,k$. Therefore by (3.6), we have
\begin{align}
(  &  \mid\mu_{j}-\mid(m-l_{1}-l_{2}-...-l_{q}+v)\delta\mid^{2}\mid
)^{-1}=O(\mid m\delta\mid^{-2}),\tag{3.13}\\
(  &  \mid\mu_{m}-\mid(j-l_{1}-l_{2}-...-l_{q}+v)\delta\mid^{2}\mid
)^{-1}=O(\mid m\delta\mid^{-2}), \tag{3.14}%
\end{align}
for $q=0,1,...,k$. Iterating (3.11) $k$ times, by using (3.13), we get
\begin{equation}
(\varphi_{j},e^{i(m+v)\zeta})=\sum_{\mid l_{1}\delta\mid,\mid l_{2}\delta
\mid,...,\mid l_{k+1}\delta\mid<\frac{\mid m\delta\mid}{2s}}q_{l_{1}\delta
}q_{l_{2}\delta}...q_{l_{k+1}\delta}\times\tag{3.15}%
\end{equation}%
\[
\frac{(\varphi_{j},e^{i(m-l_{1}-l_{2}-...-l_{k+1}+v)\zeta})}{\sqcap_{q=0}%
^{k}(\mu_{j}-\mid(m-l_{1}-l_{2}-...-l_{q}+v)\delta\mid^{2})}+O(\mid
m\delta\mid^{-s-1}).
\]
Now (3.9) follows from (3.13), (3.15), and (1.7). Formula (3.10) can be proved
in the same way by using (3.14) instead of (3.13). Note that in (3.9), and
(3.10) instead of $O(\mid m\delta\mid^{-s-1})$ we can write $O(\rho
^{-(s+1)\alpha}),$ since $\mid m\delta\mid\geq r\geq r_{1}>2\rho^{\alpha}$ (
see (3.5))
\end{proof}

\begin{lemma}
If $\mid j\delta\mid<r$ and $(n_{1},\beta_{1})\in\Gamma^{^{\prime}}%
(\rho^{\alpha}),$ then
\begin{align}
&  \ e^{i(n_{1}-(2\pi)^{-1}(\beta_{1},\delta^{\ast}))\zeta}\varphi
_{j,v(\beta)}(\zeta)\tag{3.16}\\
\  &  =\sum_{\mid j_{1}\delta\mid<9r}a(n_{1},\beta_{1},j,\beta,j+j_{1,}%
\beta+\beta_{1})\varphi_{j+j_{1},v(\beta+\beta_{1})}(\zeta)+O(\rho
^{-(s-1)\alpha}),\nonumber
\end{align}

where $r,$ $\Gamma^{^{\prime}}(\rho^{\alpha})$ are defined in Lemma 3.2(a),
(3.7), and

$a(n_{1},\beta_{1},j,\beta,j+j_{1,}\beta+\beta_{1})=(e^{i(n_{1}-(2\pi
)^{-1}(\beta_{1},\delta^{\ast}))\zeta}\varphi_{j,v(\beta)}(\zeta
),\varphi_{j+j_{1},v(\beta+\beta_{1})}(\zeta)).$
\end{lemma}

\begin{proof}
Since $\ e^{i(n_{1}-(2\pi)^{-1}(\beta_{1},\delta^{\ast}))\zeta}\varphi
_{j,v(\beta)}(\zeta)$ is equal to its Fourier series with the orthonormal
basis $\{\varphi_{j+j_{1},v(\beta+\beta_{1})}(\zeta):j_{1}\in\mathbb{Z}\}$ it
suffices to show that%
\[
\sum_{j_{1}:\mid j_{1}\delta\mid\geq9r}\mid a(n_{1},\beta_{1},j,\beta
,j+j_{1,}\beta+\beta_{1})\mid=O(\rho^{-(s-1)\alpha}).
\]
For this we prove
\begin{equation}
\mid a(n_{1},\beta_{1},j,\beta,j+j_{1,}\beta+\beta_{1})\mid=O(\mid j_{1}%
\delta\mid^{-s}) \tag{3.17}%
\end{equation}
for all $j_{1}$ satisfying $\mid j_{1}\delta\mid\geq9r$ and take into account
that $r\geq r_{1}>\rho^{\alpha}$ ( see the last inequality in (3.5)).
Decomposing $\varphi_{j,v(\beta)}$ over $\{e^{i(m+v)\zeta}:m\in\mathbb{Z}\}$
and using the last inequality in (3.8), we have
\begin{equation}
e^{i(n_{1}-(2\pi)^{-1}(\beta_{1},\delta^{\ast}))\zeta}\varphi_{j}(\zeta
)=\sum_{m\in\mathbb{Z}}(\varphi_{j},e^{i(m+v)\zeta})e^{i(m+n+v(\beta+\beta
_{1}))\zeta}, \tag{3.18}%
\end{equation}
where $n\in\mathbb{Z}$ and $\mid n\delta\mid<r$. This and the decomposition%
\[
\varphi_{j+j_{1}}(\zeta)=\sum_{m\in\mathbb{Z}}(\varphi_{j+j_{1}}%
,e^{i(m+v(\beta+\beta_{1}))\zeta})e^{i(m+v(\beta+\beta_{1}))v}%
\]
imply that
\begin{equation}
a(n_{1},\beta_{1},j,\beta,j+j_{1,}\beta+\beta_{1})=\sum_{m\in\mathbb{Z}%
}(\varphi_{j},e^{i(m-n+v)\zeta})(\varphi_{j+j_{1}},e^{i(m+v(\beta+\beta
_{1}))\zeta}) \tag{3.19}%
\end{equation}
where $j,$ $j_{1,}$ $n$ satisfy the conditions$\mid j\delta\mid<r,$ $\mid
j_{1}\delta\mid\geq9r,$ $\mid n\delta\mid<r$ due to the conditions in Lemma
3.3, \ (3.17), (3.18) respectively. Consider two cases:

Case 1: $\mid m\delta\mid>\frac{1}{3}\mid j_{1}\delta\mid\geq3r$ . In this
case using the conditions of (3.19), we get $\mid(m-n)\delta\mid>2r$ and $\mid
m-n\mid>\mid j\mid$. Therefore (3.9) implies that
\[
(\varphi_{j},e^{i(m-n+v)\zeta})=O(\mid m\delta\mid^{-s-1}),\text{ }\sum_{\mid
m\mid>\frac{1}{3}\mid j_{1}\mid}\mid(\varphi_{j},e^{i(m-n+v)\zeta})\mid=O(\mid
j_{1}\delta\mid^{-s}).
\]
Case 2: $\mid m\mid\leq\frac{1}{3}\mid j_{1}\mid.$ Again using the conditions
of (3.19) we obtain that $\mid j_{1}+j\mid>2\mid m\mid$. Therefore it follows
from (3.10) that
\begin{align*}
(\varphi_{j+j_{1}},e^{i(m+v(\beta+\beta_{1}))\zeta})  &  =O(\mid
(j_{1}+j)\delta\mid^{-(s-1)})=O(\mid j_{1}\delta\mid^{-s-1}),\\
\sum_{\mid m\mid\leq\frac{1}{3}\mid j_{1}\mid}  &  \mid(\varphi_{j+j_{1}%
}(\zeta),e^{i(m+v(\beta+\beta_{1}))\zeta})\mid=O(\mid j_{1}\delta\mid^{-s}).
\end{align*}
These estimations for these two cases together with (3.19) yield (3.17)
\end{proof}

Now it follows from (3.7) and (3.16) that
\[
(q(x)-Q(\zeta))\Phi_{j^{^{\prime}},\beta^{^{\prime}}}(x)=O(\rho^{-p\alpha
})+\sum\limits_{(n_{1,}j_{_{1}},\beta_{1})\in G(\rho^{\alpha},9r)}%
c(n_{1},\beta_{1})\times
\]%
\begin{equation}
a(n_{1},\beta_{1},j,\beta^{^{\prime}},j^{^{\prime}}+j_{1,}\beta^{^{\prime}%
}+\beta_{1})e^{i(\beta_{1}+\beta^{^{\prime}}+\tau,x)}\varphi_{j^{^{\prime}%
}+j_{1},v(\beta^{^{\prime}}+\beta_{1})}(\zeta) \tag{3.20}%
\end{equation}
for all $j^{^{\prime}}$ satisfying $\mid j^{^{\prime}}\delta\mid<r,$ where%
\[
G(\rho^{\alpha},9r)=\{(n,j,\beta):\mid j\delta\mid<9r,(n,\beta)\in
\Gamma^{^{\prime}}(\rho^{\alpha}),\beta\neq0\}.
\]
In (3.20) the multiplicand $e^{i(\beta_{1}+\beta^{^{\prime}}+\tau,x)}%
\varphi_{j^{^{\prime}}+j_{1},v(\beta+\beta_{1})}(\zeta)=\Phi_{j^{^{\prime}%
}+j_{1,}\beta^{^{\prime}}+\beta_{1}}(x)$ does not depend on $n_{1}.$ Its
coefficient is
\begin{equation}
\overline{A(j^{^{\prime}},\beta^{^{\prime}},j^{^{\prime}}+j_{1,}%
\beta^{^{\prime}}+\beta_{1}}=\sum\limits_{n_{1}:(n_{1},\beta_{1})\in
\Gamma^{^{\prime}}(\rho^{\alpha})}c(n_{1},\beta_{1})a(n_{1},\beta
_{1},j^{^{\prime}},\beta^{^{\prime}},j^{^{\prime}}+j_{1,}\beta^{^{\prime}%
}+\beta_{1}). \tag{3.21}%
\end{equation}

\begin{lemma}
If $\mid\beta^{^{\prime}}\mid\sim\rho$ and $\mid j^{^{\prime}}\delta\mid<r,$
where $r\geq r_{1}=\frac{\rho^{\alpha_{1}}}{\mid2\delta\mid}+\mid2\delta\mid$,
then
\begin{align}
&  (q(x)-Q(\zeta))\Phi_{j^{^{\prime}},\beta^{^{\prime}}}(x)=\tag{3.22}\\
&  \sum\limits_{(j_{_{1}},\beta_{1})\in Q(\rho^{\alpha},9r)}\overline
{A(j^{^{\prime}},\beta^{^{\prime}},j^{^{\prime}}+j_{1,}\beta^{^{\prime}}%
+\beta_{1}})\Phi_{j^{^{\prime}}+j_{1,}\beta^{^{\prime}}+\beta_{1}}%
(x)+O(\rho^{-p\alpha}),\nonumber
\end{align}
where $Q(\rho^{\alpha},9r)=\{(j,\beta):\mid j\delta\mid<9r,$ $0<\mid\beta
\mid<\rho^{\alpha}\}.$ Moreover,
\begin{equation}
\sum\limits_{(j_{1},\beta_{1})\in Q(\rho^{\alpha},9r)}\mid A(j^{^{\prime}%
},\beta^{^{\prime}},j^{^{\prime}}+j_{1,}\beta^{^{\prime}}+\beta_{1})\mid
<c_{9}, \tag{3.23}%
\end{equation}
where $c_{9}$ does not depend on $(j^{^{\prime}},\beta^{^{\prime}}).$
\end{lemma}

\begin{proof}
The formula (3.22) follows from (3.20), (3.21). Now we prove (3.23). Since
$c(n_{1},\beta_{1})=q_{\gamma_{1}}$ ( see (3.7)), it follows from the first
inequality of (1.7) and (3.21) that we need to prove the inequality
\begin{equation}
\sum\limits_{j_{_{1}}}\mid a(n_{1},\beta_{1},j^{^{\prime}},\beta^{^{\prime}%
},j^{^{\prime}}+j_{1,}\beta^{^{\prime}}+\beta_{1})\mid<c_{9}(c_{3})^{-1}.
\tag{3.24}%
\end{equation}
For this we use (3.19) and prove the inequalities:
\begin{align}
\sum_{m\in\mathbb{Z}}  &  \mid(\varphi_{j^{^{\prime}}},e^{i(m-n+v(\beta
^{^{\prime}})\zeta})\mid<c_{10},\tag{3.25}\\
\sum_{j_{1}\in\mathbb{Z}}  &  \mid(\varphi_{j^{^{\prime}}+j_{1}}%
,e^{i(m+v(\beta_{1}+\beta^{^{\prime}}))\zeta})\mid<c_{11}. \tag{3.26}%
\end{align}
Since the distance between numbers $\mid v\delta\mid^{2},\mid(1+v)\delta
\mid^{2},...,$ and similarly the distance between numbers $\mid(-1+v)\delta
\mid^{2},\mid(-2+v)\delta\mid^{2},...,$ where $v\in\lbrack0,1],$ is not less
than $c_{12}$, it follows from (3.6) that the number of elements of the sets%
\begin{align*}
A  &  =\{m:\mid(m-n+v(\beta^{^{\prime}}))\delta\mid^{2}\in\lbrack
\mu_{j^{^{\prime}}}(v(\beta^{^{\prime}}))-1,\mu_{j^{^{\prime}}}(v(\beta
^{^{\prime}}))+1]\},\\
B  &  =\{j_{1}:\mu_{j^{^{\prime}}+j_{1}}(v(\beta_{1}+\beta^{^{\prime}}%
))\in\lbrack\mid(m+v(\beta_{1}+\beta^{^{\prime}}))\delta\mid^{2}%
-1,\mid(m+v)\delta\mid^{2}+1]\}
\end{align*}
is less than $c_{13}.$ Now in (3.25) and (3.26) isolating the term with $m\in
A$ and $j_{1}\in B$ respectively, applying (3.11) to other terms and then
using
\begin{align*}
\sum_{m\notin A}\frac{1}{\mid\mu_{j^{^{\prime}}}(v^{^{\prime}})-\mid
(m-n+v^{^{\prime}})\delta\mid^{2}\mid}  &  <c_{14},\\
\sum_{j_{1}\notin B}\frac{1}{\mid\mu_{j^{^{\prime}}+j_{1}}(v_{1}^{^{\prime}%
})-\mid(m+v_{1}^{^{\prime}})\delta\mid^{2}\mid}  &  <c_{14}%
\end{align*}
we get the proof of (3.25) and (3.26). Thus (3.24) and hence (3.23) is proved.
Clearly the constants $c_{14},c_{13},c_{12},c_{11},c_{10}$ can be chosen
independently on $(j^{^{\prime}},\beta^{^{\prime}}).$ Therefore $c_{9}$ does
not depend on $(j^{^{\prime}},\beta^{^{\prime}})$
\end{proof}

Replacing $(j,\beta)$ by $(j^{^{\prime}},\beta^{^{\prime}})$ in (1.21) and
using (3.22), we get%

\[
(\Lambda_{N}-\lambda_{j^{^{\prime}},\beta^{^{\prime}}})b(N,j^{^{\prime}}%
,\beta^{^{\prime}})=(\Psi_{N}(x),(q(x)-Q(\zeta))\Phi_{j^{^{\prime}}%
,\beta^{^{\prime}}}(x))=O(\rho^{-p\alpha})
\]

\begin{equation}
+\sum\limits_{(j_{1},\beta_{1})\in Q(\rho^{\alpha},9r)}A(j^{^{\prime}}%
,\beta^{^{\prime}},j^{^{\prime}}+j_{1,}\beta^{^{\prime}}+\beta_{1}%
)b(N,j^{^{\prime}}+j_{1},\beta^{^{\prime}}+\beta_{1}) \tag{3.27}%
\end{equation}
for $\mid\beta^{^{\prime}}\mid\sim\rho$ and $\mid j^{^{\prime}}\delta\mid<r,$
where $b(N,j,\beta)=(\Psi_{N}(x),\Phi_{j,\beta}(x)).$ Note that if $\mid
j^{^{\prime}}\delta\mid<r,$ then the summation in (3.27) is taken over
$Q(\rho^{\alpha},9r).$ Therefore if $\mid j\delta\mid<r_{1},$ where is defined
in (3.5), then we have the formula
\[
(\Lambda_{N}-\lambda_{j,\beta})b(N,j,\beta)=O(\rho^{-p\alpha})
\]%
\begin{equation}
+\sum\limits_{(j_{1},\beta_{1})\in Q(\rho^{\alpha},9r_{1})}A(j,\beta
,j+j_{1,}\beta+\beta_{1})b(N,j+j_{1},\beta+\beta_{1}). \tag{3.28}%
\end{equation}
Thus (3.28) is obtained from (3.27) by interchanging $j^{^{\prime}}%
,\beta^{^{\prime}},r,$ and $j,\beta,r_{1}$. Now to find the eigenvalue
$\Lambda_{N}(t),$ which is close to $\lambda_{j,\beta}$ , where $\mid
j\delta\mid<r_{1}$, we are going to iterate (3.28) as follows. Since $\mid
j\delta\mid<r_{1}$ and $(j_{1},\beta_{1})\in Q(\rho^{\alpha},9r_{1}),$ we have
$\mid(j+j_{1})\delta\mid<10r_{1}.$ Therefore in (3.27) interchanging
$j^{^{\prime}},\beta^{^{\prime}},r,$ and $j+j_{1,}\beta+\beta_{1},10r_{1}$ and
then introducing the notations $r_{2}=10r_{1},$ $j^{2}=j+j_{1}+j_{2},$
$\beta^{2}=\beta+\beta_{1}+\beta_{2},$ we obtain%
\[
(\Lambda_{N}-\lambda_{j+j_{1},\beta_{1}+\beta})b(N,j+j_{1},\beta+\beta
_{1})=O(\rho^{-p\alpha})+
\]%
\begin{equation}%
{\displaystyle\sum_{(j_{2},\beta_{2})\in Q(\rho^{\alpha},9r_{2})}}
b(N,j^{2},\beta^{2})A(j+j_{1},\beta+\beta_{1},j^{2},\beta^{2}). \tag{3.29}%
\end{equation}
Clearly, there exist an eigenvalue $\Lambda_{N}(t)$ satisfying $\mid
\lambda_{j,\beta}-\Lambda_{N}(t)\mid\leq2M,$ where

$M=\sup\mid q(x)\mid.$ Moreover, in the next lemma (Lemma 3.5 ), we will prove
that \ if $\mid\beta\mid\sim\rho,$ and $(j_{1},\beta_{1})\in Q(\rho^{\alpha
},9r_{1})$, then%
\begin{equation}
\mid\lambda_{j,\beta}-\lambda_{j+j_{1},\beta+\beta_{1}}\mid>\frac{5}{9}%
\rho^{\alpha_{2}},\text{ }\mid\Lambda_{N}(t)-\lambda_{j+j_{1},\beta+\beta_{1}%
}\mid>\frac{1}{2}\rho^{\alpha_{2}}. \tag{3.30}%
\end{equation}
Therefore dividing both side of (3.29) by $\Lambda_{N}-\lambda_{j+j_{1}%
,\beta+\beta_{1}}$, we get
\begin{equation}
b(N,j+j_{1},\beta_{1}+\beta)=%
{\displaystyle\sum_{(j_{2},\beta_{2})\in Q(\rho^{\alpha},9r_{2})}}
\dfrac{A(j+j_{1},\beta+\beta_{1},j^{2},\beta^{2})b(N,j^{2},\beta^{2})}%
{\Lambda_{N}-\lambda_{j+j_{1},\beta_{1}+\beta}}+O(\frac{1}{\rho^{p\alpha
+\alpha_{2}}}) \tag{3.31}%
\end{equation}
Putting the obtained formula for $b(N,j+j_{1},\beta_{1}+\beta)$ into (3.28),
we obtain
\begin{equation}
(\Lambda_{N}-\lambda_{j,\beta})b(N,j,\beta)=O(\rho^{-p\alpha})+ \tag{3.32}%
\end{equation}%
\[%
{\displaystyle\sum_{\substack{(j_{1},\beta_{1})\in Q(\rho^{\alpha}%
,9r_{1})\\(j_{2},\beta_{2})\in Q(\rho^{\alpha},9r_{2})}}}
\dfrac{A(j,\beta,j+j_{1,}\beta+\beta_{1})A(j+j_{1},\beta+\beta_{1},j^{2}%
,\beta^{2})b(N,j^{2},\beta^{2})}{\Lambda_{N}-\lambda_{j+j_{1},\beta+\beta_{1}%
}}.
\]
Thus we got the one time iteration of (3.28). It will give the first term of
asymptotic formula for $\Lambda_{N}.$ For this we find the index $N$ such that
$b(N,j,\beta)$ is not very small (see Lemma 3.6) and (3.30) is satisfied,
i.e., the denominator of the fraction in (3.32) is a big number. Then dividing
both sides of (3.32) by $b(N,j,\beta),$ we get the asymptotic formula for
$\Lambda_{N}(t)$ (see Theorem 3.1).

\begin{lemma}
Let $\gamma+t\equiv\beta+\tau+(j+v)\delta\in V_{\delta}^{^{\prime}}%
(\rho^{\alpha_{1}})\equiv V_{\delta}(\rho^{\alpha_{1}})\backslash E_{2}$ (see
(3.4), Remark 3.1 and Definition 1.1), and $(j_{1},\beta_{1})\in
Q(\rho^{\alpha},9r_{1}),$ $(j_{k},\beta_{k})\in Q(\rho^{\alpha},9r_{k}),$
where $r_{1}$ is defined in (3.5) and $r_{k}=10r_{k-1}$ for $k=2,3,...,p-1$.
Then
\begin{equation}
\mid j\delta\mid=O(\rho^{\alpha_{1}}),\text{ }\mid j_{k}\delta\mid
=O(\rho^{\alpha_{1}}),\text{ }\mid\beta_{k}\mid<\rho^{\alpha},\forall
k=1,2,...,p-1 \tag{3.33}%
\end{equation}

Moreover if $\mid j^{^{\prime}}\delta\mid<\frac{1}{2}\rho^{\frac{1}{2}%
\alpha_{2}},\mid\beta^{^{\prime}}-\beta\mid<(p-1)\rho^{\alpha},$
$\beta^{^{\prime}}\in\Gamma_{\delta}$, $j^{k}=j+j_{1}+...+j_{k},$ $\beta
^{k}=\beta+\beta_{1}+...+\beta_{k},$ where $k=1,2,...,p-1,$ then
\begin{align}
&  \mid\lambda_{j,\beta}-\lambda_{j^{^{\prime}},\beta^{^{\prime}}}\mid
>\frac{5}{9}\rho^{\alpha_{2}},\text{ }\forall\beta^{^{\prime}}\neq
\beta,\tag{3.34}\\
&  \mid\lambda_{j,\beta}(v,\tau)-\lambda_{j^{k},\beta^{k}}\mid>\frac{5}{9}%
\rho^{\alpha_{2}},\text{ }\forall\beta^{k}\neq\beta. \tag{3.35}%
\end{align}

\end{lemma}

\begin{proof}
The relations in (3.33) follows from (3.5) and the definitions of $r_{1}%
,r_{k},$ $Q(\rho^{\alpha},9r_{k})$ (see Lemma 3.4). Inequality (3.35) follows
from (3.34) and (3.33). It remains to prove (3.34). Since
\begin{equation}
\mid\lambda_{j,\beta}-\lambda_{j^{^{\prime}},\beta^{^{\prime}}}\mid\geq
\mid\mid\beta^{^{\prime}}+\tau\mid^{2}-\mid\beta+\tau\mid^{2}\mid-\mid\mu
_{j}-\mu_{j^{^{\prime}}}\mid, \tag{3.36}%
\end{equation}
it is enough to prove the following two inequalities $\mid\mu_{j}%
-\mu_{j^{^{\prime}}}\mid<\frac{1}{3}\rho^{\alpha_{2}},$
\begin{equation}
\mid\mid\beta+\tau\mid^{2}-\mid\beta^{^{\prime}}+\tau\mid^{2}\mid>\frac{8}%
{9}\rho^{\alpha_{2}}. \tag{3.37}%
\end{equation}
The first inequality follows from $\mid j^{^{\prime}}\delta\mid<\frac{1}%
{2}\rho^{\frac{1}{2}\alpha_{2}},\mid j\delta\mid=O(\rho^{\alpha_{1}})$ ( see
the conditions of this lemma and (3.33)) and (3.6), since $\alpha_{2}%
=3\alpha_{1}.$ Now we prove (3.37). The conditions $\mid\beta^{^{\prime}%
}-\beta\mid<(p-1)\rho^{\alpha},\mid\delta\mid<\rho^{\alpha}$ imply that there
exist $n\in\mathbb{Z}$ and $\gamma^{^{\prime}}\in\Gamma$ such that
\begin{equation}
\gamma^{^{\prime}}=\beta^{^{\prime}}-\beta+(n+(2\pi)^{-1}(\beta^{^{\prime}%
}-\beta,\delta^{\ast}))\delta\in\Gamma(p\rho^{\alpha}). \tag{3.38}%
\end{equation}
Since $\beta^{^{\prime}}-\beta\neq0$ ( see (3.34)) and $\beta^{^{\prime}%
}-\beta\in\Gamma_{\delta}$ , that is , $(\beta^{^{\prime}}-\beta,\delta)=0$
the relation (3.38) imply that $\gamma^{^{\prime}}\in\Gamma(p\rho^{\alpha
})\backslash\delta R.$ This together with the condition%
\[
\gamma+t=\beta+\tau+(j+v)\delta\in V_{\delta}(\rho^{\alpha_{1}})\backslash
E_{2}%
\]
( see assumption of the lemma and the definition of $E_{2}$ in Definition 1.1) gives

$\gamma+t\notin V_{\gamma^{^{\prime}}}(\rho^{\alpha_{2}}),$ that is, $\mid
\mid\gamma+t\mid^{2}-\mid\gamma+t+\gamma^{^{\prime}}\mid^{2}\mid\geq
\rho^{\alpha_{2}}.$ From this using the orthogonal decompositions (3.4) and
(3.38) of $\gamma+t$ and $\gamma^{^{\prime}}$ respectively, taking into
account that $\beta,\tau,\beta^{^{\prime}}$ are orthogonal to $\delta$ and
then using the relations $\mid j\delta\mid=O(\rho^{\alpha_{1}})$ (see (3.33)),
$\mid n+(2\pi)^{-1}(\beta^{^{\prime}}-\beta,\delta^{\ast}))\delta\mid
=O(\rho^{\alpha})$ ( see the inclusion in the orthogonal decompositions (3.38)
of $\gamma^{^{\prime}}$) and $\alpha_{2}>2\alpha$ ( see Definition 1.1), we
obtain (3.37)
\end{proof}

\begin{lemma}
Suppose $h_{1}(x),h_{2}(x),...,h_{m}(x)\in L_{2}(F),$ where $m=p_{1}-1,$
$p_{1}=[\frac{p}{3}]+1$. Then for every eigenvalue $\lambda_{j,\beta}\sim
\rho^{2}$ of the operator $L_{t}(q^{\delta})$ there exists an eigenvalue
$\Lambda_{N}(t)$ and a corresponding normalized eigenfunction $\Psi_{N,t}(x)$
of the operator $L_{t}(q)$ such that:

$(i)$ $\ \mid\lambda_{j,\beta}-\Lambda_{N}(t)\mid\leq2M,$ where $M=\sup\mid
q(x)\mid,$

$(ii)$ $\ \mid b(N,j,\beta)\mid>c_{15}\rho^{-\frac{1}{2}(d-1)},$

$(iii)$ $\ \mid b(N,j,\beta)\mid^{2}>\frac{1}{2m}%
{\displaystyle\sum_{i=1}^{m}}
\mid(\Psi_{N},\frac{h_{i}}{\mid\mid h_{i}\mid\mid})\mid^{2}>\frac{1}{2m}%
\mid(\Psi_{N},\frac{h_{i}}{\mid\mid h_{i}\mid\mid})\mid^{2},$ $\forall i.$
\end{lemma}

\begin{proof}
Let $A,B,C$ be the set of indexes $N$ satisfying $(i),(ii),(iii)$
respectively. Using (1.21), the Bessel inequality, and the Parseval equality,
we get
\begin{align*}
\sum_{N\notin A}  &  \mid b(N,j,\beta)\mid^{2}=\sum_{N\notin A}\mid
\dfrac{(\Psi_{N}(x),(q(x)-Q(\zeta))\Phi_{j,\beta}(x))}{\Lambda_{N}%
-\lambda_{j,\beta}}\mid^{2}<\\
(2M)^{-2}  &  \mid\mid q(x)-Q(\zeta))\Phi_{j,\beta}(x))\mid\mid^{2}\leq
\frac{1}{4},\text{ }\sum_{N\in A}\mid b(N,j,\beta)\mid^{2}\geq\frac{3}{4}.
\end{align*}
On the other hand the inequality $\mid A\mid<c_{16}\rho^{(d-1)}$ ( see
(1.37a)) and the definition of $B$ imply that if $c_{15}^{2}<\frac{1}{4c_{16}%
},$ then
\[
\sum_{N\in A\backslash B}\mid b(N,j,\beta)\mid^{2}<\frac{1}{4}.
\]
Therefore using the relation $A=(A\backslash B)\cup(A\cap B),$ we obtain
\[
\text{ }\sum_{N\in A\cap B}\mid b(N,j,\beta)\mid^{2}>\frac{1}{2}.
\]
Now to prove the lemma we show that there exists $N\in A\cap B$ satisfying
$(iii)$. Assume that the assertion $(iii)$ does not hold for all $N\in A\cap
B$ . Using the last inequality, the assumption that $(iii)$ does not holds for
$N\in A\cap B$ , and then the Bessel inequality, we get%
\[
\frac{1}{2}<\sum_{N\in A\cap B}\mid b(N,j,\beta)\mid^{2}\leq\frac{1}{2m}%
\sum_{i=1}^{m}\sum_{N\in A}\mid(\Psi_{N},\frac{h_{i}}{\mid\mid h_{i}\mid\mid
})\mid^{2}%
\]%
\[
\leq\frac{1}{2m}\sum_{i=1}^{m}\mid\mid\frac{h_{i}}{\mid\mid h_{i}\mid\mid}%
\mid\mid^{2}=\frac{1}{2}.
\]
This contradiction complete the proof of the lemma
\end{proof}

\begin{theorem}
For every eigenvalue $\lambda_{j,\beta}(v,\tau)$ of $L_{t}(q^{\delta})$ such that

$\beta+\tau+(j+v)\delta\in V_{\delta}^{^{\prime}}(\rho^{\alpha_{1}})$ there
exists an eigenvalue $\Lambda_{N}$ of $L_{t}(q)$, denoted by $\Lambda
_{N}(\lambda_{j,\beta}(v,\tau)),$ satisfying%
\begin{equation}
\Lambda_{N}(\lambda_{j,\beta}(v,\tau))=\lambda_{j,\beta}(v,\tau)+O(\rho
^{-\alpha_{2}}). \tag{3.39}%
\end{equation}

\end{theorem}

\begin{proof}
In the proof of this theorem we use the inequalities
\begin{equation}
p_{1}\alpha_{2}>p\alpha,\text{ }p\alpha-\frac{1}{2}(d-1)>\alpha_{2} \tag{3.40}%
\end{equation}
which follows from the definitions of $p,\alpha,\alpha_{2}$ and $p_{2}$ given
in (1.6), Definition 1.1, and (2.5). By Lemma 3.6 there is an eigenvalue
$\Lambda_{N}(t)$ satisfying $(i)$-$(iii)$ for
\[
h_{i}(x)=%
{\displaystyle\sum_{\substack{(j_{1},\beta_{1})\in Q(\rho^{\alpha}%
,9r_{1}),\\(j_{2},\beta_{2})\in Q(\rho^{\alpha},9r_{2})}}}
\dfrac{\overline{A(j,\beta,j^{1},\beta^{1}})\overline{A(j^{1},\beta^{1}%
,j^{2},\beta^{2}})\Phi_{j^{2},\beta^{2}}(x)}{(\lambda_{j,\beta}-\lambda
_{j+j_{1},\beta+\beta_{1}})^{i}},
\]
where $i=1,2,...,m;$ $m=p_{1}-1$ and $Q(\rho^{\alpha},9r)$ is defined in Lemma
3.4. By definition of $Q(\rho^{\alpha},9r_{1})$ we have $\beta_{1}\neq0$.
Therefore the inequality (3.34) and assertion $(i)$ of lemma 3.6 yield (3.30).
Hence, in brief notations $a=\lambda_{j,\beta},$ $z=\lambda_{j+j_{1}%
,\beta+\beta_{1}},$ we have $\mid\Lambda_{N}-a\mid<2M,$ $\mid z-a\mid>\frac
{1}{2}\rho^{\alpha_{2}}.$ Using the relations
\[
\frac{1}{\Lambda_{N}-z}=-\sum_{i=1}^{\infty}\frac{(\Lambda_{N}-a)^{i-1}%
}{(z-a)^{i}}=-\sum_{i=1}^{m}\frac{(\Lambda_{N}-a)^{i-1}}{(z-a)^{i}}%
+O(\rho^{-p_{1}\alpha_{2}})
\]
and the first inequality of (3.40), we see that formula (3.32) can be written
as
\[
(\Lambda_{N}-\lambda_{j,\beta})b(N,j,\beta,)=\sum_{i=1}^{m}(\Lambda
_{N}-a)^{i-1}(\Psi_{N},\frac{h_{i}}{\mid\mid h_{i}\mid\mid})\parallel
h_{i}\parallel+O(\rho^{-p\alpha}).
\]
Dividing both sides by $b(N,j,\beta)$, using assertions $(ii),(iii)$ of Lemma
3.6, and the second inequality of (3.40), we get
\[
\mid(\Lambda_{N}-\lambda_{j,\beta})\mid<(2m)^{\frac{1}{2}}\sum_{i=1}^{m}%
\mid\Lambda_{N}-a\mid^{i-1}\parallel h_{i}\parallel+O(\rho^{-\alpha_{2}})
\]
On the other hand the inequalities (3.23) and (3.35) imply that $\parallel
h_{i}\parallel=O(\rho^{-\alpha_{2}}).$ These relations and the above
inequality $\mid\Lambda_{N}-a\mid<2M,$ yield the proof of the theorem
\end{proof}

Thus we iterated (3.28) one time and got (3.32) from which the formula (3.39)
is obtained. Now to obtain the asymptotic formulas of the arbitrary order we
repeat this iteration $2p_{1}$ times. For this we need to estimate the
distance of $\lambda_{j,\beta}(v,\tau)$ and $\lambda_{j^{^{\prime}},\beta
}(v,\tau)$ for $j^{^{\prime}}\neq j,$ namely we use the \ following lemma.

\begin{lemma}
\bigskip\ There exists a positive function $\varepsilon(\rho)$ such that
$\varepsilon(\rho)\rightarrow0$ as $\rho\rightarrow\infty$ and the set
$A(\varepsilon(\rho))\equiv(\varepsilon(\rho),\frac{1}{2}-\varepsilon
(\rho))\cup(\frac{1}{2}+\varepsilon(\rho),1-\varepsilon(\rho))$ is a subset
of
\[
W(\rho)\equiv\{v\in(0,1):\mid\mu_{j}(v)-\mu_{j^{^{\prime}}}(v)\mid>\frac
{2}{\ln\rho},\forall j^{^{\prime}},j\in\mathbb{Z},j^{^{\prime}}\neq j\}.
\]
If $v(\beta)\in W(\rho)$, then
\begin{equation}
\mid\lambda_{j,\beta}(v,\tau)-\lambda_{j^{^{\prime}},\beta}(v,\tau)\mid
>2(\ln\rho)^{-1},\text{ }\forall j^{^{\prime}}\neq j. \tag{3.41}%
\end{equation}

\end{lemma}

\begin{proof}
\bigskip Denote by $\widetilde{\mu}_{1}(v),\widetilde{\mu}_{2}(v),...,$ the
eigenvalues of $T_{v}(Q(\zeta))$ numbered in nondecreasing order:
$\widetilde{\mu}_{1}(v)\leq\widetilde{\mu}_{2}(v)\leq...$.It is well-known
that the spectrum of Hill's operator $T(Q(\zeta))$ consists of the intervals%
\[
\Delta_{2j-1\text{ }}\equiv\lbrack\widetilde{\mu}_{2j-1}(0),\widetilde{\mu
}_{2j-1}(\frac{1}{2})],\text{ }\Delta_{2j\text{ }}\equiv\lbrack\widetilde{\mu
}_{2j}(\frac{1}{2}),\widetilde{\mu}_{2j}(1)]
\]
for $j=1,2,....$ The length of the $j$th interval $\Delta_{j\text{ }}$of the
spectrum tends to infinity as $j$ tends to infinity. The distance between
neighboring intervals, that is the length of gaps in spectrum, tends to zero.
The eigenvalues $\widetilde{\mu}_{2j-1}(v)$ and $\widetilde{\mu}_{2j}(v)$ are
increasing continuous functions in the intervals $(0,\frac{1}{2})$ and
$(\frac{1}{2},1)$ respectively and $\widetilde{\mu}_{j}(1+v)=\widetilde{\mu
}_{j}(v)=\widetilde{\mu}_{j}(1-v).$ Since $(\ln\rho)^{-1}\rightarrow0$ as
$\rho\rightarrow\infty,$ the length of the interval $\Delta_{j\text{ }}$ is
sufficiently greater than $(\ln\rho)^{-1}$ for $\rho\gg1$ and there are
numbers $\varepsilon_{j}^{^{\prime}}(\rho),\varepsilon_{j}^{^{\prime\prime}%
}(\rho)$ in $(0,\frac{1}{2})$ such that
\begin{align}
\widetilde{\mu}_{2j-1}(\varepsilon_{2j-1}^{^{\prime}}(\rho))  &
=\widetilde{\mu}_{2j-1}(0)+(\ln\rho)^{-1},\nonumber\\
\widetilde{\mu}_{2j-1}(\frac{1}{2}-\varepsilon_{j}^{^{\prime\prime}}(\rho))
&  =\widetilde{\mu}_{2j-1}(\frac{1}{2})-(\ln\rho)^{-1},\tag{3.42}\\
\widetilde{\mu}_{2j}(\frac{1}{2}+\varepsilon_{2j}^{^{\prime}}(\rho))  &
=\widetilde{\mu}_{2j}(\frac{1}{2})+(\ln\rho)^{-1},\nonumber\\
\widetilde{\mu}_{2j}(1-\varepsilon_{j}^{^{\prime\prime}}(\rho))  &
=\widetilde{\mu}_{2j}(1)-(\ln\rho)^{-1}.\nonumber
\end{align}
Denote $\varepsilon^{^{\prime}}(\rho)=\sup_{j}\varepsilon_{j}^{^{\prime}}%
(\rho),$ $\varepsilon^{^{\prime\prime}}(\rho)=\sup_{j}\varepsilon
_{j}^{^{\prime\prime}}(\rho),$ $\varepsilon(\rho)=\max\{\varepsilon^{^{\prime
}}(\rho),$ $\varepsilon^{^{\prime\prime}}(\rho)\}.$ To prove that
$\varepsilon(\rho)\rightarrow0$ as $\rho\rightarrow\infty$ we show that both
$\varepsilon^{^{\prime}}(\rho)$ and $\varepsilon^{^{\prime\prime}}(\rho)$ tend
to zero as $\rho\rightarrow\infty.$ If $\rho_{1}<\rho_{2}$ then $\varepsilon
_{j}^{^{\prime}}(\rho_{2})<\varepsilon_{j}^{^{\prime}}(\rho_{1}),$
$\varepsilon^{^{\prime}}(\rho_{2})<\varepsilon^{^{\prime}}(\rho_{1}),$ since
$\widetilde{\mu}_{2j-1}(v)$ and $\widetilde{\mu}_{2j}(v)$ are increasing
functions in intervals $(0,\frac{1}{2})$ and $(\frac{1}{2},1)$ respectively.
Hence $\varepsilon^{^{\prime}}(\rho)\rightarrow a\in\lbrack0,\frac{1}{2}]$ as
$\rho\rightarrow\infty.$ Suppose that $a>0.$ Then there is sequence $\rho
_{k}\rightarrow\infty$ as $k\rightarrow\infty$ such that $\varepsilon
^{^{\prime}}(\rho_{k})>\frac{a}{2}$ for all $k.$ This implies that there is a
sequence $\{i_{k}\}$ and without loss of generality it can be assumed that
there is a sequence $\{2j_{k}-1\}$ of odd numbers such that $\varepsilon
_{2j_{k}-1}^{^{\prime}}(\rho_{k})>\frac{a}{4}$ for all $k.$ Since
$\widetilde{\mu}_{2j-1}(v)$ increases in $(0,\frac{1}{2})$ and $\widetilde
{\mu}_{2j_{k}-1}(\varepsilon_{2j_{k}-1}^{^{\prime}}(\rho_{k}))-\widetilde{\mu
}_{2j_{k}-1}(0)=(\ln\rho_{k})^{-1}$ we have%
\[
\mid\widetilde{\mu}_{2j_{k}-1}(\frac{a}{4})-\widetilde{\mu}_{2j_{k}-1}%
(0)\mid\leq(\ln\rho_{k})^{-1}\rightarrow0
\]
as $k\rightarrow\infty,$ which contradicts the well-known asymptotic formulas
for eigenvalues $\widetilde{\mu}_{j}(v),$ for $v=0$ and $v=\frac{a}{4},$ where
$a\in(0,\frac{1}{2}].$ Thus we proved that $\varepsilon^{^{\prime}}%
(\rho)\rightarrow0$ as $\rho\rightarrow\infty$. In the same way we prove this
for $\varepsilon^{^{\prime\prime}}(\rho)$, and hence for $\varepsilon(\rho).$
Now suppose $v\in A(\varepsilon(\rho)).$ Using (3.42), the definition of
$\varepsilon(\rho),$ and taking into account that $\widetilde{\mu}_{2j-1}(v)$
and $\widetilde{\mu}_{2j}(v)$ increase in $(0,\frac{1}{2})$ and $(\frac{1}%
{2},1)$ respectively, we obtain that the eigenvalues $\widetilde{\mu}%
_{1}(v),\widetilde{\mu}_{2}(v),...,$ are in intervals%
\[
\lbrack\widetilde{\mu}_{2j-1}(0)+(\ln\rho)^{-1},\widetilde{\mu}_{2j-1}%
(\frac{1}{2})-(\ln\rho)^{-1}],\text{ }[\widetilde{\mu}_{2j}(\frac{1}{2}%
)+(\ln\rho)^{-1},\widetilde{\mu}_{2j}(1)-(\ln\rho)^{-1}]
\]
for $j=1,2,...,$ and in each interval there exists a unique eigenvalue of
$T_{v}.$ Therefore the distance between eigenvalues of $T_{v}$ for $v\in
A(\varepsilon(\rho))$ is not less than the distance between these intervals,
which is not less than $2(\ln\rho)^{-1}.$ Hence the inequality in the
definition of $W(\rho)$ holds, i.e., $A(\varepsilon(\rho))\subset W(\rho).$
Inequality (3.41) follows from the definition of $W(\rho)$
\end{proof}

It follow from formulas (3.35), (3.41) and (3.39) that
\begin{equation}
\mid\Lambda_{N}(\lambda_{j,\beta})-\lambda_{j^{k},\beta^{k}}(v,\tau
)\mid>c(\beta^{k},\rho),\forall\text{ }v(\beta)\in W(\rho), \tag{3.43}%
\end{equation}
where $(j_{k},\beta_{k})\in Q(\rho^{\alpha},9r_{k}),$ $k=1,2,...,p-1$;
$c(\beta^{k},\rho)=(\ln\rho)^{-1}$ when $\beta^{k}=\beta,$ $j^{k}\neq j$ and
$c(\beta^{k},\rho)=\frac{1}{2}\rho^{\alpha_{2}}$ when $\beta^{k}\neq\beta.$
Now to obtain the asymptotic formulas of the arbitrary order for $\Lambda
_{N}(t)$ we iterate the formula (3.28) $2p_{1}$ times, by using (3.43), as
follows. Since $\mid j\delta\mid<r_{1}$ ( see (3.5)), $(j_{1},\beta_{1})\in
Q(\rho^{\alpha},9r_{1}),(j_{2},\beta_{2})\in Q(\rho^{\alpha},9r_{2})$ (see
(3.32)) and $j^{2}=j+j_{1}+j_{2}$ (see (3.29) for this notation), we have
$\mid j^{2}\delta\mid<10r_{2}$. Therefore in (3.27) interchanging
$j^{^{\prime}},\beta^{^{\prime}},r,$ and $j^{2},\beta^{2},10r_{2}$ and using
the notations $r_{3}=10r_{2},$ $j^{3}=j^{2}+j_{3},$ $\beta^{3}=\beta^{2}%
+\beta_{3}$ ( see Lemma 3.5), we obtain%
\[
(\Lambda_{N}-\lambda_{j^{2},\beta^{2}})b(N,j^{2},\beta^{2})=O(\rho^{-p\alpha
})+
\]%
\begin{equation}%
{\displaystyle\sum_{(j_{3},\beta_{3})\in Q(\rho^{\alpha},9r_{3})}}
b(N,j^{3},\beta^{3})A(j^{2},\beta^{2},j^{3},\beta^{3}). \tag{3.44}%
\end{equation}
Dividing both side of (3.44) by $\Lambda_{N}-\lambda_{j^{2},\beta^{2}}$ and
using (3.43), we get
\[
b(N,j^{2},\beta^{2})=O(\rho^{-p\alpha}(c(\beta^{2},\rho))^{-1})+
\]%
\begin{equation}%
{\displaystyle\sum_{(j_{3},\beta_{3})\in Q(\rho^{\alpha},9r_{3})}}
\dfrac{b(N,j^{3},\beta^{3})A(j^{2},\beta^{2},j^{3},\beta^{3})}{\Lambda
_{N}-\lambda_{j^{2},\beta^{2}}}. \tag{3.45}%
\end{equation}
for $(j^{2},\beta^{2})\neq(j,\beta).$ In the same way we obtain
\[
b(N,j^{k},\beta^{k})=O(\rho^{-p\alpha}(c(\beta^{k},\rho))^{-1})+
\]%
\begin{equation}%
{\displaystyle\sum_{(j_{k+1},\beta_{k+1})\in Q(\rho^{\alpha},9r_{k+1})}}
\dfrac{b(N,j^{k+1},\beta^{k+1})A(j^{k},\beta^{k},j^{k+1},\beta^{k+1})}%
{\Lambda_{N}-\lambda_{j^{k},\beta^{k}}}. \tag{3.46}%
\end{equation}
for $(j^{k},\beta^{k})\neq(j,\beta),$ $k=3,4,....$ Now we isolate the terms in
the right-hand side of (3.32) with multiplicand $b(N,j,\beta)$ , i.e.,the case
$(j^{2},\beta^{2})=(j,\beta)$, and replace $b(N,j^{2},\beta^{2})$ in (3.32) by
the right-hand side of (3.45) when $(j^{2},\beta^{2})\neq(j,\beta)$ and use
(3.30), (3.43) to get
\begin{equation}
(\Lambda_{N}-\lambda_{j,\beta})b(N,j,\beta)=%
{\displaystyle\sum_{(j_{1},\beta_{1})\in Q(\rho^{\alpha},9r_{1})}}
\dfrac{A(j_{,}\beta,j^{1},\beta^{1})A(j^{1},\beta^{1},j,\beta)}{\Lambda
_{N}-\lambda_{j+j_{1},\beta+\beta_{1}}}b(N,j,\beta)+ \tag{3.47}%
\end{equation}%
\[%
{\displaystyle\sum_{(j_{1},\beta_{1}),(j_{2},\beta_{2}),(j_{3},\beta_{3})}}
\dfrac{A(j_{,}\beta,j^{1},\beta^{1})A(j_{,}^{1}\beta^{1},j^{2},\beta
^{2})A(j^{2},\beta^{2},j^{3},\beta^{3})b(N,j^{3},\beta^{3})}{(\Lambda
_{N}-\lambda_{j+j_{1},\beta+\beta_{1}})(\Lambda_{N}-\lambda_{j^{2},\beta^{2}%
})}+O(\rho^{-p\alpha}),
\]
where the last summation is taken under conditions $(j^{2},\beta^{2}%
)\neq(j,\beta)$ and $(j_{i},\beta_{i})\in Q(\rho^{\alpha},9r_{i}),$ for
$i=1,2,3$. The formula (3.47) is the two times iteration of (3.28). Again
isolating the terms with multiplicand $b(N,j,\beta)$ (i.e., the case
$(j^{3},\beta^{3})=(j,\beta))$ and replacing $b(N,j^{3},\beta^{3})$ by the
right-hand side of (3.46) (for $k=3)$ when $(j^{3},\beta^{3})\neq(j,\beta)$,
we obtain
\begin{equation}
(\Lambda_{N}-\lambda_{j,\beta})b(N,j,\beta)=(S_{1}^{^{\prime}}(\Lambda
_{N},\lambda_{j,\beta})+S_{2}^{^{\prime}}(\Lambda_{N},\lambda_{j,\beta
}))b(N,j,\beta)+C_{3}^{^{\prime}}+O(\rho^{-p\alpha}), \tag{3.48}%
\end{equation}
where $\ S_{1}^{\prime}(\Lambda_{N},\lambda_{j,\beta})=%
{\displaystyle\sum_{(j_{1},\beta_{1})\in Q(\rho^{\alpha},9r_{1})}}
\dfrac{A(j^{1},\beta^{1},j,\beta)}{\Lambda_{N}-\lambda_{j+j_{1},\beta
+\beta_{1}}}A(j_{,}\beta,j^{1},\beta^{1}),$%
\[
S_{2}^{\prime}(\Lambda_{N},\lambda_{j,\beta})=%
{\displaystyle\sum_{\substack{(j_{1},\beta_{1})\in Q(\rho^{\alpha}%
,9r_{1}),\\(j_{2},\beta_{2})\in Q(\rho^{\alpha},9r_{2}),(j^{2},\beta^{2}%
)\neq(j,\beta)}}}
\dfrac{A(j^{1},\beta^{1},j^{2},\beta^{2})A(j^{2},\beta^{2},j,\beta)}%
{(\Lambda_{N}-\lambda_{j+j_{1},\beta+\beta_{1}})(\Lambda_{N}-\lambda
_{j^{2},\beta^{2}})}A(j_{,}\beta,j^{1},\beta^{1}).
\]%
\[
C_{3}^{\prime}=\sum_{\substack{(j_{1},\beta_{1}),(j_{2},\beta_{2}%
),\\(j_{3},\beta_{3}),(j_{4},\beta_{4})}}\dfrac{A(j^{1},\beta^{1},j^{2}%
,\beta^{2})A(j^{2},\beta^{2},j^{3},\beta^{3})A(j^{3},\beta^{3},j^{4},\beta
^{4})}{(\Lambda_{N}-\lambda_{j+j_{1},\beta+\beta_{1}})(\Lambda_{N}%
-\lambda_{j^{2},\beta^{2}})(\Lambda_{N}-\lambda_{j^{3},\beta^{3}})}%
A(j_{,}\beta,j^{1},\beta^{1})b(N,j^{4},\beta^{4}),
\]
and the summation for $C_{3}^{^{\prime}}$ are taken under the conditions
$(j_{i},\beta_{i})\in Q(\rho^{\alpha},9r_{i}),$ for

$i=1,2,3,4$ and $(j^{i},\beta^{i})\neq(j,\beta)$ for $i=2,3.$ The formula
(3.48) is the three times iteration of (3.28). Repeating these process
$2p_{1}$ times, i.e., in (3.47) isolating the terms with multiplicand
$b(N,j,\beta)$ (i.e., the case $(j^{4},\beta^{4})=(j,\beta))$ and replacing
$b(N,j^{4},\beta^{4})$ by the right-hand side of (3.46) (for $k=4)$ when
$(j^{4},\beta^{4})\neq(j,\beta)$ etc., we obtain
\begin{equation}
(\Lambda_{N}-\lambda_{j,\beta})b(N,j,\beta)=A_{p_{1}-1}^{^{\prime}}%
(\Lambda_{N},\lambda_{j,\beta})b(N,j,\beta)+C_{2p_{1}-1}^{^{\prime}}%
+O(\rho^{-p\alpha}), \tag{3.49}%
\end{equation}
where $A_{n}^{^{\prime}}(\Lambda_{N},\lambda_{j,\beta})=\sum_{k=1}^{2n}%
S_{k}^{^{\prime}}(\Lambda_{N},\lambda_{j,\beta}),$ $S_{1}^{^{\prime}},$
$S_{2}^{^{\prime}}$ are defined in (3.48) and
\[
S_{k}^{\prime}(\Lambda_{N},\lambda_{j,\beta})=%
{\displaystyle\sum}
(\prod_{i=2}^{k}\dfrac{A(j_{,}^{i-1}\beta^{i-1},j^{i},\beta^{i})}{(\Lambda
_{N}-\lambda_{j^{i},\beta^{i}})})\dfrac{A(j_{,}\beta,j^{1},\beta^{1})}%
{\Lambda_{N}-\lambda_{j+j_{1},\beta+\beta_{1}}}A(j^{k},\beta^{k},j,\beta),
\]%
\[
C_{k}^{\prime}=%
{\displaystyle\sum}
(\prod_{i=2}^{k}\dfrac{A(j_{,}^{i-1}\beta^{i-1},j^{i},\beta^{i})}{(\Lambda
_{N}-\lambda_{j^{i},\beta^{i}})})\dfrac{A(j_{,}\beta,j^{1},\beta^{1})}%
{\Lambda_{N}-\lambda_{j+j_{1},\beta+\beta_{1}}}A(j^{k},\beta^{k},j^{k+1}%
,\beta^{k+1})b(N,j^{k+1},\beta^{k+1})
\]
for $k\geq2.$ Here the summations for $S_{k}^{\prime},$ and $C_{k}^{^{\prime}%
}$ are taken under the conditions

$(j_{i},\beta_{i})\in Q(\rho^{\alpha},9r_{i}),$ $(j^{i},\beta^{i})\neq
(j,\beta),$ for $i=1,2,...,k-1$ and for $i=1,2,...,k$ respectively. Besides by
definition of $Q(\rho^{\alpha},9r_{i})$ we have $\beta_{k}\neq0$ for
$k=1,2,....$ Therefore $\beta^{1}\neq\beta$ and the equality $\beta^{i}=\beta$
implies that $\beta^{i\pm1}\neq\beta.$ Hence the number of the multiplicands
$\Lambda_{N}-\lambda_{j^{i},\beta^{i}}$ in the denominators of $S_{k}%
^{^{\prime}}$ and $C_{2p_{1}-1}^{^{\prime}}$ satisfying
\[
\mid\Lambda_{N}(\lambda_{j,\beta})-\lambda_{j^{i},\beta^{i}}\mid>\frac{1}%
{2}\rho^{\alpha_{2}}%
\]
( see (3.43)) is not less than $\frac{k}{2}$ and $p_{1}$ respectively. Now
using (3.23) and the first inequality of \ (3.40), we obtain
\begin{align}
C_{2p_{1}-1}^{^{\prime}}  &  =O((\rho^{-\alpha_{2}}\ln\rho)^{p_{1}}%
)=O(\rho^{-p\alpha}),\text{ }S_{1}^{^{\prime}}(\Lambda_{N},\lambda_{j,\beta
})=O(\rho^{-\alpha_{2}}),\tag{3.50}\\
S_{2k-1}^{^{\prime}}(\Lambda_{N},\lambda_{j,\beta})  &  =O((\rho^{-\alpha_{2}%
}\ln\rho)^{k}),S_{2k}^{^{\prime}}(\Lambda_{N},\lambda_{j,\beta})=O((\rho
^{-\alpha_{2}}\ln\rho)^{k}).\nonumber
\end{align}
To prove this estimation we used (3.43). Moreover, if a real number $a$ satisfies

$\mid a-\lambda_{j,\beta}\mid<(\ln\rho)^{-1}$ then, by (3.35), (3.37) we have
\[
\mid a-\lambda_{j^{k},\beta^{k}}(v,\tau)\mid>c(\beta^{k},\rho).
\]
Therefore using this instead of (3.43) and repeating the proof of (3.50) we
obtain
\begin{equation}
S_{1}^{^{\prime}}(a,\lambda_{j,\beta})=O(\rho^{-\alpha_{2}}),\text{ }%
S_{2k-1}^{^{\prime}}(a,\lambda_{j,\beta})=O(\rho^{-\alpha_{2}}\ln\rho
)^{k}),\text{ }S_{2k}^{^{\prime}}(a,\lambda_{j,\beta})=O(\rho^{-\alpha_{2}}%
\ln\rho)^{k}). \tag{3.51}%
\end{equation}

\begin{theorem}
For every eigenvalue $\lambda_{j,\beta}(v,\tau)$ of the operator
$L_{t}(q^{\delta})$ such that

$\beta+\tau+(j+v)\delta\in V_{\delta}^{^{\prime}}(\rho^{\alpha_{1}})$,
$v(\beta)\in W(\rho)$ there exists an eigenvalue $\Lambda_{N},$ denoted by
$\Lambda_{N}(\lambda_{j,\beta}(v,\tau)),$ of $L_{t}(q)$ satisfying the
formulas
\begin{equation}
\Lambda_{N}(\lambda_{j,\beta}(v,\tau))=\lambda_{j,\beta}(v,\tau)+E_{k-1}%
(\lambda_{j,\beta})+O(\rho^{-k\alpha_{2}}(\ln\rho)^{2k}), \tag{3.52}%
\end{equation}
where $E_{0}=0,$ $E_{s}=A_{s}^{^{\prime}}(\lambda_{j,\beta}+E_{s-1}%
,\lambda_{j,\beta})$ for $s=1,2,...,$
\begin{equation}
E_{k-1}(\lambda_{j,\beta})=O(\rho^{-\alpha_{2}}(\ln\rho)) \tag{3.53}%
\end{equation}
for $k=1,2,...,[\frac{1}{9}(p-\frac{1}{2}\varkappa(d-1)],$ and $A_{s}%
^{^{\prime}}$ is defined in (3.49).
\end{theorem}

\begin{proof}
The proof of this Theorem is similar to the proof of Theorem 2.1(a). By
Theorem 3.1 formula (3.52) for the case $k=1$ is proved and $E_{0}=0$. Hence
(3.53) for $k=1$ is also proved. The proof of (3.53), for arbitrary $k,$
follows from (3.51) and the definition of $E_{s}$ by induction. Now we prove
(3.52) by induction. Assume that (3.52) is true for $k=s<[\frac{1}{9}%
(p-\frac{1}{2}\varkappa(d-1)]$ ,i.e.,%
\[
\Lambda_{N}=\lambda_{j,\beta}+E_{s-1}+O(\rho^{-s\alpha_{2}}(\ln\rho)^{2s})).
\]
Putting this expression for $\Lambda_{N\text{ }}$ into $A_{p_{1}-1}^{^{\prime
}}(\Lambda_{N},\lambda_{j,\beta})$, dividing both sides of (3.49) by
$b(N,j,\beta),$ taking into account that $A_{p_{1}-1}^{^{\prime}}(\Lambda
_{N},\lambda_{j,\beta})=A_{s}^{^{\prime}}(\Lambda_{N},\lambda_{j,\beta
})+O(\rho^{-(s+1)\alpha_{2}}(\ln\rho)^{(s+1)})$ ( see definition of
$A_{s}^{^{\prime}}$ and (3.51)), using (3.50), (3.51), assertion $(ii)$ of
Lemma 3.6 and the equality $\alpha_{2}=9\alpha,$ we get%
\[
\Lambda_{N}=\lambda_{j,\beta}+A_{p_{1}-1}^{^{\prime}}(\lambda_{j,\beta
}+E_{s-1}+O(\frac{(\ln\rho)^{2s}}{\rho^{s\alpha_{2}}}),\lambda_{j,\beta
})+O(\rho^{-\frac{1}{9}(p-\frac{1}{2}\tau(d-1))\alpha_{2}})
\]%
\begin{align*}
&  =\lambda_{j,\beta}+A_{s}^{^{\prime}}(\lambda_{j,\beta}+E_{s-1}%
,\lambda_{j,\beta})+O(\rho^{-(s+1)\alpha_{2}}(\ln\rho)^{2(s+1)}))+O(\rho
^{-\frac{1}{9}(p-\frac{1}{2}\tau(d-1))\alpha_{2}})\\
&  \{A_{s}^{^{\prime}}(\lambda_{j,\beta}+E_{s-1}+O(\rho^{-s\alpha_{2}}(\ln
\rho)^{2s}),\lambda_{j,\beta})-A_{s}^{^{\prime}}(\lambda_{j,\beta}%
+E_{s-1},\lambda_{j,\beta})\}.
\end{align*}
To prove (3.52) for $k=s+1$ we need to show that the expression in the curly
brackets is equal to $O((\rho^{-(s+1)\alpha_{2}}(\ln\rho)^{2s+1}).$ This can
be checked by using the estimations (3.24), (3.53), (3.35), (3.37) and the
obvious relation
\begin{align*}
&  \dfrac{1}{\prod_{i=1}^{n}(\lambda_{j,\beta}+E_{s-1}+O(\rho^{-s\alpha_{2}%
}(\ln\rho)^{2s})-\lambda_{j^{i},\beta^{i}})}-\dfrac{1}{\prod_{i=1}^{n}%
(\lambda_{j,\beta}+E_{s-1}-\lambda_{j^{i},\beta^{i}})}\\
&  =\dfrac{1}{\prod_{i=1}^{n}(\lambda_{j,\beta}+E_{s-1}-\lambda_{j^{i}%
,\beta^{i}})}(\frac{1}{1+O(\rho^{-s\alpha_{2}}(\ln\rho)^{2s}\ln\rho)}-1)\\
&  =O(\rho^{-(s+1)\alpha_{2}}(\ln\rho)^{2(s+1)})
\end{align*}
for $n=1,2,...,2p_{1}$
\end{proof}

\begin{remark}
Here we note some properties of the known parts $\lambda_{j,\beta}+E_{k}$ (see
(3.52)), where $\lambda_{j,\beta}=\mu_{j}(v)+\mid\beta+\tau\mid^{2}$ ( see
Lemma 3.1), of the eigenvalues of $L_{t}(q)$. We prove that
\begin{equation}
\frac{\partial(E_{k}(\mu_{j}(v)+\mid\beta+\tau\mid^{2}))}{\partial\tau_{i}%
}=O(\rho^{-2\alpha_{2}+\alpha}\ln\rho) \tag{3.54}%
\end{equation}
for $i=1,2,...,d-1,$ where $\tau=(\tau_{1},\tau_{2},...,\tau_{d-1}%
),k<[\frac{1}{9}(p-\frac{1}{2}\tau(d-1)]$, and $v(\beta)\in W(\rho).$ To prove
(3.54) for $k=1$ we evaluate the derivatives of
\[
H(\beta^{k},j^{k},\tau,v)\equiv(\mu_{j}(v)+\mid\beta+\tau\mid^{2}-\mu_{j^{k}%
}(v)-\mid\beta^{k}+\tau\mid^{2})^{-1}.
\]
Since $\mu_{j}(v)$, and $\mu_{j^{^{\prime}}}(v)$ do not depend on $\tau_{i},$
the function $H(\beta^{k},j^{k},\tau,v)$ for $\beta^{k}=\beta$ do not depend
on $\tau_{i}.$ Besides it follows from the definition of $W(\rho)$ ( see Lemma
3.7) that $H(\beta,j^{k},\tau,v)=O(\ln\rho).$ For $\beta^{k}\neq\beta$ using
(3.35), and equality
\[
\mid\beta^{k}-\beta\mid=\mid\beta_{1}+\beta_{2}...+\beta_{i}\mid
=O(\rho^{\alpha})
\]
(see last inequality in (3.33)), we obtain that the derivatives of
$H(\beta^{k},j^{k},\tau,v)$ is equal to $O(\rho^{-2\alpha_{2}+\alpha}).$
Therefore using (3.23) and the definition of $E_{1}(\lambda_{j,\beta})$ ( see
(3.52) and (3.49)), by direct calculation, we get (3.54) for $k=1.$ Now
suppose that (3.54) holds for $k=s-1.$ Using this, replacing $\mu_{j}%
+\mid\beta+\tau\mid^{2}$ by $\mu_{j}+\mid\beta+\tau\mid^{2}+E_{s-1}$ in
$H(\beta^{k},j^{k},\tau,v)$, arguing as above we get (3.54) for $k=s$.
\end{remark}

\section{Asymptotic Formulas for the Bloch Functions}

In this section using the asymptotic formulas for the eigenvalues and the
simplicity conditions (1.28), (1.29), we obtain the asymptotic formulas for
the Bloch functions with a quasimomentum of the simple set $B$ defined in
Definition 1.2.

\begin{theorem}
If $\gamma+t\in B,$ then there exists a unique eigenvalue $\Lambda_{N}(t)$
satisfying (1.14) for $k=1,2,...,[\frac{p}{3}],$ where $p$ is defined in
(1.6). This eigenvalue is a simple eigenvalue of $L_{t}(q)$ and the
corresponding eigenfunction $\Psi_{N,t}(x),$ denoted by $\Psi_{\gamma+t}(x),$
satisfies (1.32) if $q(x)\in W_{2}^{s_{0}}(F),$ where $s_{0}$ is defined in (1.2).
\end{theorem}

\begin{proof}
By Theorem 2.1(b) if $\gamma+t\in B\subset U(\rho^{\alpha_{1}},p),$ then there
exists an eigenvalue $\Lambda_{N}(t)$ satisfying (1.14) for $k=1,2,...,[\frac
{1}{3}(p-\frac{1}{2}\varkappa(d-1))]$ and by the first inequality of (1.40)
formula (1.14) holds for $k=k_{1}.$ Therefore using (1.14) for $k=k_{1},$ the
relation $3k_{1}\alpha>d+2\alpha$ ( see the second inequality of (1.40)), and
the notations of \ (1.26), we obtain that the eigenvalue $\Lambda_{N}(t)$
satisfies the asymptotic formula (1.27). Let $\Psi_{N,t}(x)$ be any normalized
eigenfunction corresponding to $\Lambda_{N}(t)$. Since the normalized
eigenfunction is defined up to constant of modulus $1,$ without loss of
generality it can assumed that $\arg b(N,\gamma)=0,$ where $b(N,\gamma
)=(\Psi_{N,t}(x),e^{i(\gamma+t,x)}).$ Therefore to prove (1.32) it suffices to
show that (1.31) holds. To prove (1.31) we estimate the following summations
\begin{equation}
\ \sum_{\gamma^{^{\prime}}\notin K}\mid b(N,\gamma^{^{\prime}})\mid^{2},\text{
}\sum_{\gamma^{^{\prime}}\in K\backslash\{\gamma\}}\mid b(N,\gamma^{^{\prime}%
})\mid^{2} \tag{4.1}%
\end{equation}
separately, where $K$ is defined by (1.30). Using (1.27) and (1.30), we get
\begin{equation}
\mid\Lambda_{N}(t)-\mid\gamma^{^{\prime}}+t\mid^{2}\mid>\frac{1}{4}%
\rho^{\alpha_{1}},\text{ }\forall\gamma^{^{\prime}}\notin K, \tag{4.2}%
\end{equation}%
\begin{equation}
\mid\Lambda_{N}(t)-\mid\gamma^{^{\prime}}+t\mid^{2}\mid<\frac{1}{2}%
\rho^{\alpha_{1}},\text{ }\forall\gamma^{^{\prime}}\in K. \tag{4.3}%
\end{equation}
It follows from (1.8) and (4.2) that%
\begin{equation}
\sum_{\gamma^{^{\prime}}\notin K}\mid b(N,\gamma^{^{\prime}})\mid
^{2}=\parallel q\Psi_{N,t}\parallel^{2}O(\rho^{-2\alpha_{1}})=O(\rho
^{-2\alpha_{1}}). \tag{4.4}%
\end{equation}
Now let us estimate the second summation in (4.1). For this, we prove that the
simplicity conditions (1.28), (1.29) imply
\begin{equation}
\mid b(N,\gamma^{^{\prime}})\mid\leq c_{5}\rho^{-c\alpha},\text{ }%
\forall\gamma^{^{\prime}}\in K\backslash\{\gamma\}, \tag{4.5}%
\end{equation}
where $c=p-d\varkappa-\frac{1}{4}d3^{d}-3.$ The conditions $\gamma^{^{\prime}%
}\in K,$ $\gamma+t\in B$ ( see (1.30) and the Definition 1.2), the notation
(1.26) and the equality (2.8) yield the inclusion $\gamma^{^{\prime}}+t\in
R(\frac{3}{2}\rho)\backslash R(\frac{1}{2}\rho).$ By (2.33) there are two cases.

Case 1: $\gamma^{^{\prime}}+t\in U(\rho^{\alpha_{1}},p).$ Case 2:
$\gamma^{^{\prime}}+t\in(E_{s}\backslash E_{s+1}),$ where $s=1,2,...,d-1.$ To
prove (4.5) in Case 1 and Case 2, we suppose that (4.5) does not hold, use
Theorem 2.1(a) and Theorem 2.2(a) respectively to get a contradiction.

Case 1. If the inequality in (4.5) is not true, then by (4.3) the conditions
of Theorem 2.1(a) hold and hence we have
\begin{equation}
\Lambda_{N}(t)=\mid\gamma^{^{\prime}}+t\mid^{2}+F_{k-1}(\gamma^{^{\prime}%
}+t)+O(\rho^{-3k\alpha}) \tag{4.6}%
\end{equation}
for $k\leq\lbrack\frac{1}{3}(p-c)]=[\frac{1}{3}(d\varkappa+\frac{1}{4}%
d3^{d}+3)].$ On the other hand it follows from the definitions $k_{1}%
\equiv\lbrack\frac{d}{3\alpha}]+2$ ( see (1.26)), $\alpha\equiv\frac
{1}{\varkappa}$ ( see (1.6)) of $k_{1}$ and $\alpha$ that
\[
k_{1}\leq\frac{1}{3}d\varkappa+2<\frac{1}{3}(d\varkappa+\frac{1}{4}d3^{d}+3),
\]
\ that is, formula (4.6) holds for $k=k_{1}.$ Therefore arguing as in the
prove of (1.27) ( see the beginning of the proof of this theorem), we get
\[
\Lambda_{N}(t)-F(\gamma^{^{\prime}}+t)=o(\varepsilon_{1}).
\]
\ This with (1.27) contradicts (1.28). Thus (4.5) in Case 1 is proved.
Similarly, if the inequality in (4.5) does not hold in Case 2 ,that is, for
$\gamma^{^{\prime}}+t\in(E_{s}\backslash E_{s+1})$ and $\gamma^{^{\prime}}\in
K,$ then by (4.3) the conditions of Theorem 2.2(a) hold and
\begin{equation}
\Lambda_{N}(t)=\lambda_{j}(\gamma^{^{\prime}}+t)+O(\rho^{-(p-c-\frac{1}%
{4}d3^{d})\alpha}), \tag{4.7}%
\end{equation}
where $(p-c-\frac{1}{4}d3^{d})\alpha=(d\varkappa+3)\alpha>d+2\alpha$ . Hence
we have%
\[
\Lambda_{N}(t)-\lambda_{j}(\gamma^{^{\prime}}+t)=o(\varepsilon_{1}).
\]
This with (1.27) contradicts (1.29). Thus the inequality in (4.5) holds.
Therefore, using $\mid K\mid=O(\rho^{d-1})$ (see (1.37))$,$ $\varkappa
\alpha=1$ ( see (1.6)), we get
\begin{equation}
\sum_{\gamma^{^{\prime}}\in K\backslash\{\gamma\}}\mid b(N,\gamma^{^{\prime}%
})\mid^{2}=O(\rho^{-(2c-\varkappa(d-1))\alpha})=O(\rho^{-(2p-(3d-1)\varkappa
-\frac{1}{2}d3^{d}-6)\alpha}). \tag{4.8}%
\end{equation}
If $s=s_{0},$ that is, $p=s_{0}-d,$ then $2p-(3d-1)\varkappa-\frac{1}{2}%
d3^{d}-6=6.$ Since $\alpha_{1}=3\alpha,$ the equalities (4.4) and (4.8) imply
(1.31). Thus we proved that the equality (1.32) holds for any normalized
eigenfunction $\Psi_{N,t}(x)$ corresponding to any eigenvalue $\Lambda_{N}(t)$
\ satisfying (1.14). If there exist two different eigenvalues or multiple
eigenvalue satisfying (1.14), then there exist two orthogonal normalized
eigenfunctions satisfying (1.32), which is impossible. Therefore $\Lambda
_{N}(t)$\ is a simple eigenvalue. It follows from Theorem 2.1(a) that
$\Lambda_{N}(t)$ satisfies (1.14) for $k=1,2,...,[\frac{p}{3}],$ since (1.32)
holds and hence (1.16) holds for $c=0$
\end{proof}

\begin{remark}
Since for $\gamma+t\in B$ \ there exists a unique eigenvalue satisfying
(1.14), (1.27), we denote this eigenvalue by $\Lambda(\gamma+t).$ Since this
eigenvalue is simple, we denote the \ corresponding eigenfunction by
$\Psi_{\gamma+t}(x).$ By Theorem 4.1 this eigenfunction satisfies (1.32).
Clearly, for $\gamma+t\in B$ \ there exists a unique index $N\equiv
N(\gamma+t)$ such that $\Lambda(\gamma+t)=\Lambda_{N(\gamma+t)}(t)$ and
$\Psi_{\gamma+t}(x)=\Psi_{N(\gamma+t),t}(x).$
\end{remark}

Now we prove the asymptotic formulas of arbitrary order for $\Psi_{\gamma
+t}(x).$

\begin{theorem}
If $\gamma+t\in B,$ then the eigenfunction $\Psi_{\gamma+t}(x)\equiv\Psi
_{N,t}(x)$ corresponding to the eigenvalue $\Lambda(\gamma+t)\equiv\Lambda
_{N}(t)$ satisfies formulas (1.33), for $k=1,2,...,n$, where

$n=[\frac{1}{6}(2p-(3d-1)\varkappa-\frac{1}{2}d3^{d}-6)],$%
\begin{align*}
F_{0}^{\ast}  &  =e^{i(\gamma+t,x)},\text{ }F_{1}^{\ast}=e^{i(\gamma+t,x)}+%
{\displaystyle\sum_{\gamma_{1}\in\Gamma(\rho^{\alpha})}}
\dfrac{q_{\gamma_{1}}e^{i(\gamma+t+\gamma_{1},x)}}{\mid\gamma+t\mid^{2}%
-\mid\gamma+\gamma_{1}+t\mid^{2}},\\
F_{k}^{\ast}(\gamma+t)  &  =(1+\parallel\widetilde{F}_{k}\parallel
)^{-1}(e^{i(\gamma+t,x)}+\widetilde{F}_{k}(\gamma+t)),
\end{align*}
$\widetilde{F}_{k}$ is obtained from $F_{k}$ by replacing $q_{\gamma_{1}}$
with $e^{i(\gamma-\gamma_{1}+t,x)},$ and $F_{k}$ is defined by (2.10).
\end{theorem}

\begin{proof}
By Theorem 4.1, formula (1.33) for $k=1$ is proved. To prove formula (1.33)
for $2\leq$ $k\leq n,$ first we prove the following equivalent relations
\begin{equation}
\sum_{\gamma^{^{\prime}}\in\Gamma^{c}(k-1)}\mid b(N,\gamma+\gamma^{^{\prime}%
})\mid^{2}=O(\rho^{-2k\alpha_{1}}), \tag{4.9}%
\end{equation}%
\begin{equation}
\Psi_{N,t}(x)=b(N,\gamma)e^{i(\gamma+t,x)}+\sum_{\gamma^{^{\prime}}\in
\Gamma(\frac{k-1}{n}\rho^{\alpha})}b(N,\gamma+\gamma^{^{\prime}}%
)e^{i(\gamma+t+\gamma^{^{\prime}},x)}+H_{k}(x), \tag{4.10}%
\end{equation}
where $\Gamma^{c}(k-1)\equiv\Gamma\backslash(\Gamma(\frac{k-1}{n}\rho^{\alpha
})\cup\{0\})$ and $\parallel H_{k}\parallel=O(\rho^{-k\alpha_{1}}).$ The case
$k=1$ is proved due to (1.31). Assume that (4.9) is true for $k=m<n$ . Then
using (4.10) for$\ k=m,$ and the obvious decomposition
\[
q(x)=\sum_{\gamma_{1}\in\Gamma(\frac{1}{n}\rho^{\alpha})}q_{\gamma_{1}%
}e^{i(\gamma_{1},x)}+O(\rho^{-p\alpha})
\]
(see (1.6)), we have $\Psi_{N,t}(x)q(x)=H(x)+O(\rho^{-m\alpha_{1}}),$ where
$H(x)$ is a linear combination of $e^{i(\gamma+t+\gamma^{^{\prime}},x)}$ for
$\gamma^{^{\prime}}\in\Gamma(\frac{m}{n}\rho^{\alpha})\cup\{0\}.$ Hence
$(H(x),e^{i(\gamma+t+\gamma^{^{\prime}},x)})=0$ for $\gamma^{^{\prime}}%
\in\Gamma^{c}(m).$ Thus, using (1.8), (4.2), and Bessel's inequality, we get%
\begin{align}
\sum_{\gamma^{^{\prime}}:\gamma^{^{\prime}}\in\Gamma^{c}(m),\text{ }%
\gamma+\gamma^{^{\prime}}\notin K}  &  \mid b(N,\gamma+\gamma^{^{\prime}}%
)\mid^{2}=\nonumber\\
\sum_{\gamma^{^{\prime}}:\gamma^{^{\prime}}\in\Gamma^{c}(m),\text{ }%
\gamma+\gamma^{^{\prime}}\notin K}  &  \mid\dfrac{(H(x)+O(\rho^{-m\alpha_{1}%
}),e^{i(\gamma+t+\gamma^{^{\prime}},x)})}{\Lambda_{N}-\mid\gamma
+\gamma^{^{\prime}}+t\mid^{2}}\mid^{2}= \tag{4.11}%
\end{align}%
\[
\sum_{\gamma^{^{\prime}}:\gamma^{^{\prime}}\in\Gamma^{c}(m),\text{ }%
\gamma+\gamma^{^{\prime}}\notin K}\mid\dfrac{(O(\rho^{-m\alpha_{1}%
}),e^{i(\gamma+t+\gamma^{^{\prime}},x)})}{\Lambda_{N}-\mid\gamma
+\gamma^{^{\prime}}+t\mid^{2}}\mid^{2}=O(\rho^{-2(m+1)\alpha_{1}}).
\]
On the other hand, using $\alpha_{1}=3\alpha,$ (4.8), and the definition of
$n$, we obtain
\[
\sum_{\gamma^{^{\prime}}:\gamma^{^{\prime}}\in\Gamma^{c}(m),\text{ }%
\gamma+\gamma^{^{\prime}}\in K}\mid b(N,\gamma+\gamma^{^{\prime}})\mid^{2}%
\leq\sum_{\gamma^{^{\prime}}\in K\backslash\{\gamma\}}\mid b(N,\gamma
^{^{\prime}})\mid^{2}=O(\rho^{-2n\alpha_{1}}).
\]
This with (4.11) implies (4.9) and hence (4.10) for $k=m+1$. It follows from
(4.9) that
\[
\parallel\sum_{\gamma^{^{\prime}}\in(\Gamma(\rho^{\alpha})\backslash
\Gamma(\frac{k-1}{n}\rho^{\alpha}))}b(N,\gamma+\gamma^{^{\prime}}%
)e^{i(\gamma+t+\gamma^{^{\prime}},x)}\parallel=O(\rho^{-k\alpha_{1}})
\]
Therefore the formula (4.10) for $k\leq n$ can be written in the form
\begin{equation}
\Psi_{N,t}-b(N,\gamma)e^{i(\gamma+t,x)}-\widetilde{H}_{k}=\sum_{\gamma_{1}%
\in\Gamma(\rho^{\alpha})}b(N,\gamma-\gamma_{1})e^{i(\gamma-\gamma_{1}+t,x)},
\tag{4.12}%
\end{equation}
where $\parallel\widetilde{H}_{k}\parallel=O(\rho^{-k\alpha_{1}})$. It is
clear that the right-hand side of (4.12) can be obtained from the right-hand
side of the equality%
\[
(\Lambda_{N}-\mid\gamma+t\mid^{2})b(N,\gamma)+O(\rho^{-p\alpha})=\sum
_{\gamma_{1}\in\Gamma(\rho^{\alpha})}q_{\gamma_{1}}b(N,\gamma-\gamma_{1}),
\]
which is (1.9), by replacing $q_{\gamma_{1}}$ with $e^{i(\gamma-\gamma
_{1}+t,x)}$. Therefore in (4.12) doing the iteration which was done in order
to obtain (2.5) from (1.9), we get%
\begin{equation}
\Psi_{N,t}(x)-b(N,\gamma)e^{i(\gamma+t,x)}-\widetilde{H}_{k}(x)=\widetilde
{A}_{k-1}(\Lambda_{N},\gamma+t)b(N,\gamma)+\widetilde{C}_{k}+O(\rho^{-p\alpha
}), \tag{4.13}%
\end{equation}
where $\widetilde{A}_{k}(\Lambda_{N},\gamma+t)$ and $\widetilde{C}_{k}$ is
obtained from $A_{k}(\Lambda_{N},\gamma+t)$ and $C_{k}$ by replacing
$q_{\gamma_{1}}$ with $e^{i(\gamma-\gamma_{1}+t,x)}$ respectively and the term
$O(\rho^{-p\alpha})$ in the right-hand side of (4.13) is a function whose norm
is $O(\rho^{-p\alpha}).$ It follows from the definitions of the functions
$\widetilde{F}_{k},$ $\widetilde{A}_{k},$ $\widetilde{C}_{k}$ that the
estimations similar to the estimations of $F_{k},$ $A_{k},$ $C_{k}$ holds for
these functions and the proof of these estimations are the same. Namely,
repeating the proof of (2.6), (2.8) we see that
\begin{equation}
\parallel\widetilde{A}_{k-1}\parallel=O(\rho^{-\alpha_{1}}),\text{ }%
\parallel\widetilde{C}_{k}\parallel=O(\rho^{-k\alpha_{1}}),\text{ }%
\parallel\widetilde{F}_{k-1}(\gamma+t)\parallel=O(\rho^{-\alpha_{1}}).
\tag{4.14}%
\end{equation}
Now using the equalities
\begin{equation}
b(N,\gamma)=1+O(\rho^{-2\alpha_{1}}), \tag{4.15}%
\end{equation}%
\begin{align*}
\widetilde{A}_{k-1}(\Lambda_{N},\gamma+t)  &  =\widetilde{A}_{k-1}%
(F_{k-2}(\gamma+t),\gamma+t)+O(\rho^{-k\alpha_{1}})\\
&  =\widetilde{F}_{k-1}(\gamma+t)+O(\rho^{-k\alpha_{1}})
\end{align*}
( see (1.31a),\ (1.14), (2.12) and the definition of $\widetilde{F}_{k}$) ,
dividing both side of (4.13) by $b(N,\gamma),$ we get
\begin{equation}
\frac{1}{b(N,\gamma)}\Psi_{N,t}(x)=e^{i(\gamma+t,x)}+\widetilde{F}%
_{k-1}(\gamma+t)+ \tag{4.16}%
\end{equation}%
\[
O(\rho^{-k\alpha_{1}})+\frac{1}{b(N,\gamma)}(\widetilde{H}_{k}(x)+\widetilde
{C}_{k}+O(\rho^{-p\alpha})).
\]
Moreover the relations $\parallel\widetilde{H}_{k}\parallel=O(\rho
^{-k\alpha_{1}})$ ( see (4.12)), the formulas (4.14), (4.15), and the
inequality $p\alpha\geq n\alpha_{1}\geq k\alpha_{1}$ (see definition of $n$)
imply that
\begin{equation}
\parallel O(\rho^{-k\alpha_{1}})+\frac{1}{b(N,\gamma)}(\widetilde{H}%
_{k}+\widetilde{C}_{k}+O(\rho^{-p\alpha}))\parallel=O(\rho^{-k\alpha_{1}}).
\tag{4.17}%
\end{equation}
Therefore using the equality $\parallel\Psi_{N,t}\parallel=1,$ the assumption
$\arg b(N,\gamma)=0,$ the last equality of (4.14) and taking into account that
$\widetilde{F}_{k-1}(\gamma+t)$ is a linear combination of $e^{i(\gamma
+t-\gamma_{1},x)}$ for $\gamma_{1}\in\Gamma(\rho^{\alpha})$ ( since
$\widetilde{F}_{k-1}(\gamma+t)$ is obtained from the right-hand side of
(4.12)) and hence the functions $e^{i(\gamma+t,x)}$, $\widetilde{F}%
_{k-1}(\gamma+t)$ are orthogonal, from (4.16), we obtain%
\begin{equation}
\frac{1}{b(N,\gamma)}=(1+\parallel\widetilde{F}_{k-1}(\gamma+t)\parallel
))+O(\rho^{-k\alpha_{1}})), \tag{4.18}%
\end{equation}%
\begin{equation}
\Psi_{N,t}(x)=(1+\parallel\widetilde{F}_{k-1}\parallel)^{-1}(e^{i(\gamma
+t,x)}+\widetilde{F}_{k-1}(\gamma+t)+O(\rho^{-k\alpha_{1}})). \tag{4.19}%
\end{equation}
Thus (1.33) is proved. Let us consider the case $k=2.$ Using (4.15) and (4.17)
in (4.16) for $k=2$ and recalling the definitions of $\widetilde{F}_{1},$
$F_{1}$ ( see (2.13)), we get
\begin{equation}
\Psi_{N,t}(x)=e^{i(\gamma+t,x)}+%
{\displaystyle\sum_{\gamma_{1}\in\Gamma(\rho^{\alpha})}}
\dfrac{q_{\gamma_{1}}e^{i(\gamma+t+\gamma_{1},x)}}{\mid\gamma+t\mid^{2}%
-\mid\gamma+\gamma_{1}+t\mid^{2}}+O(\rho^{-2\alpha_{1}}), \tag{4.20}%
\end{equation}
that is, we obtain the proof of the equality for $F_{1}^{\ast}(\gamma+t)$
\end{proof}

\section{Simple Sets and Isoenergetic Surfaces}

In this section we consider the simple sets $B$ defined in Definition 1.2 and
construct a large part of the isoenergetic surfaces
\[
I_{\rho}(q)=\{t\in F^{\ast}:\exists N,\Lambda_{N}(t)=\rho^{2}\}.
\]
corresponding to $\rho^{2}$ for large $\rho.$ In the case $q(x)=0$ \ the
isoenergetic surface
\[
I_{\rho}(0)=\{t\in F^{\ast}:\exists\gamma\in\Gamma,\mid\gamma+t\mid^{2}%
=\rho^{2}\}
\]
is the translation of the sphere $B(\rho)=\{\gamma+t:t\in F^{\ast},\gamma
\in\Gamma,\mid\gamma+t\mid^{2}=\rho^{2}\}$ by the vectors $\gamma\in\Gamma.$
For simplicity of formulation of the main results of this section we start
with a conversation about this results and introduce the needed notations.

\begin{notation}
We construct a part of isoenergetic surfaces by using the Property 3 ( see
introduction) of the simple set $B,$ that is, by investigation of the function
$\Lambda(\gamma+t)$ in the set $B,$ where $\Lambda(\gamma+t)$ is defined in
Remark 4.1. In other word, we consider the part
\[
\text{ }PI_{\rho}(q)\equiv\{t\in F^{\ast}:\exists\gamma\in\Gamma,\text{
}\Lambda(\gamma+t)=\rho^{2}\},
\]
of the isoenergetic surfaces $I_{\rho}(q).$ The set $PI_{\rho}(q)$ is
translation of
\[
TPI_{\rho}(q)\equiv\{\gamma+t:\Lambda(\gamma+t)=\rho^{2}\}.
\]
We say that $TPI_{\rho}(q)$ is the part of the translated (on the simple set
$B$) isoenergetic surfaces. In this section we construct the subsets $I_{\rho
}^{^{\prime}}$ and $I_{\rho}^{^{\prime\prime}}$ of $TPI_{\rho}(q)$ and
$PI_{\rho}(q)$ respectively and prove that the measures of these subsets are
asymptotically equal to the measure of the isoenergetic surfaces $I_{\rho}(0)$
of $L(0)$. In other word we construct a large (in some sense) part $I_{\rho
}^{^{\prime\prime}}$ of isoenergetic surfaces $I_{\rho}(q)$ of $L(q)$. Since
$\Lambda(\gamma+t)$ approximately equal to $F(\gamma+t)$ ( see (1.27) and
Remark 4.1) it is natural to call
\[
S_{\rho}=\{x\in U(2\rho^{\alpha_{1}},p):F(x)=\rho^{2}\},
\]
where $U$ , and $F(x)$ are defined in Definition 1.1 and in (1.26),
approximated isoenergetic surfaces in the non-resonance domain. Here we
construct a part of the simple set $B$ in neighborhood of $S_{\rho}$ that
contains $I_{\rho}^{^{\prime}}$. For this we consider the surface $S_{\rho}$.
As we noted in introduction \ ( see Step 2 and (1.28)) the eigenvalue
$\Lambda(\gamma+t)$ does not coincide with the eigenvalues $\Lambda
(\gamma+t+b)$ if $\mid F(\gamma+t)-F(\gamma+t+b)\mid>2\varepsilon_{1}$ for
$\gamma+t+b\in U(\rho^{\alpha_{1}},p)$ and $b\in\Gamma\backslash\{0\}$.
Therefore we eliminate
\begin{equation}
P_{b}=\{x:x\in S_{\rho},\text{ }x+b\in U(\frac{1}{2}\rho^{\alpha_{1}},p),\mid
F(x)-F(x+b)\mid<3\varepsilon_{1}\} \tag{5.1}%
\end{equation}
for $b\in\Gamma$ from $S_{\rho},$ denote the remaining part of $S_{\rho}$ by
$S_{\rho}^{^{\prime}},$ and consider its $\varepsilon$-neighborhood:
\[
S_{\rho}^{^{\prime}}=S_{\rho}\backslash(\cup_{b\in\Gamma}P_{b}),\text{
}U_{\varepsilon}(S_{\rho}^{^{\prime}})=\cup_{a\in S_{\rho}^{^{\prime}}%
}U_{\varepsilon}(a)\},
\]
where $\varepsilon=\frac{\varepsilon_{1}}{7\rho},$ $U_{\varepsilon}(a)=\{x\in
R^{d}:\mid x-a\mid<\varepsilon\},$ $\varepsilon_{1}=\rho^{-d-2\alpha}.$ In
Theorem 5.1 we prove that in the set $U_{\varepsilon}(S_{\rho}^{^{\prime}})$
the simplicity condition (1.28) holds. Denote by
\[
Tr(E)=\{\gamma+x\in U_{\varepsilon}(S_{\rho}^{^{\prime}}):\gamma\in\Gamma,x\in
E\},\text{ }Tr_{F^{\star}}(E)\equiv\{\gamma+x\in F^{\star}:\gamma\in
\Gamma,x\in E\}
\]
the translations of $E\subset R^{d}$ into $U_{\varepsilon}(S_{\rho}^{^{\prime
}})$ and $F^{\star}$ respectively. In order that the simplicity condition
(1.29) holds, we discard from $U_{\varepsilon}(S_{\rho}^{^{\prime}})$ the
translation $Tr(A(\rho))$ of
\begin{equation}
A(\rho)\equiv\cup_{k=1}^{d-1}(\cup_{\gamma_{1},\gamma_{2},...,\gamma_{k}%
\in\Gamma(p\rho^{\alpha})}(\cup_{i=1}^{b_{k}}A_{k,i}(\gamma_{1},\gamma
_{2},...,\gamma_{k}))), \tag{5.2}%
\end{equation}
where $A_{k,i}(\gamma_{1},...,\gamma_{k})=\{x\in(\cap_{i=1}^{k}V_{\gamma_{i}%
}(\rho^{\alpha_{k}})\backslash E_{k+1})\cap K_{\rho}:\lambda_{i}(x)\in
(\rho^{2}-3\varepsilon_{1},\rho^{2}+3\varepsilon_{1})\},$ $\lambda_{i}(x),$
$b_{k}$ is defined in Theorem 2.2, and $K_{\rho}=\{x\in\mathbb{R}^{d}:\mid\mid
x\mid^{2}-\rho^{2}\mid<\rho^{\alpha_{1}}\}.$ As a result we construct the part
$U_{\varepsilon}(S_{\rho}^{^{\prime}})\backslash Tr(A(\rho))$ of the simple
set $B$ (see Theorem 5.1(a)) which contains the set $I_{\rho}^{^{\prime}}$
(see Theorem 5.1(c)).
\end{notation}

To prove the main result ( Theorem 5.1) of this section we use the following
property, namely (5.3) and Lemma 5.1, of the set constructed in Notation 5.1:
\begin{align}
\rho-\rho^{\alpha_{1}-1}  &  <\mid x\mid<\rho+\rho^{\alpha_{1}-1},\text{
}\forall x\in U_{\varepsilon}(K_{\rho}),\nonumber\\
\text{ }  &  \mid\frac{\partial F}{\partial x_{i}}\mid<3\rho,\text{ }\forall
x\in U(\rho^{\alpha_{1}},p)\cap U_{\varepsilon}(K_{\rho}),\tag{5.3}\\
U_{\varepsilon}(S_{\rho}^{^{\prime}})  &  \subset U(\rho^{\alpha_{1}},p)\cap
K_{\rho}.\nonumber
\end{align}
To prove (5.3) recall that
\begin{align}
F(x)  &  =\mid x\mid^{2}+F_{k_{1}-1}(x),\text{ }\forall x\in U(c_{4}%
\rho^{\alpha_{1}},p)\text{ }\tag{5.4}\\
F_{k_{1}-1}(x)  &  =O(\rho^{-\alpha_{1}}),\text{ }\forall x\in U(c_{4}%
\rho^{\alpha_{1}},p)\tag{5.4(a)}\\
\frac{\partial F_{k_{1}-1}(x)}{\partial x_{i}}  &  =O(\rho^{-2\alpha
_{1}+\alpha})=O(\rho^{-5\alpha}),\text{ }\forall x\in U(c_{4}\rho^{\alpha_{1}%
},p)\tag{5.4(b)}\\
F(x)  &  =\rho^{2},\text{ }\mid x\mid=\rho+O(\rho^{-\alpha_{1}-1}),\text{
}\forall x\in S_{\rho} \tag{5.4(c)}%
\end{align}
( see (1.26), (2.8), (2.34)) and the definition of $S_{\rho}$). One can
readily see that the inequalities in (5.3) follows from the definitions of
$K_{\rho}$ and (5.4), (5.4(a)), (5.4(b)). Since $S_{\rho}^{^{\prime}}\subset
S_{\rho},$ using (5.4(c)), we obtain the inclusion $U_{\varepsilon}(S_{\rho
}^{^{\prime}})\subset K_{\rho}$. This inclusion with $S_{\rho}^{^{\prime}%
}\subset U(2\rho^{\alpha_{1}},p)$ ( see definition of \ $S_{\rho}^{^{\prime}}$
and $S_{\rho}$) imply the inclusion in (5.3).

\begin{lemma}
$(a)$ If $x\in U_{\varepsilon}(S_{\rho}^{^{\prime}})$ and $x+b\in
U(\rho^{\alpha_{1}},p)\cap K_{\rho},$ where $b\in\Gamma,$ then

$\ \ \ \ \ \ \mid F(x)-F(x+b)\mid>2\varepsilon_{1},$where $\varepsilon
=\frac{\varepsilon_{1}}{7\rho},\varepsilon_{1}=\rho^{-d-2\alpha}.$

$(b)$ If $x\in U_{\varepsilon}(S_{\rho}^{^{\prime}}),$ then $x+b\notin
U_{\varepsilon}(S_{\rho}^{^{\prime}})$ for all $b\in\Gamma$ $.$

$(c)$ If $E$ is a bounded subset of $\mathbb{R}^{d}$, then $\mu(Tr(E))\leq
\mu(E)$.

$(d)$ If $E\subset U_{\varepsilon}(S_{\rho}^{^{\prime}}),$ then $\mu
(Tr_{F^{\star}}(E))=\mu(E).$
\end{lemma}

\begin{proof}
$(a)$ If $x\in U_{\varepsilon}(S_{\rho}^{^{\prime}}),$ then there exists a
point $a$ such that $a\in S_{\rho}^{^{\prime}}$ and $x\in U_{\varepsilon}(a)$.
Since $a+b$ lies in $\varepsilon$ neighborhood of $x+b,$ where $x+b\in
U(\rho^{\alpha_{1}},p)\cap K_{\rho},$ we have $a+b\in U(\frac{1}{2}%
\rho^{\alpha_{1}},p).$ Therefore using the definitions of $S_{\rho}^{^{\prime
}},$ and $P_{b}$ ( see (5.1)), we obtain $a\notin P_{b}$ and%
\begin{equation}
\mid F(a)-F(a+b)\mid\geq3\varepsilon_{1}. \tag{5.5}%
\end{equation}
On the other hand, using the last inequality of (5.3) and the obvious
relations $\mid x-a\mid<\varepsilon,$ $\mid x+b-a-b\mid<\varepsilon,$ we
obtain
\begin{equation}
\mid F(x)-F(a)\mid<3\rho\varepsilon,\text{ }\mid F(x+b)-F(a+b)\mid
<3\rho\varepsilon. \tag{5.6}%
\end{equation}
These inequalities with (5.5) give the proof of Lemma 5.1(a), since
$6\rho\varepsilon<\varepsilon_{1}.$

$(b)$ If $x$ and $x+b$ lie in $U_{\varepsilon}(S_{\rho}^{^{\prime}}),$ then
there exist points $a$ and $c$ in $S_{\rho}^{^{\prime}}$ such that $x\in
U_{\varepsilon}(a)$ and $x+b\in U_{\varepsilon}(c).$ Repeating the proof of
(5.6), we get

$\mid F(c)-F(x+b)\mid<3\rho\varepsilon.$ This, the first inequality in (5.6),
and the relations $F(a)=\rho^{2},$ $F(c)=\rho^{2}$ for $a\in S_{\rho},$ $c\in
S_{\rho}$ give $\mid F(x)-F(x+b)\mid<\varepsilon_{1},$ where $x\in
U_{\varepsilon}(S_{\rho}^{^{\prime}})$ and $x+b\in U_{\varepsilon}(S_{\rho
}^{^{\prime}})\subset U(\rho^{\alpha_{1}},p)\cap K_{\rho}$ ( see (5.3)), which
contradicts the Lemma 5.1(a).

$(c)$ Clearly, for any bounded set $E$ there exist only a finite number of
vectors $\gamma_{1},\gamma_{2},...,\gamma_{s}$ such that $E(k)\equiv
(E+\gamma_{k})\cap U_{\varepsilon}(S_{\rho}^{^{\prime}})\neq\emptyset$ for
$k=1,2,...,s$ and $Tr(E)$ is the union of the sets $E(k)$. By definition of
$E(k)$ we have $E(k)-\gamma_{k}\subset E,$ $\mu(E(k)-\gamma_{k})=\mu(E(k)).$
Moreover, by $(b),$ $(E(k)-\gamma_{k})\cap(E(j)-\gamma_{j})=\emptyset$ for
$k\neq j.$ Therefore $(c)$ is true.

$(d)$ Now let $E\subset U_{\varepsilon}(S_{\rho}^{^{\prime}}).$ Then by $(b)$
the set $E$ can be divided into a finite number of the pairwise disjoint sets
$E_{1},E_{2},...,E_{n}$ such that there exist the vectors $\gamma_{1}%
,\gamma_{2},...,\gamma_{n}$ satisfying
\[
(E_{k}+\gamma_{k})\subset F^{\star},\text{ }(E_{k}+\gamma_{k})\cap
(E_{j}+\gamma_{j})\neq\emptyset
\]
for $k,j=1,2,...,n$ and $k\neq j.$ Using $\mu(E_{k}+\gamma_{k})=\mu(E_{k}),$
we get the proof of $(d),$ since $Tr_{F^{\star}}(E)$ and $E$ are union of the
pairwise disjoint sets $E_{k}+\gamma_{k}$ and $E_{k}$ for $k=1,2,...,n$ respectively
\end{proof}

In the following Theorem we use the sets defined in Notation 5.1.

\begin{theorem}
$(a)$ The set $U_{\varepsilon}(S_{\rho}^{^{\prime}})\backslash Tr(A(\rho))$ is
a subset of the simple set $B$ defined in Definition 1.2. For every connected
open subset $E$ of $U_{\varepsilon}(S_{\rho}^{^{\prime}})\backslash
Tr(A(\rho)$ there exists a unique index $N$ such that $\Lambda_{N}%
(t)=\Lambda(\gamma+t)$ for $\gamma+t\in E,$ where $\Lambda(\gamma+t)$ is
defined in Remark 4.1. Moreover,
\begin{equation}
\frac{\partial}{\partial t_{j}}\Lambda(\gamma+t)=\frac{\partial}{\partial
t_{j}}\mid\gamma+t\mid^{2}+O(\rho^{1-2\alpha_{1}}),\forall j=1,2,...,d.
\tag{5.7}%
\end{equation}

$(b)$ For the part $V_{\rho}\equiv S_{\rho}^{^{\prime}}\backslash
U_{\varepsilon}(Tr(A(\rho)))$ of the approximated isoenergetic surface
$S_{\rho}$ the following holds
\begin{equation}
\mu(V_{\rho})>(1-c_{17}\rho^{-\alpha}))\mu(B(\rho)). \tag{5.8}%
\end{equation}
Moreover, $U_{\varepsilon}(V_{\rho})$ lies in the subset $U_{\varepsilon
}(S_{\rho}^{^{\prime}})\backslash Tr(A(\rho))$ of the simple set $B.$

$(c)$ The isoenergetic surface $I(\rho)$ contains the set $I_{\rho}%
^{^{\prime\prime}},$ which consists of the smooth surfaces and has the
measure
\begin{equation}
\mu(I_{\rho}^{^{\prime\prime}})=\mu(I_{\rho}^{^{\prime}})>(1-c_{18}%
\rho^{-\alpha})\mu(B(\rho)), \tag{5.9}%
\end{equation}
where $I_{\rho}^{^{\prime}}$ is a part of the translated isoenergetic surfaces
$TPI_{\rho}(q)$ of $L(q),$ which is contained in the subset $U_{\varepsilon
}(S_{\rho}^{^{\prime}})\backslash Tr(A(\rho))$ of the simple set $B.$

In particular the number $\rho^{2}$ for $\rho\gg1$ lies in the spectrum of
$L(q),$ that is, the number of the gaps in the spectrum of $L(q)$ is finite,
where $q(x)\in W_{2}^{s_{0}}(\mathbb{R}^{d}/\Omega),$ $d\geq2,$

$s_{0}=\frac{3d-1}{2}(3^{d}+d+2)+\frac{1}{4}d3^{d}+d+6,$ and $\Omega$ is an
arbitrary lattice.
\end{theorem}

\begin{proof}
$(a)$ To prove that $U_{\varepsilon}(S_{\rho}^{^{\prime}})\backslash
Tr(A(\rho))\subset B$ we need to show that for each point $\gamma+t$ of
$U_{\varepsilon}(S_{\rho}^{^{\prime}})\backslash Tr(A(\rho))$ the following
assertions are true:

\textbf{As.1} \ $\gamma+t\in U(\rho^{\alpha_{1}},p)\cap(R(\frac{3}{2}\rho
-\rho^{\alpha_{1}-1})\backslash R(\frac{1}{2}\rho+\rho^{\alpha_{1}-1})).$

\textbf{As.2 \ }If $\gamma^{^{\prime}}\in K,$ where $K$ is defined by (1.30),
and $\gamma^{^{\prime}}+t\in U(\rho^{\alpha_{1}},p),$ then (1.28) holds.

\textbf{As.3 \ }If$\ \gamma^{^{\prime}}\in K$ and $\gamma^{^{\prime}}+t\in
E_{k}\backslash E_{k+1},$ then (1.29) holds.

The proof of \ \textbf{As.1 }follows from the inclusion in (5.3).

The proof of \ \textbf{As.2. \ }If $\gamma^{^{\prime}}\in K,$ then (1.30)
holds. Since $\gamma+t\in U_{\varepsilon}(S_{\rho}^{^{\prime}}),$ there exists
$a\in S_{\rho}^{^{\prime}}\subset S_{\rho}$ such that $\gamma+t\in
U_{\varepsilon}(a).$ Then (5.6), the equalities $F(a)=\rho^{2}$ ( see
definition of $S_{\rho}$ in Notation 5.1) and $\varepsilon_{1}=7\rho
\varepsilon$ ( see Lemma 5.1(a)) give
\begin{equation}
F(\gamma+t)\in(\rho^{2}-\varepsilon_{1},\rho^{2}+\varepsilon_{1}). \tag{5.10}%
\end{equation}
This with (1.30) imply that $\gamma^{^{\prime}}+t\in U(\rho^{\alpha_{1}%
},p)\cap K_{\rho}.$ Now in Lemma 5.1(a) considering $x$ as $\gamma+t$ and
$x+b$ as $\gamma^{^{\prime}}+t$ we get (1.28).

The proof of \ \textbf{As.3. }As in case\textbf{ As.2 }the inclusion
$\gamma^{^{\prime}}\in K$ yields

$\gamma^{^{\prime}}+t\in(E_{k}\backslash E_{k+1})\cap K_{\rho}.$ On the other
hand $\gamma+t\notin Tr(A(\rho))$ which means that $\gamma^{^{\prime}}+t\notin
A(\rho).$ Therefore it follows from the definition of $A(\rho)$ ( see (5.2)) that

$\lambda_{i}(\gamma^{^{\prime}}+t)\notin(\rho^{2}-3\varepsilon_{1},\rho
^{2}+3\varepsilon_{1}).$ This with (5.10) implies (1.29).

Now let $E$ be a connected open subset of $U_{\varepsilon}(S_{\rho}^{^{\prime
}})\backslash Tr(A(\rho)\subset B.$ By Theorem 4.1 and Remark 4.1 for $a\in
E\subset U_{\varepsilon}(S_{\rho}^{^{\prime}})\backslash Tr(A(\rho)$ there
exists a unique index $N(a)$ such that
\[
\Lambda(a)=\Lambda_{N(a)}(a),\Psi_{a}(x)=\Psi_{N(a),a}(x),\mid(\Psi
_{N(a),a}(x),e^{i(a,x)})\mid^{2}>\frac{1}{2}%
\]
and $\Lambda(a)$ is a simple eigenvalue. On the other hand, for fixed $N$ the
functions $\Lambda_{N}(t)$ and $(\Psi_{N,t}(x),e^{i(t,x)})$ are continuous in
a neighborhood of $a$ if $\Lambda_{N}(a)$ is a simple eigenvalue. Therefore
for each $a\in E$ there exists a neighborhood $U(a)\subset E$ of $a$ such
that
\[
\mid(\Psi_{N(a),y}(x),e^{i(y,x)})\mid^{2}>\frac{1}{2}%
\]
for $y\in U(a).$ Since for $y\in E$ there is a unique integer $N(y)$
satisfying
\[
\mid(\Psi_{N(y),y}(x),e^{i(y,x)})\mid^{2}>\frac{1}{2},
\]
we have $N(y)=N(a)$ for $y\in U(a).$ Hence we proved that%
\begin{equation}
\forall a\in E,\exists U(a)\subset E:N(y)=N(a),\forall y\in U(a). \tag{5.11}%
\end{equation}

Now let $a_{1}$ and $a_{2}$ be two points of $E$ , and let $C\subset E$ be the
arc that joins these points. Let $U(y_{1}),U(y_{2}),...,U(y_{k})$ be a finite
subcover of the open cover $\{U(a):a\in C\}$ of the compact $C,$ where $U(a)$
is the neighborhood of $a$ satisfying (5.11). By (5.11), we have
$N(y)=N(y_{i})=N_{i}$ for $y\in U(y_{i}).$ Clearly, if $U(y_{i})\cap
U(y_{j})\neq\emptyset,$ then $N_{i}=N(z)=N_{j},$ where $z\in U(y_{i})\cap
U(y_{j})$. Thus $N_{1}=N_{2}=...=N_{k}$ and $N(a_{1})=N(a_{2}).$

To calculate the partial derivatives of the function $\Lambda(\gamma
+t)=\Lambda_{N}(t)$ we write the operator $L_{t}$ in the form $-\triangle
-(2it,\nabla)+(t,t).$ Then, it is clear that
\begin{align}
\frac{\partial}{\partial t_{j}}\Lambda_{N}(t)  &  =2t_{j}(\Phi_{N,t}%
(x),\Phi_{N,t}(x))-2i(\frac{\partial}{\partial x_{j}}\Phi_{N,t}(x),\Phi
_{N,t}(x)),\tag{5.12}\\
\Phi_{N,t}(x)  &  =\sum_{\gamma^{^{\prime}}\in\Gamma}b(N,\gamma^{^{\prime}%
})e^{i(\gamma^{^{\prime}},x)}, \tag{5.13}%
\end{align}
where $\Phi_{N,t}(x)=e^{-i(t,x)}\Psi_{N,t}(x).$ If $\mid\gamma^{^{\prime}}%
\mid\geq2\rho,$ then using%
\[
\Lambda_{N}\equiv\Lambda(\gamma+t)=\rho^{2}+O(\rho^{-\alpha}),
\]
( see (1.27), (5.10)), and the obvious inequality%
\[
\mid\Lambda_{N}-\mid\gamma^{^{\prime}}-\gamma_{1}-\gamma_{2}-...-\gamma
_{k}+t\mid^{2}\mid>c_{19}\mid\gamma^{^{\prime}}\mid^{2}%
\]
for $k=0,1,...,p,$ where $\mid\gamma_{1}\mid<\frac{1}{4p}\mid\gamma^{^{\prime
}}\mid,$ and iterating (1.8) $p$ times by using decomposition
\[
q(x)=\sum_{\mid\gamma_{1}\mid<\frac{1}{4p}\mid\gamma^{^{\prime}}\mid}%
q_{\gamma_{1}}e^{i(\gamma_{1},x)}+O(\mid\gamma^{^{\prime}}\mid^{-p}),
\]
we get%
\begin{align}
b(N,\gamma^{^{\prime}})  &  =\sum_{\gamma_{1},\gamma_{2},...}\dfrac
{q_{\gamma_{1}}q_{\gamma_{2}}...q_{\gamma_{p}}b(N,\gamma^{^{\prime}}%
-\sum_{i=1}^{p}\gamma_{i})}{\prod_{j=0}^{p-1}(\Lambda_{N}-\mid\gamma
^{^{\prime}}-\sum_{i=1}^{j}\gamma_{i}+t\mid^{2})}+O(\mid\gamma^{^{\prime}}%
\mid^{-p}),\tag{5.14}\\
b(N,\gamma^{^{\prime}})  &  =O(\mid\gamma^{^{\prime}}\mid^{-p}),\text{
}\forall\mid\gamma^{^{\prime}}\mid\geq2\rho. \tag{5.15}%
\end{align}
By (5.15) the series in (5.13) can be differentiated term by term. Hence
\begin{equation}
-i(\frac{\partial}{\partial x_{j}}\Phi_{N,t},\Phi_{N,t})=\sum_{\gamma
^{^{\prime}}\in\Gamma}\gamma^{^{\prime}}(j)\mid b(N,\gamma^{^{\prime}}%
)\mid^{2}=\gamma(j)\mid b(N,\gamma)\mid^{2}+a_{1}+a_{2}, \tag{5.16}%
\end{equation}
where
\[
a_{1}=\sum_{\mid\gamma^{^{\prime}}\mid\geq2\rho}\gamma^{^{\prime}}(j)\mid
b(N,\gamma^{^{\prime}})\mid^{2},\text{ }a_{2}=\sum_{\mid\gamma^{^{\prime}}%
\mid<2\rho,\gamma^{^{\prime}}\neq\gamma}\gamma^{^{\prime}}(j)\mid
b(N,\gamma^{^{\prime}})\mid^{2}.
\]
By (1.31), (1.31a) $a_{2}=O(\rho^{-2\alpha_{1}+1}),$ $\gamma(j)\mid
b(N,\gamma)\mid^{2}=\gamma(j)(1+O(\rho^{-2\alpha_{1}}),$ and by (5.15),
$a_{1}=O(\rho^{-2\alpha_{1}}).$ Therefore (5.12) and (5.16) imply (5.7).

$(b)$ To prove the inclusion $U_{\varepsilon}(V_{\rho})\subset$
$U_{\varepsilon}(S_{\rho}^{^{\prime}})\backslash Tr(A(\rho))$ we need to show
that if $a\in V_{\rho},$ then $U_{\varepsilon}(a)\subset U_{\varepsilon
}(S_{\rho}^{^{\prime}})\backslash Tr(A(\rho)).$ This is clear, since the
relations $a\in V_{\rho}\subset S_{\rho}^{^{\prime}}$ imply that
$U_{\varepsilon}(a)\subset U_{\varepsilon}(S_{\rho}^{^{\prime}})$ and the
relation $a\notin U_{\varepsilon}(Tr(A(\rho)))$ implies that $U_{\varepsilon
}(a)\cap Tr(A(\rho))=\emptyset.$ To prove (5.8) first we estimate the measure
of $S_{\rho},S_{\rho}^{^{\prime}},U_{2\varepsilon}(A(\rho))$, namely we prove
\begin{align}
\mu(S_{\rho})  &  >(1-c_{20}\rho^{-\alpha})\mu(B(\rho)),\tag{5.17}\\
\mu(S_{\rho}^{^{\prime}})  &  >(1-c_{21}\rho^{-\alpha})\mu(B(\rho
)),\tag{5.18}\\
\mu(U_{2\varepsilon}(A(\rho)))  &  =O(\rho^{-\alpha})\mu(B(\rho))\varepsilon
\tag{5.19}%
\end{align}
( see below, Estimations 1, 2, 3). The estimation (5.8) of the measure of the
set $V_{\rho}$ is done in Estimation 4 by using Estimations 1, 2, 3.

$(c)$ In Estimation 5 we prove the formula (5.9). The Theorem is proved
\end{proof}

In Estimations 1-5 we use the notations:

$G(+i,a)=\{x\in G,$ $x_{i}>a\},$ $G(-i,a)=\{x\in G,$ $x_{i}<-a\},$ where
$x=(x_{1},x_{2},...,x_{d}),$ $a>0.$ Recalling the definitions of the sets
$S_{\rho}^{^{\prime}},$ $A(\rho),$ and using (5.3), it is not hard to verify
that for any subset $G$ of $U_{\varepsilon}(S_{\rho}^{^{\prime}})\cup
U_{2\varepsilon}(A(\rho))$ , that is, for all considered sets $G$ in these
estimations, and for any $x\in G$ the followings hold
\begin{equation}
\rho-1<\mid x\mid<\rho+1,\text{ }G\subset(\cup_{i=1}^{d}(G(+i,\rho d^{-1})\cup
G(-i,\rho d^{-1})) \tag{5.20}%
\end{equation}
Indeed, (5.3) imply the inequalities in (5.20) and the inclusion in (5.20)
follows from these inequalities. If $G$ $\subset S_{\rho},$ then by (5.4),
(5.4(b)) we have $\frac{\partial F(x)}{\partial x_{k}}>0$ for $x\in
G(+k,\rho^{-\alpha})$. Therefore to calculate the measure of $G(+k,a)$ for
$a\geq\rho^{-\alpha}$ we use the formula
\begin{equation}
\mu(G(+k,a))=\int\limits_{\Pr_{k}(G(+k,a))}(\frac{\partial F}{\partial x_{k}%
})^{-1}\mid grad(F)\mid dx_{1}...dx_{k-1}dx_{k+1}...dx_{d}, \tag{5.21}%
\end{equation}
where $\Pr_{k}(G)\equiv\{(x_{1},x_{2},...,x_{k-1},x_{k+1},x_{k+2}%
,...,x_{d}):x\in G\}$ is the projection of $G$ on the hyperplane $x_{k}=0.$
Instead of $\Pr_{k}(G)$ we write $\Pr(G)$ if $k$ is unambiguous. If $D$ is
$m-$dimensional subset of $\mathbb{R}^{m},$ then to estimate $\mu(D),$ we use
the formula
\begin{equation}
\mu(D)=\int\limits_{\Pr_{k}(D)}\mu(D(x_{1},...x_{k-1},x_{k+1},...,x_{m}%
))dx_{1}...dx_{k-1}dx_{k+1}...dx_{m}, \tag{5.22}%
\end{equation}
where $D(x_{1},...x_{k-1},x_{k+1},...,x_{m})=\{x_{k}:(x_{1},x_{2}%
,...,x_{m})\in D\}.$

ESTIMATION 1. Here we prove (5.17) by using (5.21). During this estimation the
set $S_{\rho}$ is redenoted by $G.$ First we estimate $\mu(G(+1,a))$ for
$a=\rho^{1-\alpha}$ by using (5.21) for $k=1$ and \ the relations%
\begin{equation}
\frac{\partial F}{\partial x_{1}}>\rho^{1-\alpha},(\frac{\partial F}{\partial
x_{1}})^{-1}\mid grad(F)\mid=\frac{\rho}{\sqrt{\rho^{2}-x_{2}^{2}-x_{3}%
^{2}-...-x_{d}^{2}}}+O(\rho^{-\alpha}), \tag{5.23}%
\end{equation}%
\begin{equation}
\Pr(G(+1,a))\supset\Pr(A(+1,2a)), \tag{5.24}%
\end{equation}
where $x\in G(+1,a),$ $A=B(\rho)\cap U(3\rho^{\alpha_{1}},p),$ and
$B(\rho)=\{x\in\mathbb{R}^{d}:\mid x\mid=\rho\}.$ Here (5.23) follows from
(5.4), (5.4(b)), and (5.4(c)). Now we prove (5.24). If

$(x_{2},...,x_{d})\in\Pr_{1}(A(+1,2a)),$ then by definition of $A(+1,2a)$
there exists $x_{1}$ such that
\begin{equation}
x_{1}>2a=2\rho^{1-\alpha},\text{ }x_{1}^{2}+x_{2}^{2}+...+x_{d}^{2}=\rho
^{2},\text{ }\mid\sum_{i\geq1}(2x_{i}b_{i}-b_{i}^{2})\mid\geq3\rho^{\alpha
_{1}} \tag{5.25}%
\end{equation}
for all $(b_{1},b_{2},...,b_{d})\in\Gamma(p\rho^{\alpha}).$ Therefore for
$h=\rho^{-\alpha}$ we have%
\[
(x_{1}+h)^{2}+x_{2}^{2}+...+x_{q}^{2}>\rho^{2}+\rho^{-\alpha},(x_{1}%
-h)^{2}+x_{2}^{2}+...+x_{q}^{2}<\rho^{2}-\rho^{-\alpha}.
\]
This, (5.4) and (5.4(a)) give $F(x_{1}+h,x_{2},...,x_{d})>\rho^{2}$,
$F(x_{1}-h,x_{2},...,x_{d})<\rho^{2}$. Since $F$ is a continuous function (
see Remark 2.2) on $U(c_{4}\rho^{\alpha_{1}},p)$ there is $y_{1}\in
(x_{1}-h,x_{1}+h)$ such that (see (5.25))$y_{1}>a,$ $F(y_{1},x_{2}%
,...,x_{d})=\rho^{2}.$ Moreover
\begin{equation}
\mid2y_{1}b_{1}-b_{1}^{2}+\sum_{i\geq2}(2x_{i}b_{i}-b_{i}^{2})\mid
>\rho^{\alpha_{1}}, \tag{5.26}%
\end{equation}
because the expression under the absolute value in (5.26) differ from the
expression under the absolute value in (5.25) by $2(y_{1}-x_{1})b_{1},$ where
$\mid y_{1}-x_{1}\mid<h=\rho^{-\alpha},$ $\mid b_{1}\mid<p\rho^{\alpha},$
$\mid2(y_{1}-x_{1})b_{1}\mid<2p<$ $\rho^{\alpha_{1}}.$ Now recalling the
definition of \ $G(+1,a)$ and $S_{\rho}$ we see that these relations imply the
inclusion $(x_{2},...,x_{d})\in\Pr_{1}G(+1,a).$ Hence (5.24) is proved. Now
(5.23), (5.24), and the obvious relation $\mu(\Pr_{1}G(+1,a))=O(\rho^{d-1})$ (
see (5.20)) give%
\[
\mu(G(+1,a))=\int\limits_{\Pr(G(+1,a))}\frac{\rho}{\sqrt{\rho^{2}-x_{2}%
^{2}-x_{3}^{2}-...-x_{d}^{2}}}dx_{2}dx_{3}...dx_{d}+O(\frac{1}{\rho^{\alpha}%
})\mu(B(\rho))
\]%
\[
\geq\int\limits_{\Pr(A(+1,2a))}\frac{\rho}{\sqrt{\rho^{2}-x_{2}^{2}-x_{3}%
^{2}-...-x_{d}^{2}}}dx_{2}dx_{3}...dx_{d}-c_{22}\rho^{-\alpha}\mu(B(\rho))
\]%
\[
=\mu(A(+1,2a))-c_{22}\rho^{-\alpha}\mu(B(\rho)).
\]
Similarly, $\mu(G(-1,a))\geq\mu(A(-1,2a))-c_{22}\rho^{-\alpha}\mu(B(\rho)).$
Now using \ the inequality

$\mu(G)\geq\mu(G(+1,a))+\mu(G(-1,a)),$ we get

$\mu(G)\geq\mu(A(-1,2a))+\mu(A(+1,2a))-2c_{22}\rho^{-\alpha}\mu(B(\rho)).$ On
the other hand it follows from the obvious relation\ \ $\mu(\{x\in
B(\rho):-2a\leq x_{1}\leq2a\})=O(\rho^{-\alpha})\mu(B(\rho))$ that

$\mu(A(-1,2a))+\mu(A(+1,2a))\geq\mu(A)-c_{22}\rho^{-\alpha}\mu(B(\rho)).$ Therefore

$\mu(G)>\mu(A)-3c_{22}\rho^{-\alpha}\mu(B(\rho)).$ It implies (5.17), since
$\mu(A))=(1+O(\rho^{-\alpha}))\mu(B(\rho))$ (see (2.32) ).

ESTIMATION 2. Here we prove (5.18). For this we estimate the measure of the
set $S_{\rho}\cap P_{b}$ by using (5.21). During this estimation the set
$S_{\rho}\cap P_{b}$ is redenoted by $G$. We choose the coordinate axis so
that the direction of $b$ coincides with the direction of $(1,0,0,...,0),$
i.e., $b=(b_{1},0,0,...,0)$ and $b_{1}>0$. It follows from the definition of
$P_{b}$ ( see (5.1)), (5.4), (5.4(c)) that if $(x_{1},x_{2},...,x_{d})\in G,$
then
\begin{align}
x_{1}^{2}+x_{2}^{2}+...+x_{d}^{2}+F_{k_{1}-1}(x)  &  =\rho^{2},\tag{5.27}\\
(x_{1}+b_{1})^{2}+x_{2}^{2}+x_{3}^{2}+...+x_{d}^{2}+F_{k_{1}-1}(x+b)  &
=\rho^{2}+h, \tag{5.28}%
\end{align}
where $h\in(-3\varepsilon_{1},3\varepsilon_{1}),$ $\varepsilon_{1}%
=\rho^{-d-2\alpha}.$ Therefore subtracting (5.27) from (5.28) and using
(5.4(a)), we get
\begin{equation}
(2x_{1}+b_{1})b_{1}=O(\rho^{-\alpha_{1}}). \tag{5.29}%
\end{equation}
This and the inequalities in (5.20) imply
\begin{equation}
\mid b_{1}\mid<2\rho+3,\text{ }x_{1}=\frac{b_{1}}{2}+O(\rho^{-\alpha_{1}}%
b_{1}^{-1}),\mid x_{1}^{2}-(\frac{b_{1}}{2})^{2}\mid=O(\rho^{-\alpha_{1}}).
\tag{5.30}%
\end{equation}
Consider two cases. Case 1: $b\in\Gamma_{1},$ where $\Gamma_{1}=\{b\in
\Gamma:\mid\rho^{2}-\mid\frac{b}{2}\mid^{2}\mid<3d\rho^{-2\alpha}\}.$ In this
case using the last equality in (5.30), (5.27), (5.4(a)), and taking into
account that $b=(b_{1},0,0,...,0),$ $\alpha_{1}=3\alpha,$ we obtain
\begin{equation}
x_{1}^{2}=\rho^{2}+O(\rho^{-2\alpha}),\mid x_{1}\mid=\rho+O(\rho^{-2\alpha
-1}),x_{2}^{2}+x_{3}^{2}+...+x_{d}^{2}=O(\rho^{-2\alpha}). \tag{5.31}%
\end{equation}
Therefore $G\subset G(+1,a)\cup G(-1,a),$ where $a=\rho-\rho^{-1}$. Using
(5.21), the obvious relation $\mu(\Pr_{1}(G(\pm1,a))=O(\rho^{-(d-1)\alpha})$
(see (5.31)) and taking into account that the expression under the integral in
(5.21) for $k=1$ is equal to $1+O(\rho^{-\alpha})$ (see (5.4(b)) and (5.31)),
we get $\mu(G(\pm1,a))=O(\rho^{-(d-1)\alpha}).$ Thus $\mu(G)=O(\rho
^{-(d-1)\alpha}).$ Since $\mid\Gamma_{1}\mid=O(\rho^{d-1})$ (see (1.37)), we
have
\begin{equation}
\text{ }\mu(\cup_{b\in\Gamma_{1}}(S_{\rho}\cap P_{b})=O(\rho^{-(d-1)\alpha
+d-1})=O(\rho^{-\alpha})\mu(B(\rho)). \tag{5.32}%
\end{equation}
Case 2: $\mid\rho^{2}-\mid\frac{b}{2}\mid^{2}\mid\geq3d\rho^{-2\alpha}.$
Repeating the proof of (5.31), we get%
\begin{equation}
\mid x_{1}^{2}-\rho^{2}\mid>2d\rho^{-2\alpha},\text{ }\sum_{k=2}^{d}x_{k}%
^{2}>d\rho^{-2\alpha},\text{ }\max_{k\geq2}\mid x_{k}\mid>\rho^{-\alpha}.
\tag{5.33}%
\end{equation}
Therefore $G\subset\cup_{k\geq2}(G(+k,\rho^{-\alpha})\cup G(-k,\rho^{-\alpha
})).$ Now we estimate $\mu(G(+d,\rho^{-\alpha}))$ by using (5.21). If $x\in
G(+d,\rho^{-\alpha}),$ then according to (5.27) and (5.4(b)) the under
integral expression in (5.21) for $k=d$ is $O(\rho^{1+\alpha}).$ Therefore the
first equality in
\begin{equation}
\mu(D)=O(\varepsilon_{1}\mid b\mid^{-1}\rho^{d-2}),\text{ }\mu(G(+d,\rho
^{-\alpha}))=O(\rho^{d-1+\alpha}\varepsilon_{1}\mid b\mid^{-1}), \tag{5.34}%
\end{equation}
where the set $\Pr_{d}G(+d,\rho^{-\alpha})$ is redenoted by $D$ , implies the
second equality in (5.34). To prove the first equality in (5.34) we use (5.22)
for $m=d-1$ and $k=1$ and prove the relations $\mu(\Pr_{1}D)=O(\rho^{d-2}),$
\begin{equation}
\text{ }\mu(D(x_{2},x_{3},...,x_{d-1}))<6\varepsilon_{1}\mid b\mid^{-1}
\tag{5.35}%
\end{equation}
for $(x_{2},x_{3},...,x_{d-1})\in\Pr_{1}D.$ First relation follows from the
inequalities in (5.20)). So we need to prove (5.35). If $x_{1}\in
D(x_{2},x_{3},...,x_{d-1}),$ then by definition of $D(x_{2},x_{3}%
,...,x_{d-1})$ and $D$ we have $(x_{1}x_{2},,...,x_{d-1})\in D$ and
$(x_{1},x_{2},...,x_{d})\in G(+d,\rho^{-\alpha})\subset G\equiv S_{\rho}\cap
P_{b}.$ Therefore (5.27) and (5.28) hold. Subtracting (5.27) from (5.28), we
get
\begin{equation}
2x_{1}b_{1}+(b_{1})^{2}+F_{k_{1}-1}(x+b)-F_{k_{1}-1}(x)=h, \tag{5.36}%
\end{equation}
where $x_{2},x_{3},...,x_{d-1}$ are fixed . Hence we have two equations (5.27)
and (5.36) with respect two unknown $x_{1}$ and $x_{d}$. Using (5.4(b)), the
implicit function theorem, and the inequalities $\mid x_{d}\mid>\rho^{-\alpha
},$ $\alpha_{1}>2\alpha$ from (5.27), we obtain
\begin{equation}
x_{d}=f(x_{1}),\text{ }\frac{df}{dx_{1}}=-\frac{2x_{1}+O(\rho^{-2\alpha
_{1}+\alpha})}{2x_{d}+O(\rho^{-2\alpha_{1}+\alpha})}. \tag{5.37}%
\end{equation}
Substituting this in (5.36), we get
\begin{equation}
2x_{1}b_{1}+b_{1}^{2}+F_{k_{1}-1}(x_{1}+b_{1},x_{2},...,x_{d-1},f(x_{1}%
))-F_{k_{1}-1}(x_{1},...,x_{d-1},f)=h. \tag{5.38}%
\end{equation}
Using (5.4(b), (5.37), the first equality in (5.30), and $x_{d}>\rho^{-\alpha
}$ we see that the absolute value of the derivative (w.r.t. $x_{1}$) of the
left-hand side of (5.38) satisfies the inequality%
\[
\mid2b_{1}+O(\rho^{-2\alpha_{1}+\alpha})+O(\rho^{-2\alpha_{1}+\alpha}%
)\frac{x_{1}+O(\rho^{-2\alpha_{1}+\alpha})}{x_{d}+O(\rho^{-2\alpha_{1}+\alpha
})})\mid>b_{1}%
\]
for $x_{1}=\frac{b_{1}}{2}+O(\rho^{-\alpha_{1}})$ (see (5.30)). Therefore from
(5.38), using the implicit function theorem, we get
\[
\mid\frac{dx_{1}}{dh}\mid<\frac{1}{\mid b\mid},\text{ }\forall h\in
(-3\varepsilon_{1},3\varepsilon_{1})
\]
This inequality implies that the image $\{x_{1}(h):h\in(-3\varepsilon
_{1},3\varepsilon_{1})\}$ of the interval $(-3\varepsilon_{1},3\varepsilon
_{1})$ ( see (5.28)) under differentiable function $x_{1}(h)$ is an interval
$I$ with the length less than $6\varepsilon_{1}\mid b\mid^{-1}.$ Since
$D(x_{2},x_{3},...,x_{d-1})$ is a measurable subset of $I,$ (5.35) holds. Thus
(5.34) is proved. In the same way we get the same estimation for the sets
$G(-d,\rho^{-\alpha}),$ $G(+k,\rho^{-\alpha})$ and $G(-k,\rho^{-\alpha}),$
where $k\geq2$. Hence%
\[
\mu(S_{\rho}\cap P_{b})=O(\rho^{d-1+\alpha}\varepsilon_{1}\mid b\mid^{-1})
\]
for $b\notin\Gamma_{1}.$ Since $\mid b\mid<2\rho+3$ ( see (5.30)) and
$\varepsilon_{1}=\rho^{-d-2\alpha}$, taking into account that the number of
the vectors of $\Gamma$ satisfying $\mid b\mid<2\rho+3$ is $O(\rho^{d}),$ we
obtain%
\[
\mu(\cup_{b\notin\Gamma_{1}}(S_{\rho}\cap P_{b}))=O(\rho^{2d-1+\alpha
}\varepsilon_{1})=O(\rho^{-\alpha})\mu(B(\rho)).
\]
This, (5.32) and (5.17) give the proof of (5.18).

ESTIMATION 3. Here we prove (5.19). Denote $U_{2\varepsilon}(A_{k,j}%
(\gamma_{1,}\gamma_{2},...,\gamma_{k}))$ by $G,$ where $\gamma_{1,}\gamma
_{2},...,\gamma_{k}\in\Gamma(p\rho^{\alpha}),k\leq d-1,$ and $A_{k,j}$ is
defined in (5.2). To estimate $\mu(G)$ we turn the coordinate axis so that%
\[
Span\{\gamma_{1,}\gamma_{2},...,\gamma_{k}\}=\{x=(x_{1},x_{2},...,x_{k}%
,0,0,...,0):x_{1},x_{2},...,x_{k}\in\mathbb{R}\}.
\]
Then by (2.22), we have $x_{i}=O(\rho^{\alpha_{k}+(k-1)\alpha})$ for $i\leq
k,$ $x\in G$. This, (5.20), and

$\alpha_{k}+(k-1)\alpha<1$ ( see the first inequality in (1.39)) give%

\[
G\subset(\cup_{i>k}(G(+i,\rho d^{-1})\cup G(-i,\rho d^{-1})),
\]%
\begin{equation}
\mu(\Pr(G(+i,\rho d^{-1})))=O(\rho^{k(\alpha_{k}+(k-1)\alpha)+(d-1-k)})
\tag{5.39}%
\end{equation}
for $i>k.$ Now using this and (5.22) for $m=d,$ we prove that
\begin{equation}
\mu(G(+i,\rho d^{-1}))=O(\varepsilon\rho^{k(\alpha_{k}+(k-1)\alpha
)+(d-1-k)}),\forall i>k. \tag{5.40}%
\end{equation}
For this we redenote by $D$ the set $G(+i,\rho d^{-1})$ and prove that
\begin{equation}
\mu((D(x_{1},x_{2},...x_{i-1},x_{i+1},...x_{d}))\leq(42d^{2}+4)\varepsilon
\tag{5.41}%
\end{equation}
for $(x_{1},x_{2},...x_{i-1},x_{i+1},...x_{d})\in\Pr_{i}(D)$ and $i>k,$ since
using (5.41) and (5.39) in (5.22) one can easily get the proof of (5.40).
Hence we need to prove (5.41). To prove (5.41) it is sufficient to show that
if both $x=(x_{1},x_{2},...,x_{i},...x_{d})$ and $x^{^{\prime}}=(x_{1}%
,x_{2},...,x_{i}^{^{\prime}},...,x_{d})$ are in $D,$ then $\mid x_{i}%
-x_{i}^{^{\prime}}\mid\leq(42d^{2}+4)\varepsilon.$ Assume the converse. Then
$\mid x_{i}-x_{i}^{^{\prime}}\mid>(42d^{2}+4)\varepsilon$. Without loss of
generality it can be assumed that $x_{i}^{^{\prime}}>x_{i}.$ Then we have the
inequalities
\begin{equation}
x_{i}^{^{\prime}}>x_{i}+(42d^{2}+4)\varepsilon,\text{ }x_{i}>\rho d^{-1}
\tag{5.42}%
\end{equation}
since $x=(x_{1},x_{2},...,x_{i},...x_{d})\in D\equiv G(+i,\rho d^{-1})$. By
the definition of $G$ the points $x$ and $x^{^{\prime}}$ lie in the
$2\varepsilon$ neighborhood of $A_{k,j}.$ Therefore there exist points $a$ and
$a^{^{\prime}}$ in $A_{k,j}$ such that $\mid x-a\mid<2\varepsilon$ and $\mid
x^{^{\prime}}-a^{^{\prime}}\mid<2\varepsilon.$ These inequalities with (5.42)
imply that%
\begin{align}
\rho d^{-1}-2\varepsilon &  <a_{i}<a_{i}^{^{\prime}},\text{ }a_{i}^{^{\prime}%
}-a_{i}>42d^{2}\varepsilon,\tag{5.43}\\
(a_{i}^{^{\prime}})^{2}-(a_{i})^{2}  &  >2(\rho d^{-1}-2\varepsilon
)(a_{i}^{^{\prime}}-a_{i}), \tag{5.43(a)}%
\end{align}%
\begin{equation}
a_{s},a_{s}^{^{\prime}}\in(x_{s}-2\varepsilon,x_{s}+2\varepsilon,),\text{
}\mid\mid a_{s}\mid-\mid a_{s}^{^{\prime}}\mid\mid<4\varepsilon\tag{5.44}%
\end{equation}
for $s\neq i,$ since $x_{s}^{^{\prime}}=x_{s}$ for $s\neq i.$ On the other
hand the inequalities in (5.20) hold for the points of $A_{k,j}$ , that is, we
have $\mid a_{s}\mid<\rho+1,\mid a_{s}^{^{\prime}}\mid<\rho+1.$ These
inequalities and (5.44) imply $\mid\mid a_{s}\mid^{2}-\mid a_{s}^{^{\prime}%
}\mid^{2}\mid<12\rho\varepsilon$ for $s\neq i$, and by (5.43)%
\begin{equation}
\sum_{s\neq i}\mid\mid a_{s}\mid^{2}-\mid a_{s}^{^{\prime}}\mid^{2}%
\mid<12d\rho\varepsilon<\frac{2}{7}\rho d^{-1}(a_{i}^{^{\prime}}-a_{i}).
\tag{5.45}%
\end{equation}
Using this and (5.43(a)), we get%
\begin{equation}
\mid\mid a\mid^{2}-\mid a^{^{\prime}}\mid^{2}\mid>\frac{3}{2}\rho d^{-1}\mid
a_{i}^{^{\prime}}-a_{i}\mid. \tag{5.46}%
\end{equation}
At last, the inequalities $a_{i}^{^{\prime}}-a_{i}>42d^{2}\varepsilon$ ( see
(5.43)), $\mid a_{s}-a_{s}^{^{\prime}}\mid<4\varepsilon$ \ for $s\neq i$ (see
the inclusion in (5.44)) shows that
\begin{equation}
\mid a-a^{^{\prime}}\mid<2\mid a_{i}^{^{\prime}}-a_{i}\mid\tag{5.46(a)}%
\end{equation}
Now we prove that (5.46) and (5.46(a)) contradict the inclusions $a\in
A_{k,j}$ and $a^{^{\prime}}\in A_{k,j}.$ Using the inequality (2.36), the
obvious relation $\frac{1}{2}\alpha_{d}<1$ ( see definitions of $\alpha$ and
$\alpha_{d}$ in (1.6) and in Definition 1.1) and (5.46(a)), we get%
\[
\mid r_{j}(a)-r_{j}(a^{^{\prime}})\mid<\rho^{\frac{1}{2}\alpha_{d}}\mid
a-a^{^{\prime}}\mid<\frac{1}{2}\rho d^{-1}\mid a_{i}^{^{\prime}}-a_{i}\mid,
\]
where $r_{j}(x)=\lambda_{j}(x)-\mid x\mid^{2}$ ( see Remark 2.2). This
inequality, (5.46), the inequality $a_{i}^{^{\prime}}-a_{i}>42d^{2}%
\varepsilon$ ( see (5.43)), and the relation $\varepsilon_{1}=7\rho
\varepsilon$ ( see Lemma 5.1(a)) imply%
\[
\mid\lambda_{j}(a)-\lambda_{j}(a^{^{\prime}})\mid\geq\mid\mid a\mid^{2}-\mid
a^{^{\prime}}\mid^{2}\mid-\mid r_{j}(a)-r_{j}(a^{^{\prime}})\mid>\rho
d^{-1}\mid a_{i}^{^{\prime}}-a_{i}\mid>42d\rho\varepsilon>6\varepsilon_{1}.
\]
The obtained inequality $\mid\lambda_{j}(a)-\lambda_{j}(a^{^{\prime}}%
)\mid>6\varepsilon_{1}$ contradicts with the inclusions $a$ $\in A_{k,j},$
$a^{^{\prime}}\in A_{k,j},$ since by definition of $A_{k,j}$ ( see (5.2)) both
$\lambda_{j}(a)$ and $\lambda_{j}(a^{^{\prime}})$ lie in $(\rho^{2}%
-3\varepsilon_{1},\rho^{2}+3\varepsilon_{1}).$ Thus (5.41) and hence (5.40) is
proved. In the same way we get the same estimation for $G(-i,\frac{\rho}{d}).$
Thus%
\[
\mu(U_{2\varepsilon}(A_{k,j}(\gamma_{1,}\gamma_{2},...,\gamma_{k}%
)))=O(\varepsilon\rho^{k(\alpha_{k}+(k-1)\alpha)+d-1-k}).
\]
Now taking into account that $U_{2\varepsilon}(A(\rho))$ is union of
$U_{2\varepsilon}(A_{k,j}(\gamma_{1,}\gamma_{2},...,\gamma_{k})$ for
$k=1,2,..,d-1;$ $j=1,2,...,b_{k}(\gamma_{1},\gamma_{2},...,\gamma_{k}),$ and
$\gamma_{1},\gamma_{2},...,\gamma_{k}\in\Gamma(p\rho^{\alpha})$ \ ( see (5.2))
and using that
\[
b_{k}=O(\rho^{d\alpha+\frac{k}{2}\alpha_{k+1}})
\]
( see (2.30)), the number of the vectors $(\gamma_{1},\gamma_{2}%
,...,\gamma_{k})$ for $\gamma_{1},\gamma_{2},...,\gamma_{k}\in\Gamma
(p\rho^{\alpha})$ is $O(\rho^{dk\alpha}),$ we obtain
\[
\mu(U_{2\varepsilon}(A(\rho)))=O(\varepsilon\rho^{d\alpha+\frac{k}{2}%
\alpha_{k+1}+dk\alpha+k(\alpha_{k}+(k-1)\alpha)+d-1-k}).
\]
Therefore to prove (5.19), it remains to show that

$d\alpha+\frac{k}{2}\alpha_{k+1}+dk\alpha+k(\alpha_{k}+(k-1)\alpha)+d-1-k\leq
d-1-\alpha$ or%
\[
(d+1)\alpha+\frac{k}{2}\alpha_{k+1}+dk\alpha+k(\alpha_{k}+(k-1)\alpha)\leq k
\]
for $1\leq k\leq d-1$. Dividing both sides by $k\alpha$ and using $\alpha
_{k}=3^{k}\alpha,$ $\alpha=\frac{1}{\varkappa},$ $\varkappa=3^{d}+d+2$ ( see
(1.6) and Definition 1.1), we get
\[
\frac{d+1}{k}+\frac{3^{k+1}}{2}+3^{k}+k-1\leq3^{d}+2.
\]
The left-hand side of this inequality gets its maximum at $k=d-1.$ Therefore
we need to show that
\[
\frac{d+1}{d-1}+\frac{5}{6}3^{d}+d\leq3^{d}+4
\]
which follows from the inequalities $\frac{d+1}{d-1}\leq3,$ $d<\frac{1}%
{6}3^{d}+1$ for $d\geq2.$

ESTIMATION 4. Here we prove (5.8). During this estimation we denote by $G$ the
set $S_{\rho}^{^{\prime}}\cap U_{\varepsilon}(Tr(A(\rho))$. Since $V_{\rho
}=S_{\rho}^{^{\prime}}\backslash G$ and (5.18) holds, it is enough to prove that

$\mu(G)=O(\rho^{-\alpha})\mu(B(\rho)).$ For this we use (5.20) and prove
\begin{equation}
\mu(G(+i,\rho d^{-1}))=O(\rho^{-\alpha})\mu(B(\rho)),\text{ }\mu(G(-i,\rho
d^{-1}))=O(\rho^{-\alpha})\mu(B(\rho)) \tag{5.47}%
\end{equation}
for $i=1,2,...,d$ by using (5.21). By (5.4(b)), if $x\in G(+i,\rho d^{-1}),$
then the under integral expression in (5.21) for $k=i$ is less than $d+1.$
Therefore to prove the first equality of (5.47) it is sufficient to prove
\begin{equation}
\mu(\Pr(G(+i,\rho d^{-1}))=O(\rho^{-\alpha})\mu(B(\rho)) \tag{5.48}%
\end{equation}
Clearly, if $(x_{1},x_{2},...x_{i-1},x_{i+1},...x_{d})\in\Pr_{i}(G(+i,\rho
d^{-1})),$ then

$\mu(U_{\varepsilon}(G)(x_{1},x_{2},...x_{i-1},x_{i+1},...x_{d}))\geq
2\varepsilon$ and by (5.22), it follows that
\begin{equation}
\mu(U_{\varepsilon}(G))\geq2\varepsilon\mu(\Pr(G(+i,\rho d^{-1})). \tag{5.49}%
\end{equation}
Hence to prove (5.48) we need to estimate $\mu(U_{\varepsilon}(G)).$ For this
we prove that
\begin{equation}
U_{\varepsilon}(G)\subset U_{\varepsilon}(S_{\rho}^{^{\prime}}),\text{
}U_{\varepsilon}(G)\subset U_{2\varepsilon}(Tr(A(\rho))),\text{ }%
U_{\varepsilon}(G)\subset Tr(U_{2\varepsilon}(A(\rho))). \tag{5.50}%
\end{equation}
The first and second inclusions follow from $G\subset S_{\rho}^{^{\prime}}$
and $G\subset U_{\varepsilon}(Tr(A(\rho)))$ respectively (see definition of
$G$ ). Now we prove the third inclusion in (5.50). If $x\in U_{\varepsilon
}(G),$ then by the second inclusion of (5.50) there exists $b$ such that $b\in
Tr(A(\rho)),$ $\mid x-b\mid<2\varepsilon.$ Then by the definition of
$Tr(A(\rho))$ there exist $\gamma\in\Gamma$ and $c\in A(\rho)$ such that
$b=\gamma+c$. Therefore
\[
\mid x-\gamma-c\mid=\mid x-b\mid<2\varepsilon,\text{ }x-\gamma\in
U_{2\varepsilon}(c)\subset U_{2\varepsilon}(A(\rho)).
\]
This together with $x\in U_{\varepsilon}(G)\subset U_{\varepsilon}(S_{\rho
}^{^{\prime}})$ (see the first inclusion of (5.50)) give $x\in
Tr(U_{2\varepsilon}(A(\rho)))$ ( see the definition of $Tr(E)$ in Notation
5.1), i.e., the third inclusion in (5.50) is proved. The third inclusion,
Lemma 5.1(c), and (5.19) imply that $\mu(U_{\varepsilon}(G))=O(\rho^{-\alpha
})\mu(B(\rho))\varepsilon.$ Now using (5.49), we get the proof of (5.48) and
hence the proof of the first equality of (5.47). The second equality of (5.47)
can be proved in the same way$\diamondsuit$

ESTIMATION 5 Here we prove (5.9). Divide the set $V_{\rho}\equiv V,$ defined
in Theorem 5.1(b), into pairwise disjoint subsets
\[
V^{^{\prime}}(\pm1,\rho d^{-1})\equiv V(\pm1,\rho d^{-1}),V^{^{\prime}}(\pm
i,\rho d^{-1})\equiv V(\pm i,\rho d^{-1})\backslash(\cup_{j=1}^{i-1}(V(\pm
j,\rho d^{-1})))
\]
for $i=2,3,...,d.$ Take any point $a\in V^{^{\prime}}(+i,\rho d^{-1})\subset
S_{\rho}$ and consider the function $F(x)$ ( see (5.4)) on the interval
$(a-\varepsilon e_{i},a+\varepsilon e_{i}),$ where $e_{1}=(1,0,0,...,0)$,
$e_{2}=(0,1,0,...,0),...$. By (5.4(c)), we have $F(a)=\rho^{2}$. It follows
from (5.4(b)) and the definition of $V^{^{\prime}}(+i,\rho d^{-1})$ that
$\frac{\partial F(x)}{\partial x_{i}}>\rho d^{-1}$ for $x\in(a-\varepsilon
e_{i},a+\varepsilon e_{i}).$ Therefore
\begin{equation}
F(a-\delta e_{i})<\rho^{2}-c_{23}\varepsilon_{1},\text{ }F(a+\delta
e_{i})>\rho^{2}+c_{23}\varepsilon_{1}, \tag{5.51}%
\end{equation}
where $\delta=\frac{\varepsilon}{2},$ $\varepsilon_{1}=7\rho\varepsilon.$
Since $[a-\delta e_{i},a+\delta e_{i}]\in U_{\varepsilon}(a)\subset
U_{\varepsilon}(V_{\rho})\subset U_{\varepsilon}(S_{\rho}^{^{\prime}%
})\backslash Tr(A(\rho))$ ( see Theorem 5.1$(b)$), it follows from Theorem
5.1$(a)$ that there exists index $N$ such that $\Lambda(y)=\Lambda_{N}(y)$ for
$y\in U_{\varepsilon}(a)$ and $\Lambda(y)$ satisfies (1.27) ( see Remark 4.1).
Hence (5.51) implies that
\begin{equation}
\Lambda(a-\delta e_{i})<\rho^{2},\Lambda(a+\delta e_{i})>\rho^{2}. \tag{5.52}%
\end{equation}
Moreover it follows from (5.7) that the derivative of $\Lambda(y)$ with
respect to $y_{i}$ is positive for $y\in\lbrack a-\delta e_{i},a+\delta
e_{i}].$ Hence $\Lambda(y)$ is a continuous and increasing function in
$[a-\delta e_{i},a+\delta e_{i}].$ Thus (5.52) implies that there exists a
unique point $y(a,i)\in\lbrack a-\delta e_{i},a+\delta e_{i}]$ such that
$\Lambda(y(a,i))=\rho^{2}.$ Define $I_{\rho}^{^{\prime}}(+i)$ by
\[
I_{\rho}^{^{\prime}}(+i)=\{y(a,i):a\in V^{^{\prime}}(+i,\rho d^{-1})\}).
\]
In the same way we define $I_{\rho}^{^{\prime}}(-i)=\{y(a,i):a\in V^{^{\prime
}}(-i,\rho d^{-1})\}$ and put
\[
I_{\rho}^{^{\prime}}=\cup_{i=1}^{d}(I_{\rho}^{^{\prime}}(+i)\cup I_{\rho
}^{^{\prime}}(-i)).
\]
To estimate the measure of $I_{\rho}^{^{\prime}}$ we compare the measure of
$V^{^{\prime}}(\pm i,\rho d^{-1})$ with the measure of $I_{\rho}^{^{\prime}%
}(\pm i)$ by using the formula (5.21) and the relations
\begin{equation}
\Pr(V^{^{\prime}}(\pm i,\rho d^{-1}))=\Pr(I_{\rho}^{^{\prime}}(\pm i)),\text{
}\mu(\Pr(I_{\rho}^{^{\prime}}(\pm i)))=O(\rho^{d-1}), \tag{5.53}%
\end{equation}%
\begin{equation}
(\frac{\partial F}{\partial x_{i}})^{-1}\mid grad(F)\mid-(\frac{\partial
\Lambda}{\partial x_{i}})^{-1}\mid grad(\Lambda)\mid=O(\rho^{-2\alpha_{1}}),
\tag{5.54}%
\end{equation}
where the first equality in (5.53) follows from the definition of $I_{\rho
}^{^{\prime}}(\pm i)$, the second equality in (5.53) follows from the
inequalities in (5.20), since $I_{\rho}^{^{\prime}}\subset U_{\varepsilon
}(S_{\rho}^{^{\prime}})$, and (5.54) follows from (5.4(b)), (5.7). Using
(5.53), (5.54), and (5.21), we get
\begin{equation}
\mu(V^{^{\prime}}(\pm i,\rho d^{-1}))-\mu(I_{\rho}^{^{\prime}}(\pm
i))=O(\rho^{d-1-2\alpha_{1}}). \tag{5.55}%
\end{equation}
On the other hand if $y\equiv(y_{1},y_{2},...,y_{d})\in I_{\rho}^{^{\prime}%
}(+i)\cap I_{\rho}^{^{\prime}}(+j)$ for $i<j$ then there exist

$a\in V^{^{\prime}}(+i,\rho d^{-1})$ and $a^{^{\prime}}\in V^{^{\prime}%
}(+j,\rho d^{-1})$ such that $y=y(a,i)=y(a^{^{\prime}},j)$ and

$y\in\lbrack a-\delta e_{i},a+\delta e_{i}],$ $y\in\lbrack a^{^{\prime}%
}-\delta e_{j},a^{^{\prime}}+\delta e_{j}].$ These inclusions and definitions
of $V^{^{\prime}}(+i,\rho d^{-1}),$ $V^{^{\prime}}(+j,\rho d^{-1})$ imply that
$\rho d^{-1}-\delta\leq y_{i}\leq\rho d^{-1}.$ Therefore using the
inequalities in (5.20), we get $\mu(\Pr_{j}(I_{\rho}^{^{\prime}}(+i)\cap
I_{\rho}^{^{\prime}}(+j)))=O(\varepsilon\rho^{d-2}).$ This equality, (5.21)
for $k=j$ and (5.7) give
\begin{equation}
\mu((I_{\rho}^{^{\prime}}(+i)\cap I_{\rho}^{^{\prime}}(+j)))=O(\varepsilon
\rho^{d-2}) \tag{5.56}%
\end{equation}
for all $i$ and $j.$ Similarly $\mu((I_{\rho}^{^{\prime}}(+i)\cap I_{\rho
}^{^{\prime}}(-j)))=O(\varepsilon\rho^{d-2})$ for all $i$ and $j.$ Now using
(5.56), (5.55), we obtain%
\[
\mu(I_{\rho}^{^{\prime}})=\sum_{i}\mu(I_{\rho}^{^{\prime}}(+i))+\sum_{i}%
\mu(I_{\rho}^{^{\prime}}(-i))+O(\varepsilon\rho^{d-2})=\sum_{i}\mu
(V^{^{\prime}}(+i,\rho d^{-1}))
\]%
\begin{equation}
+\sum_{i}\mu(V^{^{\prime}}(-i,\rho d^{-1}))+O(\rho^{d-1-2\alpha_{1}}%
)=\mu(V_{\rho})+O(\rho^{-2\alpha_{1}})\mu(B(\rho)). \tag{5.57}%
\end{equation}
This and (5.8) yield the inequality (5.9) for $I_{\rho}^{^{\prime}}$. Now we
define $I_{\rho}^{^{\prime\prime}}$ as follows. If $\gamma+t\in$ $I_{\rho
}^{^{\prime}}$ then $\Lambda(\gamma+t)=\rho^{2}$, where $\Lambda(\gamma+t)$ is
a unique eigenvalue satisfying (1.27) ( see Remark 4.1). Since%
\[
\Lambda(\gamma+t)=\mid\gamma+t\mid^{2}+O(\rho^{-\alpha_{1}})
\]
( see (1.27) and (5.4), (5.4(a))), for fixed $t$ there exist only a finite
number of vectors $\gamma_{1},\gamma_{2},...,\gamma_{s}\in\Gamma$ satisfying
$\Lambda(\gamma_{k}+t)=\rho^{2}$. Hence $I_{\rho}^{^{\prime}}$ is the union of
the pairwise disjoint sets
\[
I_{\rho,k}^{^{\prime}}\equiv\{\gamma_{k}+t\in I_{\rho}^{^{\prime}}%
:\Lambda(\gamma_{k}+t)=\rho^{2}\}\text{ }(k=1,2,...s).
\]
The translation $I_{\rho,k}^{^{\prime\prime}}=I_{\rho,k}^{^{\prime}}%
-\gamma_{k}=\{t\in F^{\ast}:\gamma_{k}+t\in I_{\rho,k}^{^{\prime}}\}$ of
$I_{\rho,k}^{^{\prime}}$ is a part of the isoenergetic surfaces $I_{\rho}$ of
$L(q).$ Put
\[
I_{\rho}^{^{\prime\prime}}=\cup_{k=1}^{s}I_{\rho,k}^{^{\prime\prime}}.
\]
If $t\in I_{\rho,k}^{^{\prime\prime}}\cap I_{\rho,m}^{^{\prime\prime}}$ for
$k\neq m,$ then $\gamma_{k}+t\in I_{\rho}^{^{\prime}}\subset U_{\varepsilon
}(S_{\rho}^{^{\prime}})$ and $\gamma_{m}+t\in U_{\varepsilon}(S_{\rho
}^{^{\prime}}),$ which contradict Lemma 5.1(b). Therefore $I_{\rho}%
^{^{\prime\prime}}$ is union of the pairwise disjoint subsets $I_{\rho
,k}^{^{\prime\prime}}$ for $k=1,2,...s.$ Thus%
\[
\mu(I_{\rho}^{^{\prime\prime}})=\sum_{k}\mu(I_{\rho,k}^{^{\prime\prime}}%
)=\sum_{k}\mu(I_{\rho,k}^{^{\prime}})=\mu(I_{\rho}^{^{\prime}}).
\]
This implies (5.9) for $I_{\rho}^{^{\prime\prime}}$, since (5.9) is proved for
$I_{\rho}^{^{\prime}}$ (see (5.57))$\square$

\section{Bloch Functions near the Diffraction Planes}

In this section we obtain the asymptotic formulas for the Bloch function
corresponding to the quasimomentum lying near the diffraction hyperplanes.
Here we assume that $q(x)\in W_{2}^{s}(F),$ where $s\geq6(3^{d}(d+1)^{2})+d$
instead of the assumption (1.2). Besides, we define the number $\varkappa$ by
$\varkappa=4(3^{d}(d+1))$ instead of the definition $\varkappa=3^{d}+d+2$ of
$\varkappa$ given in (1.6). The other numbers $p,\alpha_{k},\alpha,k_{1}%
,p_{1}$ are defined as in the introduction. Clearly these numbers satisfy all
inequalities of (1.38)-(1.40). \ Therefore the formulas obtained in previous
sections hold in this notations too. Moreover the following relations hold
\begin{equation}
k_{2}<\frac{1}{9}(p-\frac{1}{2}\varkappa(d-1)),\text{ }k_{2}\alpha
_{2}>d+2\alpha,\text{ }4(d+1)\alpha_{d}=1, \tag{6.1}%
\end{equation}
where $k_{2}=[\frac{d}{9\alpha}]+2.$ In this section we construct a subset
$B_{\delta}$ of $V_{\delta}^{^{\prime}}(\rho^{\alpha_{1}})$ such that if
$\gamma+t\equiv\beta+\tau+(j+v)\delta\in B_{\delta}$ ( see Remark 3.1 for this
notations), then there exists a unique eigenvalue $\Lambda_{N}(\lambda
_{j,\beta}(v,\tau))$ satisfying (3.52). Moreover we prove that $\Lambda
_{N}(\lambda_{j,\beta}(v,\tau))$ is a simple eigenvalue if $\beta
+\tau+(j+v)\delta$ belongs to the set $B_{\delta}.$ Therefore we call the set
$B_{\delta}$ the simple set in the resonance domain $V_{\delta}(\rho
^{\alpha_{1}}).$ Then we obtain the asymptotic formulas of arbitrary order for
the eigenfunction $\Psi_{N}(x)$ corresponding to the eigenvalue $\Lambda
_{N}(\lambda_{j,\beta}(v,\tau))$. \ At the end of this section we prove that
$B_{\delta}$ has asymptotically full measure on $V_{\delta}(\rho^{\alpha_{1}%
}).$ The construction of \ the simple set $B_{\delta}$ in the resonance domain
$V_{\delta}(\rho^{\alpha_{1}})$ is similar to the construction of the simple
set $B$ in the non-resonance domain (see Step 1 and Step 2 in introduction).
As in Step 2 we need to find the simplicity conditions for the eigenvalue
$\Lambda_{N}(\lambda_{j,\beta}).$ Since the first inequality in (6.1) holds,
$\Lambda_{N}(\lambda_{j,\beta})$ satisfies the formula (3.52) for $k=k_{2}.$
Therefore it follows from the second inequality of (6.1) that%
\begin{equation}
\Lambda_{N}(\lambda_{j,\beta}(v,\tau))=E(\lambda_{j,\beta}(v,\tau
))+o(\rho^{-d-2\alpha})=o(\varepsilon_{1}), \tag{6.2}%
\end{equation}
where $E(\lambda_{j,\beta}(v,\tau))=\lambda_{j,\beta}(v,\tau)+E_{k_{2}%
-1}(\lambda_{j,\beta}(v,\tau)),$ $\varepsilon_{1}=\rho^{-d-2\alpha},$%
\begin{align}
\lambda_{j,\beta}(v,\tau)  &  \sim\rho^{2},\text{ }E_{k_{2}-1}(\lambda
_{j,\beta})=O(\rho^{-\alpha_{2}}(\ln\rho)),\text{ }\tag{6.3}\\
\lambda_{j,\beta}(v,\tau)  &  =\mid\beta+\tau\mid^{2}+\mu_{j}(v)=\mid
\beta+\tau\mid^{2}+O(\rho^{2\alpha_{1}}),\tag{6.4}\\
E(\lambda_{j,\beta}(v,\tau))  &  =\mid\beta+\tau\mid^{2}+O(\rho^{2\alpha_{1}})
\tag{6.5}%
\end{align}
( see (3.53), Lemma 3.1(b), (3.6), (3.5), and the definition of $E(\lambda
_{j,\beta}(v,\tau))$). Due to (6.2) we call $E(\lambda_{j,\beta}(v,\tau))$ the
known part of $\Lambda_{N}(\lambda_{j,\beta}(v,\tau))$. Since known parts of
the other eigenvalues are $\lambda_{i}(\gamma^{^{\prime}}+t),$ $F(\gamma
^{^{\prime}}+t)$ ( see Step 1 in the introduction), that is, the other
eigenvalues lie in the $\varepsilon_{1}$ neighborhood of the numbers
$\lambda_{i}(\gamma^{^{\prime}}+t),$ $F(\gamma^{^{\prime}}+t),$ in order that
$\Lambda_{N}(\lambda_{j,\beta}(v,\tau))$ does not coincide with other
eigenvalues we use the following two simplicity conditions
\begin{equation}
\mid E(\lambda_{j,\beta}(v,\tau))-F(\gamma^{^{\prime}}+t)\mid\geq
2\varepsilon_{1},\forall\gamma^{^{\prime}}\in M_{1}, \tag{6.6}%
\end{equation}%
\begin{equation}
\mid E(\lambda_{j,\beta})-\lambda_{i}(\gamma^{^{\prime}}+t)\mid\geq
2\varepsilon_{1},\forall\gamma^{^{\prime}}\in M_{2};\forall i=1,2,...,b_{k},
\tag{6.7}%
\end{equation}
where $M$ is the set of $\gamma^{^{\prime}}\in\Gamma$ satisfying
\[
\mid E(\lambda_{j,\beta}(v,\tau))-\mid\gamma^{^{\prime}}+t\mid^{2}\mid
<\frac{1}{3}\rho^{\alpha_{1}},
\]
$M_{1}$ is the set of $\gamma^{^{\prime}}\in M$ satisfying $\gamma^{^{\prime}%
}+t\in U(\rho^{\alpha_{1}},p),$ $M_{2}$ is the set of $\gamma^{^{\prime}}\in
M$ such that $\gamma^{^{\prime}}+t\notin U(\rho^{\alpha_{1}},p)$ and
$\gamma^{^{\prime}}+t$ has the $\Gamma_{\delta}$ decomposition $\gamma
^{^{\prime}}+t=\beta^{^{\prime}}+\tau+(j^{^{\prime}}+v(\beta^{^{\prime}%
}))\delta$ ( see Remark 3.1) with $\beta^{^{\prime}}\neq\beta$.\ 

\begin{definition}
The simple set $B_{\delta}$ in the resonance domain $V_{\delta}(\rho
^{\alpha_{1}})$\ is the set of

$x\in V_{\delta}^{^{\prime}}(\rho^{\alpha_{1}})\cap(R(\frac{3}{2}\rho
-\rho^{\alpha_{1}-1})\backslash R(\frac{1}{2}\rho+\rho^{\alpha_{1}-1}))$ such that

$x=\gamma+t,$ $x=\beta+\tau+(j+v(\beta))\delta,$ and (6.6), (6.7) hold, where
$\gamma\in\Gamma,$ $t\in F^{\star}$, $\beta\in\Gamma_{\delta},$ $\tau\in
F_{\delta},$ $j\in\mathbb{Z},$ $v(\beta)\in W(\rho)$ ( see Remark 3.1 for
these decompositions of $x)$, and $W(\rho)$ is defined in Lemma 3.7.
\end{definition}

Using the simplicity conditions (6.6) and (6.7) we prove that $\Lambda
_{N}(\lambda_{j,\beta}(v,\tau))$ does not coincide with other eigenvalues if
$\beta+\tau+(j+v)\delta\in B_{\delta}$. The existence and properties of the
sets $B_{\delta}$ will be considered in the end of this section. Recall that
in Section 4 the simplicity conditions (1.28), (1.29) implied the asymptotic
formulas \ for the Bloch functions in the non-resonance domain. Similarly,
here the simplicity conditions (6.6), (6.7) imply the asymptotic formula for
the Bloch function in the single resonance domain $V_{\delta}^{^{\prime}}%
(\rho^{\alpha_{1}})$. For this we use the following lemma.

\begin{lemma}
Let $\Lambda_{N}(\lambda_{j,\beta}(v,\tau))$ be the eigenvalue of the operator
$L_{t}(q)$ satisfying (3.52), where $\beta+\tau+(j+v)\delta\equiv\gamma+t\in
B_{\delta}.$ If for $\gamma^{^{\prime}}+t\equiv\beta^{^{\prime}}%
+\tau+(j^{^{\prime}}+v(\beta^{^{\prime}}))\delta$ at least one of the
following conditions:%
\begin{align}
\gamma^{^{\prime}}  &  \in M,\text{ }\beta^{^{\prime}}\neq\beta,\tag{6.8}\\
&  \mid\beta-\beta^{^{\prime}}\mid>(p-1)\rho^{\alpha},\tag{6.9}\\
&  \mid\beta-\beta^{^{\prime}}\mid\leq(p-1)\rho^{\alpha},\mid j^{^{\prime}%
}\delta\mid\geq h \tag{6.10}%
\end{align}
hold, then
\begin{equation}
\mid b(N,\gamma^{^{\prime}})\mid\leq c_{5}\rho^{-c\alpha}, \tag{6.11}%
\end{equation}
where $h\equiv10^{-p}\rho^{\frac{1}{2}\alpha_{2}},$ $c=p-d\varkappa-\frac
{1}{4}d3^{d}-3,$ $b(N,\gamma^{^{\prime}})=(\Psi_{N,t},e^{i(\gamma^{^{\prime}%
}+t,x)}),$ $\Psi_{N,t}(x)$ is any normalized eigenfunction of $L_{t}(q)$
corresponding to $\Lambda_{N}(\lambda_{j,\beta}(v,\tau))$.
\end{lemma}

\begin{proof}
Repeating the proof of the inequality in (4.5) and instead of the simplicity
conditions (1.28), (1.29) and the set $K$ using the simplicity conditions
(6.6), (6.7), and the set $M,$ we obtain the proof of (6.11) under condition (6.8).

Suppose that condition (6.9) holds. Consider two cases:\ 

Case 1: $\gamma^{^{\prime}}\in M.$ It follows from (6.9) that $\beta
^{^{\prime}}\neq\beta.$ Thus, in the Case 1, the condition (6.8) holds and
hence (6.11) is true.

Case 2: $\gamma^{^{\prime}}\notin M.$ The definition of $M$ ( see (6.7))and
(6.2) imply that%
\begin{equation}
\mid\Lambda_{N}-\mid\gamma^{^{\prime}}+t\mid^{2}\mid>\frac{1}{4}\rho
^{\alpha_{1}},\text{ }\forall\gamma^{^{\prime}}\notin M. \tag{6.12}%
\end{equation}
\ Therefore using (1.9) and the definition of $c$ ( see (6.11)), we get
\begin{equation}
b(N,\gamma^{^{\prime}})=\sum_{\gamma_{1}\in\Gamma(\rho^{\alpha})}%
\dfrac{q_{\gamma_{1}}b(N,\gamma^{^{\prime}}-\gamma_{1})}{\Lambda_{N}%
-\mid\gamma^{^{\prime}}+t\mid^{2}}+o(\rho^{-c\alpha}). \tag{6.13}%
\end{equation}
Since $\gamma_{1}\in\Gamma(\rho^{\alpha})$ we have $\gamma_{1}=\beta_{1}%
+a_{1}\delta$ ( see (3.2)), where $\beta_{1}\in\Gamma_{\delta},$ $a_{1}%
\in\mathbb{R},$ $\mid\beta_{1}\mid<\rho^{\alpha}$ and $\gamma^{^{\prime}%
}-\gamma_{1}+t\equiv(\beta^{^{\prime}}-\beta_{1})+\tau+(j^{^{\prime}}%
+v(\beta^{^{\prime}})-a)\delta.$ Moreover , it follows from (6.9) that
$(\beta^{^{\prime}}-\beta_{1})\neq\beta.$ Therefore, if \ $\gamma^{^{\prime}%
}-\gamma_{1}\in M,$ then repeating the proof of (6.11) for Case 1, we obtain
\begin{equation}
\mid b(N,\gamma^{^{\prime}}-\gamma_{1})\mid\leq c_{5}\rho^{-c\alpha}
\tag{6.14}%
\end{equation}
for $\gamma^{^{\prime}}-\gamma_{1}\in M$ . Now in (6.13) instead of
$b(N,\gamma^{^{\prime}}-\gamma_{1})$ for $\gamma^{^{\prime}}-\gamma_{1}\in M$
writing $O(\rho^{-c\alpha})$, and using (1.9) for $b(N,\gamma^{^{\prime}%
}-\gamma_{1})$ when $\gamma^{^{\prime}}-\gamma_{1}\notin M$ , we get
\begin{equation}
b(N,\gamma^{^{\prime}})=\sum_{\gamma_{1},\gamma_{2}}\dfrac{q_{\gamma_{1}%
}q_{\gamma_{2}}b(N,\gamma^{^{\prime}}-\gamma_{1}-\gamma_{2})}{(\Lambda
_{N}-\mid\gamma^{^{\prime}}+t\mid^{2})(\Lambda_{N}-\mid\gamma^{^{\prime}%
}-\gamma_{1}+t\mid^{2})}+o(\rho^{-c\alpha}), \tag{6.15}%
\end{equation}
where the summation is taken under the conditions $\gamma_{1}\in\Gamma
(\rho^{\alpha}),$ $\gamma_{2}\in\Gamma(\rho^{\alpha}),$ $\gamma^{^{\prime}%
}-\gamma_{1}\notin$ $M$. Moreover, it follows from (6.12) that the
multiplicands in the denominators of (6.15) are large number , namely \
\begin{equation}
\mid\Lambda_{N}-\mid\gamma^{^{\prime}}-\sum_{i=1}^{j}\gamma_{i}+t\mid^{2}%
\mid>\frac{1}{4}\rho^{\alpha_{1}}, \tag{6.16}%
\end{equation}
for $\gamma^{^{\prime}}-\sum_{i=1}^{j}\gamma_{i}\notin$ $M,$ where $\gamma
_{i}\in\Gamma(\rho^{\alpha}),$ $j=0,1.$ Arguing as in the proof of (6.14), we
get
\begin{equation}
\mid b(N,\gamma^{^{\prime}}-\gamma_{1}-\gamma_{2})\mid\leq c_{5}\rho
^{-c\alpha} \tag{6.17}%
\end{equation}
for $(\gamma^{^{\prime}}-\gamma_{1}-\gamma_{2})\in M$ . Repeating this process
$p-1$ times, that is, in (6.15) instead of $b(N,\gamma^{^{\prime}}-\gamma
_{1}-\gamma_{2})$ for $\gamma^{^{\prime}}-\gamma_{1}-\gamma_{2}\in M$ writing
$O(\rho^{-c\alpha})$ ( see (6.17)), and using (1.9) for $b(N,\gamma^{^{\prime
}}-\gamma_{1}-\gamma_{2})$ when $\gamma^{^{\prime}}-\gamma_{1}-\gamma
_{2}\notin M$ etc. , we obtain
\begin{equation}
b(N,\gamma^{^{\prime}})=\sum_{\gamma_{1},\gamma_{2},...,\gamma_{p-1}}%
\dfrac{q_{\gamma_{1}}q_{\gamma_{2}}...q_{\gamma_{p-1}}b(N,\gamma^{^{\prime}%
}-\sum_{i=1}^{p-1}\gamma_{i})}{\prod_{j=0}^{p-2}(\Lambda_{N}-\mid
\gamma^{^{\prime}}-\sum_{i=1}^{j}\gamma_{i}+t\mid^{2})}+o(\rho^{-c\alpha}),
\tag{6.18}%
\end{equation}
where the summation is taken under the conditions $\gamma^{^{\prime}}%
-\sum_{i=1}^{j}\gamma_{i}\notin$ $M$ for $j=0,1,...,p-2$. Therefore using
(1.7) and taking into account that (6.16) holds for $j=0,1,2,...,p-2,$ we get
(6.11) for the Case 2.

Now assume that (6.10) holds. First we prove that the following implication%
\begin{equation}
\gamma^{^{\prime}}-\sum_{i=1}^{s}\gamma_{i}\in M\Longrightarrow\beta
^{^{\prime}}-\sum_{i=1}^{s}\beta_{i}\neq\beta, \tag{6.19}%
\end{equation}
where $s=0,1,...,p-1$ and
\begin{equation}
\gamma_{i}\in\Gamma(\rho^{\alpha}),\text{ }\gamma_{i}=\beta_{i}+a_{i}%
\delta,\text{ }(\beta_{i},\delta)=0,\text{ }\beta_{i}\in\Gamma_{\delta},\text{
}a_{i}\in\mathbb{R} \tag{6.20}%
\end{equation}
(see (3.2) for this orthogonal decomposition of $\gamma_{i}$) is true. Assume
the converse, i.e.,

$\beta^{^{\prime}}-\sum_{i=1}^{s}\beta_{i}=\beta.$ Then it follows from (6.20)
and $\gamma^{^{\prime}}+t\equiv\beta^{^{\prime}}+\tau+(j^{^{\prime}}%
+v(\beta^{^{\prime}}))\delta$ ( see Lemma 6.1) that
\begin{equation}
\gamma^{^{\prime}}+t-\sum_{i=1}^{s}\gamma_{i}\equiv\beta+\tau+(j^{^{\prime}%
}+v(\beta^{^{\prime}}))\delta-\sum_{i=1}^{s}a_{i}\delta. \tag{6.21}%
\end{equation}
Since $\gamma_{i}\in\Gamma(\rho^{\alpha}),\ \delta\in\Gamma(\rho^{\alpha})$,
$v(\beta^{^{\prime}})\in\lbrack0,1]$ ( see Lemma 3.1), and (6.20) is the
orthogonal decomposition of $\gamma_{i}$ we have $\mid a_{i}\delta\mid
<\rho^{\alpha},$ $\mid v(\beta^{^{\prime}})\delta\mid<\rho^{\alpha}.$ On the
other hand, by (6.10), $\mid j^{^{\prime}}\delta\mid\geq h.$ Therefore the
orthogonal decomposition (6.21) and the relations%
\begin{equation}
h=10^{-p}\rho^{\frac{1}{2}\alpha_{2}},\text{ }h^{2}\sim\rho^{\alpha_{2}%
},\alpha_{2}=3\alpha_{1}=9\alpha\tag{6.22}%
\end{equation}
imply that
\[
\mid\gamma^{^{\prime}}+t-\sum_{i=1}^{s}\gamma_{i}\mid^{2}\geq\mid\beta
+\tau\mid^{2}+\frac{1}{2}h^{2}.
\]
Using this, (6.5), and (6.22) we obtain
\[
\mid E(\lambda_{j,\beta}(v,\tau))-\mid\gamma^{^{\prime}}+t-\sum_{i=1}%
^{s}\gamma_{i}\mid^{2}\mid>\rho^{\alpha_{1}}%
\]
which contradicts $\gamma^{^{\prime}}-\sum_{i=1}^{s}\gamma_{i}\in M.$ Thus
(6.19) is proved. This implication for $s=0$ means that if $\gamma^{^{\prime}%
}\in M$ then $\beta^{^{\prime}}\neq\beta$ $.$ Therefore if (6.10) holds and
$\gamma^{^{\prime}}\in M,$ then (6.8) holds too and hence (6.11) holds. To
prove (6.11) under condition (6.10) in case $\gamma^{^{\prime}}\notin M$ we
repeat the proof of (6.11) in the Case 2, that is, use (6.18) , (6.12), and etc.
\end{proof}

\begin{theorem}
If $\gamma+t\equiv\beta+\tau+(j+v(\beta))\delta\in B_{\delta},$ then there
exists a unique eigenvalue $\Lambda_{N}(\lambda_{j,\beta}(v,\tau))$ satisfying
(3.52). This is a simple eigenvalue and the corresponding eigenfunction
$\Psi_{N,t}(x)$ satisfies the asymptotic formula
\begin{equation}
\Psi_{N,t}(x)=\Phi_{j,\beta}(x)+O(\rho^{-\alpha_{2}}\ln\rho). \tag{6.23}%
\end{equation}

\end{theorem}

\begin{proof}
The proof is similar to the proof of Theorem 4.1. Arguing as in the proof of
the Theorem 4.1 we see that to prove this theorem it is enough to show that
for any normalized eigenfunction $\Psi_{N,t}(x)$ corresponding to any
eigenvalue$\Lambda_{N}(t)$ satisfying (3.52) the following equality holds
\begin{equation}
\sum_{(j^{^{\prime}},\beta^{^{\prime}})\in K_{0}}\mid b(N,j^{^{\prime}}%
,\beta^{^{\prime}})\mid^{2}=O(\rho^{-2\alpha_{2}}(\ln\rho)^{2}), \tag{6.24}%
\end{equation}
where $K_{0}=\{(j^{^{\prime}},\beta^{^{\prime}}):j^{^{\prime}}\in
\mathbb{Z},\beta^{^{\prime}}\in\Gamma_{\delta},(j^{^{\prime}},\beta^{^{\prime
}})\neq(j,\beta)\},$ $b(N,j^{^{\prime}},\beta^{^{\prime}})=(\Psi_{N}%
,\Phi_{j^{^{\prime}},\beta^{^{\prime}}}).$ Divide $K_{0}$ into subsets:
$K_{1}^{c},$ $K_{1}\cap S(p-1),$ $K_{1}\cap S^{c}(p-1),$ where
\[
K_{1}^{c}=K_{0}\backslash K_{1},S^{c}(n)=K_{0}\backslash S(n),K_{1}%
=\{(j^{^{\prime}},\beta^{^{\prime}})\in K_{0}:\mid\Lambda_{N}(t)-\lambda
_{j^{^{\prime}},\beta^{^{\prime}}}\mid<h^{2}\},
\]
$S(n)=\{(j^{^{\prime}},\beta^{^{\prime}})\in K_{0}:\mid\beta-\beta^{^{\prime}%
}\mid\leq n\rho^{\alpha}$ $,\mid j^{^{\prime}}\delta\mid<10^{n}h\}$ and $h$ is
defined in (6.22). If $(j^{^{\prime}},\beta^{^{\prime}})\in K_{1}^{c},$ then
using (1.21), the definitions of $K_{1}^{c}$ and $h$, we have
\begin{equation}
\sum_{(j^{^{\prime}},\beta^{^{\prime}})\in K_{1}^{c}}\mid b(N,j^{^{\prime}%
},\beta^{^{\prime}})\mid^{2}=\sup_{x\in F}\mid q(x)-q^{\delta}(x)\mid
^{2}O(\frac{1}{\rho^{2\alpha_{2}}})=O(\frac{1}{\rho^{2\alpha_{2}}}).
\tag{6.25}%
\end{equation}
To consider the set $K_{1}\cap S(p-1)$ we prove that
\begin{equation}
K_{1}\cap S(n)=K_{1}\cap\{(j^{^{\prime}},\beta):j^{^{\prime}}\in
\mathbb{Z}\}\subset\{(j^{^{\prime}},\beta):\mid j^{^{\prime}}\delta\mid<2h\}
\tag{6.26}%
\end{equation}
for $n=1,2,...,p-1.$ Take any element $(j^{^{\prime}},\beta)$ from $K_{1}%
\cap\{(j^{^{\prime}},\beta):j^{^{\prime}}\in\mathbb{Z}\}$. Since%
\[
\lambda_{j^{^{\prime}},\beta}(v,\tau)=\mid\beta+\tau\mid^{2}+\mu_{j^{^{\prime
}}}(v)=\mid\beta+\tau\mid^{2}+\mid(j^{^{\prime}}+v)\delta\mid^{2}+O(1),
\]
where $v\in\lbrack0,1]$ ( see Lemma 3.1(b) and (3.6)), using the definition of
$K_{1}$, (6.2), (6.5) and \ (6.22), we obtain
\[
\mid O(\rho^{2\alpha_{1}})-\mid(j^{^{\prime}}+v)\delta\mid^{2}\mid<2h^{2},\mid
j^{^{\prime}}\delta\mid<2h.
\]
Hence the inclusion in (6.26) is proved and $K_{1}\cap\{(j^{^{\prime}}%
,\beta):j^{^{\prime}}\in\mathbb{Z}\}\subset K_{1}\cap S(n)$ for

$n=1,2,...,p-1.$ If the inclusion
\[
K_{1}\cap S(n)\subset K_{1}\cap\{(j^{^{\prime}},\beta):j^{^{\prime}}%
\in\mathbb{Z}\}
\]
does not hold, then there is an element $(j^{^{\prime}},\beta^{^{\prime}})$ of
$K_{1}\cap S(n)$ such that%
\[
0<\mid\beta-\beta^{^{\prime}}\mid\leq n\rho^{\alpha}\leq(p-1)\rho^{\alpha
},\mid j^{^{\prime}}\delta\mid<10^{n}h<\frac{1}{2}\rho^{\frac{1}{2}\alpha_{2}}%
\]
( see (6.22)). \ Hence the pairs $(j^{^{\prime}},\beta^{^{\prime}})$ and
$(j,\beta)$ satisfy the conditions of (3.34). Therefore using (3.34), (3.39)
and (6.22) we get
\begin{equation}
\mid\Lambda_{N}-\lambda_{j^{^{\prime}},\beta^{^{\prime}}}\mid>\frac{1}{2}%
\rho^{\alpha_{2}}>h^{2} \tag{6.27}%
\end{equation}
which contradicts the inclusion $(j^{^{\prime}},\beta^{^{\prime}})$ $\in
K_{1}.$ Thus (6.26) is proved. Therefore
\begin{equation}
\sum_{(j^{^{\prime}},\beta^{^{\prime}})\in K_{1}\cap S(p-1)}\mid
b(N,j^{^{\prime}},\beta^{^{\prime}})\mid^{2}\leq\sum_{j^{^{\prime}}\neq
j,\text{ }\mid j^{^{\prime}}\delta\mid<2h}\mid b(N,j^{^{\prime}},\beta
)\mid^{2} \tag{6.28}%
\end{equation}
For estimation of $b(N,j^{^{\prime}},\beta)$ for $\mid j^{^{\prime}}\delta
\mid<2h,$ we use (3.27) as follows. In (3.27) replacing $\beta^{^{\prime}}$
and $r$ by $\beta$ and $2h,$ we obtain
\[
(\Lambda_{N}-\lambda_{j^{^{\prime}},\beta})b(N,j^{^{\prime}},\beta
)=O(\rho^{-p\alpha})+
\]%
\begin{equation}
+\sum\limits_{(j_{1},\beta_{1})\in Q(\rho^{\alpha},18h)}A(j^{^{\prime}}%
,\beta,j^{^{\prime}}+j_{1},\beta+\beta_{1})b(N,j^{^{\prime}}+j_{1},\beta
+\beta_{1}). \tag{6.29}%
\end{equation}
By definition of $Q(\rho^{\alpha},18h)$ we have $\mid\beta_{1}\mid
<\rho^{\alpha},$ $\mid j_{1}\delta\mid<18h$ , and hence%
\[
\mid(j^{^{\prime}}+j_{1})\delta\mid<20h<\frac{1}{2}\rho^{\frac{1}{2}\alpha
_{2}}.
\]
Therefore in the right-hand side of (6.29) the multiplicand $b(N,j^{^{\prime}%
}+j_{1},\beta+\beta_{1})$ for

$(j^{^{\prime}}+j_{1},\beta+\beta_{1})\in D(\beta)$, where $D(\beta
)=\{(j,\beta+\beta_{1}):\mid j\delta\mid<\frac{1}{2}\rho^{\frac{1}{2}%
\alpha_{2}},0<\mid\beta_{1}\mid<\rho^{\alpha}\},$ takes part. Put
\[
\mid b(N,j_{0},\beta+\beta_{0})\mid=\max_{(j,\beta+\beta_{1})\in D(\beta)}\mid
b(N,j,\beta+\beta_{1})\mid.
\]
By definition of $D(\beta)$ and by (6.22) we have
\[
\mid\Lambda_{N}-\lambda_{j_{0},\beta+\beta_{0}}\mid>\frac{1}{2}\rho
^{\alpha_{2}}.
\]
This with (1.21) gives $\mid b(N,j_{0},\beta+\beta_{0})\mid=$ $O(\rho
^{-\alpha_{2}}).$ Using this, (6.29) and (3.23), we get
\begin{equation}
\mid b(N,j^{^{\prime}},\beta)\mid<c_{24}\mid\Lambda_{N}-\lambda_{j^{^{\prime}%
},\beta}\mid^{-1}\rho^{-\alpha_{2}} \tag{6.30}%
\end{equation}
for $j^{^{\prime}}\neq j,$ $\mid j^{^{\prime}}\delta\mid<2h,$ where%
\[
\Lambda_{N}-\lambda_{j^{^{\prime}},\beta}=\lambda_{j,\beta}-\lambda
_{j^{^{\prime}},\beta}+O(\rho^{-\alpha_{2}})=\mu_{j}(v)-\mu_{j^{^{\prime}}%
}(v)+O(\rho^{-\alpha_{2}})
\]
\ (see (3.39) and Lemma 3.1(b) ) and $v\in W(\rho)$ ( see the definition of
$B_{\delta}$). Now using the definition of $W(\rho)$ (see Lemma 3.7) and
(3.6), we obtain%
\[
\sum_{j^{^{\prime}}\neq j}\mid\Lambda_{N}-\lambda_{j^{^{\prime}},\beta}%
\mid^{-2}=O(\ln\rho).
\]
This with (6.30) and (6.28) yield
\begin{equation}
\sum_{(j^{^{\prime}},\beta^{^{\prime}})\in K_{1}\cap S(p-1)}\mid
b(N,j^{^{\prime}},\beta)\mid^{2}=O(\rho^{-2\alpha_{2}}(\ln\rho)^{2}).
\tag{6.31}%
\end{equation}
It remains to consider $K_{1}\cap S^{c}(p-1)$. Let us prove that
\begin{equation}
b(N,j^{^{\prime}},\beta^{^{\prime}})=O(\rho^{-c\alpha}) \tag{6.32}%
\end{equation}
for $(j^{^{\prime}},\beta^{^{\prime}})\in K_{1}\cap S^{c}(p-1),$ where the
number $c$ is defined in Lemma 6.1. For this using the decomposition of
$\varphi_{j^{^{\prime}},v(\beta^{^{\prime}})}(s)$ by $\{e^{i(m+v(\beta
^{^{\prime}}))s}:m\in\mathbb{Z}\},$ we get
\begin{equation}
b(N,j^{^{\prime}},\beta^{^{\prime}})=\sum_{m}(\varphi_{j^{^{\prime}}%
,v}(s),e^{i(m+v)s})(\Psi_{N,t}(x),e^{i(\beta^{^{\prime}}+\tau+(m+v)\delta
,x)}). \tag{6.33}%
\end{equation}
If $\mid\beta-\beta^{^{\prime}}\mid>(p-1)\rho^{\alpha}$ then Lemma 6.1 ( see
(6.9)) and (3.25), (6.33) give the proof (6.32). So we need to consider the
case $\mid\beta-\beta^{^{\prime}}\mid\leq(p-1)\rho^{\alpha}.$ Then by
definition of $S^{c}(p-1)$ we have $\mid j^{^{\prime}}\delta\mid\geq10^{p-1}%
h$. Write the right-hand side of (6.33) as $T_{1}+T_{2},$ where
\begin{align*}
T_{1}  &  =\sum_{m:\mid m\delta\mid\geq h}T(m),\text{ }T_{2}\sum_{m:\mid
m\delta\mid<h}T(m),\text{ }\\
T(m)  &  =(\varphi_{j^{^{\prime}},v}(s),e^{i(m+v)s})(\Psi_{N,t}(x),e^{i(\beta
^{^{\prime}}+\tau+(m+v)\delta,x)}).
\end{align*}
By (3.25) and Lemma 6.1 ( see (6.10)) we have $T_{1}=O(\rho^{-c\alpha}).$ If
$\mid m\delta\mid<h,$ then the inequality $\mid j^{^{\prime}}\mid>2\mid m\mid$
holds. Therefore using (3.10), taking into account that $\mid j^{^{\prime}%
}\delta\mid\sim\rho^{\alpha_{2}}$ ( see (6.22)) and the number of summand in
$T_{2}$ is less than $\rho^{\alpha_{2}},$ we get $T_{2}=O(\rho^{-c\alpha}).$
The estimations for $T_{1}$, $T_{2}$ give (6.32). Now using $\mid K_{1}%
\mid=O(\rho^{(d-1)\varkappa\alpha})$ ( see (1.37a)), we get
\begin{equation}
\sum_{(j^{^{\prime}},\beta^{^{\prime}})\in K_{1}\cap S^{c}(p-1)}\mid
b(N,j^{^{\prime}},\beta^{^{\prime}})\mid^{2}=O(\rho^{-(2c-(d-1)\varkappa
)\alpha}). \tag{6.34}%
\end{equation}
This, (6.25), (6.31) give the proof of (6.24), since $(2c-(d-1)\varkappa
)\alpha>\alpha_{2}.$
\end{proof}

Now using Theorem 6.1, we obtain asymptotic formulas of arbitrary order. To
formulate these formulas we need the following notations.

\begin{notation}
Define the numbers $n_{1},h_{1},h_{2},...,$ by $n_{1}=[\frac{1}{9}%
(p-\varkappa(\frac{3d-1}{2})-\frac{1}{4}d3^{d}-3)],$

$h_{1}=10^{n_{1}}h=10^{n_{1}-p}\rho^{\frac{1}{2}\alpha_{2}}$ (see (6.22)),
$h_{k}=10^{k-1}h_{1}$ for $k=2,3,....$and introduce the sets $\ Q(\rho
^{\alpha},9h_{k})=\{(j,\beta):\mid j\delta\mid<9h_{k},0<\mid\beta\mid
<\rho^{\alpha}\}$ (see Lemma 3.4 for this notations) $\widetilde{Q}(n_{1}%
\rho^{\alpha},h_{1})=\{(j,\beta):$ $\mid j\delta\mid<h_{1},\mid\beta\mid\leq
n_{1}\rho^{\alpha}\}$ and a function
\[
\widetilde{S}_{1}^{\prime}(\Lambda_{N},\lambda_{j,\beta})=%
{\displaystyle\sum_{(j_{1},\beta_{1})\in\widetilde{Q}(n_{1}\rho^{\alpha}%
,h_{1})\backslash(j,\beta)}}
\dfrac{A(j+j_{1},\beta+\beta_{1},j,\beta)\Phi_{j+j_{1,}\beta+\beta_{1}}%
(x)}{\Lambda_{N}-\lambda_{j+j_{1},\beta+\beta_{1}}}.
\]
One can easily see that $\widetilde{S}_{1}^{\prime}(\Lambda_{N},\lambda
_{j,\beta})$ is obtained from $S_{1}^{\prime}(\Lambda_{N},\lambda_{j,\beta})$
( see (3.48)) by replacing $A(j,\beta,j+j_{1,}\beta+\beta_{1})$ and
$Q(\rho^{\alpha},9r_{1})$ with $\Phi_{j+j_{1},\beta+\beta_{1}}(x),$ and
$\widetilde{Q}(n_{1}\rho^{\alpha},h_{1})\backslash(j,\beta)$\ respectively.
Similarly let $\widetilde{S}_{k}^{\prime}(\Lambda_{N},\lambda_{j,\beta})$ be a
function obtained from $S_{k}^{\prime}(\Lambda_{N},\lambda_{j,\beta})$ ( see
(3.49)) by replacing $A(j,\beta,j+j_{1,}\beta+\beta_{1})$, $Q(\rho^{\alpha
},9r_{1}),$ and $Q(\rho^{\alpha},9r_{i})$ for $i=2,3,...,k$ with
$\Phi_{j+j_{1},\beta+\beta_{1}}(x),$ $\widetilde{Q}(n_{1}\rho^{\alpha}%
,h_{1})\backslash(j,\beta),$ and $Q(\rho^{\alpha},9h_{i})$ for $i=2,3,...,k$
\ respectively. At last difine $\widetilde{A}_{n}^{\prime}$ by
\[
\widetilde{A}_{n}^{\prime}(\Lambda_{N},\lambda_{j,\beta})=\sum_{k=1}%
^{2n}\widetilde{S}_{k}^{\prime}(\Lambda_{N},\lambda_{j,\beta}).
\]

\end{notation}

\begin{theorem}
The eigenfunction $\Psi_{N,t}(x),$ defined in Theorem 6.1, satisfies the
following asymptotic formulas
\begin{equation}
\Psi_{N,t}(x)=E_{k}^{\ast}(x)+O(\rho^{-k\alpha_{2}}(\ln\rho)^{2k}) \tag{6.35}%
\end{equation}
for $k=1,2,...,n_{1},$ where
\[
E_{1}^{\ast}(x)=\Phi_{j,\beta}(x),\text{ }E_{k}^{\ast}(x)=(1+\parallel
\widetilde{E}_{k}\parallel)^{-1}(\Phi_{j,\beta}(x)+\widetilde{E}_{k}(x)),
\]
$\widetilde{E}_{k}=\widetilde{A}_{k}^{\prime}(\lambda_{j,\beta}+E_{k-1}%
,\lambda_{j,\beta})$ and $E_{k-1}$ is defined in Theorem 3.2.
\end{theorem}

\begin{proof}
The proof of this theorem is similar to the proof of the Theorem 4.2. By
Theorem 6.1 the formula (6.35) for $k=1$ is proved. To prove it for arbitrary
$k$ ( $k\leq n_{1}$) we prove the following equivalent formulas
\begin{equation}
\sum_{(j^{^{\prime}},\beta^{^{\prime}})\in S^{c}(k-1)}\mid b(N,j^{^{\prime}%
},\beta^{^{\prime}})\mid^{2}=O(\rho^{-2k\alpha_{2}}(\ln\rho)^{2}), \tag{6.36}%
\end{equation}%
\begin{equation}
\Psi_{N,t}(x)=\sum_{(j^{^{\prime}},\beta^{^{\prime}})\in S(k-1)\cup(j,\beta
)}b(N,j^{^{\prime}},\beta^{^{\prime}})\Phi_{j^{^{\prime}},\beta^{^{\prime}}%
}+O(\rho^{-k\alpha_{2}}\ln\rho), \tag{6.37}%
\end{equation}
where $S(k-1)$ and $S^{c}(k-1)$ is defined in the proof of Theorem 6.1 between
(6.24) and (6.25). First consider the set $S^{c}(k-1)\cap K_{1}.$ It follows
from the relations

$S(k-1)\cap K_{1}=S(p-1)\cap K_{1}$ (see (6.26)) and $S(k-1)\subset S(p-1)$
for $0<k<p$ ( see definition of $S(k-1)$) that $(S(p-1))\backslash S(k-1))\cap
K_{1}=\emptyset$ , and%
\[
S^{c}(k-1)=S^{c}(p-1)\cup(S(p-1)\backslash S(k-1)),S^{c}(k-1)\cap K_{1}%
=S^{c}(p-1)\cap K_{1}.
\]
Therefore using (6.34), the equalities $c=p-d\varkappa-\frac{1}{4}d3^{d}-3$ (
see Lemma 6.1), $\alpha_{2}=9\alpha,$ $n_{1}=[\frac{1}{9}(p-\varkappa
(\frac{3d-1}{2})-\frac{1}{4}d3^{d}-3)]$ ( see Theorem 6.2), we have
\[
\sum_{(j^{^{\prime}},\beta^{^{\prime}})\in S^{c}(k-1)\cap K_{1}}\mid
b(N,j^{^{\prime}},\beta^{^{\prime}})\mid^{2}=O(\rho^{-2n_{1}\alpha_{2}}).
\]
Thus it remains to prove
\begin{equation}
\sum_{(j^{^{\prime}},\beta^{^{\prime}})\in S^{c}(k-1)\cap K_{1}^{c}}\mid
b(N,j^{^{\prime}},\beta^{^{\prime}})\mid^{2}=O(\rho^{-2k\alpha_{2}}(\ln
\rho)^{2}) \tag{6.38}%
\end{equation}
for $k=2,3,...,n_{1}.$ By formula (3.22) and (6.35) we have
\[
\Psi_{N}(x)(q(x)-Q(s))=H(x)+O(\rho^{-\alpha_{2}}\ln\rho),
\]
where $H(x)$ is a linear combination of $\Phi_{j,\beta}(x)$ and $\Phi
_{j^{^{\prime}},\beta^{^{\prime}}}(x)$ for $(j^{^{\prime}},\beta^{^{\prime}%
})\in$ $S(1),$ since

$\mid j\delta\mid<r_{1}<h$ (see (3.5)). Hence $H(x)$ orthogonal to
$\Phi_{j^{^{\prime}},\beta^{^{\prime}}}(x)$ for $(j^{^{\prime}},\beta
^{^{\prime}})\in S^{c}(1)$. Therefore using (3.27) and the definition of
$K_{1}^{c}$ we have
\begin{align*}
\sum_{(j^{^{\prime}},\beta^{^{\prime}})\in S^{c}(1)\cap K_{1}^{c}}  &  \mid
b(N,j^{^{\prime}},\beta^{^{\prime}})\mid^{2}=\sum\mid\dfrac{(O(\rho
^{-\alpha_{2}}\ln\rho),\Phi_{j^{^{\prime}},\beta^{^{\prime}}})}{\Lambda
_{N}-\lambda_{j^{^{\prime}},\beta^{^{\prime}}}}\mid^{2}\\
&  =O(\rho^{-4\alpha_{2}}(\ln\rho)^{2}).
\end{align*}
Hence (6.38) for $k=2$ is proved. Assume that this is true for $k=m.$ Then
(6.37) for $k=m$ holds too. This and (3.22) for $r=10^{m-1}h$ give%
\[
\Psi_{N,t}(x)(q(x)-Q(s))=G(x)+O(\rho^{-m\alpha_{2}}\ln\rho),
\]
where $G(x)$ is a linear combination of $\Phi_{j,\beta}(x)$ and $\Phi
_{j^{^{\prime}},\beta^{^{\prime}}}(x)$ for $(j^{^{\prime}},\beta^{^{\prime}%
})\in$ $S(m).$ Thus $G(x)$ is orthogonal to $\Phi_{j^{^{\prime}}%
,\beta^{^{\prime}}}(x)$ for $(j^{^{\prime}},\beta^{^{\prime}})\in$ $S^{c}(m).$
Using this and repeating the proof of (6.38) for $k=2$ we obtain the proof of
(6.38) for $k=m+1.$ Thus (6.36) and (6.37) are proved. One can easily see that
the formula (6.37) can be written in the form
\begin{align}
&  \Psi_{N,t}(x)-b(N,j,\beta)\Phi_{j,\beta}(x)-\widetilde{G}_{k}%
(x)\tag{6.39}\\
&  =\sum\limits_{(j_{1},\beta_{1})\in\widetilde{Q}(n_{1}\rho^{\alpha}%
,h_{1})\backslash(j,\beta)}b(N,j+j_{1},\beta+\beta_{1})\Phi_{j+j_{1,}%
\beta+\beta_{1}}(x).\nonumber
\end{align}
where $\parallel\widetilde{G}_{k}\parallel=O(\rho^{-k\alpha_{2}}\ln\rho)$. It
is clear that the right-hand side of (6.39) can be obtained from the
right-hand side of the equality%
\[
(\Lambda_{N}-\lambda_{j,\beta})b(N,j,\beta)-O(\rho^{-p\alpha})=
\]%
\begin{equation}
+\sum\limits_{(j_{1},\beta_{1})\in Q(\rho^{\alpha},9r_{1})}A(j,\beta
,j+j_{1,}\beta+\beta_{1})b(N,j+j_{1},\beta+\beta_{1}), \tag{6.40}%
\end{equation}
which is (3.28), by replacing $A(j,\beta,j+j_{1,}\beta+\beta_{1})$ with
$\Phi_{j+j_{1,}\beta+\beta_{1}}(x)$. Therefore in (6.39) doing the iteration
which was done in order to obtain (3.49) from (3.28), we get%
\begin{align}
&  \Psi_{N,t}(x)-b(N,j,\beta)\Phi_{j,\beta}(x)-\widetilde{G}_{k}%
(x)=\tag{6.41}\\
&  \widetilde{A}_{k}^{\prime}(\Lambda_{N},\lambda_{j,\beta})b(N,j,\beta
)+\widetilde{C}_{2k}^{\prime}(\Lambda_{N},\lambda_{j,\beta})+O(\rho^{-p\alpha
}),\nonumber
\end{align}
where $\widetilde{C}_{2k}^{\prime}$ is obtained from $C_{2k}^{^{\prime}}$ (
see (3.49)) by replacing replacing $A(j,\beta,j+j_{1,}\beta+\beta_{1})$,
$Q(\rho^{\alpha},9r_{1}),$ and $Q(\rho^{\alpha},9r_{i})$ for $i=2,3,...,2k$
with $\Phi_{j+j_{1},\beta+\beta_{1}}(x),$ $\widetilde{Q}(n_{1}\rho^{\alpha
},h_{1})\backslash(j,\beta),$ and $Q(\rho^{\alpha},9h_{i})$ for $i=2,3,...,2k$
\ respectively and the term $O(\rho^{-p\alpha})$ in the right-hand side of
(6.41) is a function whose norm is $O(\rho^{-p\alpha}).$ It follows from the
definitions of \ $\widetilde{S}_{k}^{\prime}$ and $\widetilde{C}_{2k}^{\prime
}$ the estimations similar to the estimations (3.50), (3.51) holds for these
functions and
\[
\parallel\widetilde{C}_{2k}^{\prime}\parallel=O((\rho^{-\alpha_{2}}\ln
\rho)^{k}),\text{ }b(N,j,\beta)=1+O(\rho^{-2\alpha_{2}}(\ln\rho)^{2})
\]
( see (6.24)). Therefore repeating the part of the proof of Theorem 4.2 below
(4.15), we get the proof of this theorem
\end{proof}

Now we consider the simple set $B_{\delta}$ in the resonance domain
$V_{\delta}(\rho^{\alpha_{1}}).$ As we noted in Remark 3.1 every vectors $w$
of $\mathbb{R}^{d}$ has decomposition
\begin{equation}
w\equiv\beta+\tau+(j+v)\delta,\text{ }(\beta+\tau,\delta)=0, \tag{6.42}%
\end{equation}
where $\beta\in\Gamma_{\delta},$ $\tau\in F_{\delta},$ $j\in\mathbb{Z},$
$v\in\lbrack0,1)$. Hence the space $\mathbb{R}^{d}$ is the union of the
pairwise disjoint sets
\[
P(\beta,j)\equiv\{\beta+\tau+(j+v)\delta:\tau\in F_{\delta},v\in\lbrack0,1)\}
\]
for $\beta\in\Gamma_{\delta},$ $j\in\mathbb{Z}.$ To prove that $B_{\delta
\text{ }}$has an asymptotically full measure on $V_{\delta}(\rho^{\alpha_{1}%
}),$ that is,
\begin{equation}
\lim_{\rho\rightarrow\infty}\frac{\mu(B_{\delta})}{\mu(V_{\delta}(\rho
^{\alpha_{1}}))}=1 \tag{6.43}%
\end{equation}
we define the following sets:
\begin{align*}
R_{1}(\rho)  &  =\{j\in\mathbb{Z}:\mid j\mid<\frac{\rho^{\alpha_{1}}}%
{2\mid\delta\mid^{2}}+\frac{3}{2}\},\\
S_{1}(\rho)  &  =\{j\in\mathbb{Z}:\mid j\mid<\frac{\rho^{\alpha_{1}}}%
{2\mid\delta\mid^{2}}-\frac{3}{2}\},\\
R_{2}(\rho)  &  =\{\beta\in\Gamma_{\delta}:\beta\in R_{\delta}(\frac{3}{2}%
\rho+d_{\delta}+1)\backslash R_{\delta}(\frac{1}{2}\rho-d_{\delta}-1))\},\\
S_{2}(\rho)  &  =\{\beta\in\Gamma_{\delta}:\beta\in(R_{\delta}(\frac{3}{2}%
\rho-d_{\delta}-1)\backslash R_{\delta}(\frac{1}{2}\rho+d_{\delta
}+1))\backslash(\bigcup_{b\in\Gamma_{\delta}(\rho^{\alpha_{d}})}V_{b}^{\delta
}(\rho^{\frac{1}{2}}))\},
\end{align*}

where $R_{\delta}(\rho)=\{x\in H_{\delta}:\mid x\mid<\rho\},$ $\Gamma_{\delta
}(\rho^{\alpha_{d}})=\{b\in\Gamma_{\delta}:\mid b\mid<\rho^{\alpha_{d}}\},$%
\[
V_{b}^{\delta}(\rho^{\frac{1}{2}})=\{x\in H_{\delta}:\mid\mid x+b\mid^{2}-\mid
x\mid^{2}\mid<\rho^{\frac{1}{2}}\},
\]
and $d_{\delta}=\sup_{x,y\in F_{\delta}}\mid x-y\mid$ is the diameter of
$F_{\delta}.$

Moreover we define a subset $P^{^{\prime}}(\beta,j)$ of $P(\beta,j)$ as
follows. Introduce the sets
\begin{align*}
A(\beta,b,\rho)  &  =\{v\in\lbrack0,1):\exists j\in\mathbb{Z},\mid
2(\beta,b)+\mid b\mid^{2}+\mid(j+v)\delta\mid^{2}\mid<4d_{\delta}\rho
^{\alpha_{d}}\},\\
A(\beta,\rho)  &  =\bigcup_{b\in\Gamma_{\delta}(\rho^{\alpha_{d}})}%
A(\beta,b,\rho),S_{3}(\beta,\rho)=W(\rho)\backslash A(\beta,\rho)
\end{align*}
and put $S_{4}(\beta,j,v,\rho)=\{\tau\in F_{\delta}:\beta+\tau+(j+v)\delta\in
B_{\delta}\}$ for $j\in S_{1},$ $\beta\in S_{2},$ $v\in S_{3}(\beta,j,\rho).$
Then define $P^{^{\prime}}(\beta,j)$ by%
\[
P^{^{\prime}}(\beta,j)=\{\beta+\tau+(j+v)\delta:v\in S_{3}(\beta,\rho),\tau\in
S_{4}(\beta,j,v,\rho)\}.
\]

It is not hard to see that (6.43) follows from the following relations:%
\begin{align}
\lim_{\rho\rightarrow\infty}\frac{\mid S_{i}(\rho)\mid}{\mid R_{i}(\rho)\mid}
&  =1,\text{ }\forall i=1,2,\tag{6.44}\\
B_{\delta}  &  \supset\cup_{j\in S_{1},\beta\in S_{2}}P^{^{\prime}}%
(\beta,j),\tag{6.45}\\
V_{\delta}(\rho^{\alpha_{1}})  &  \subset\cup_{j\in R_{1},\beta\in R_{2}%
}P(\beta,j),\tag{6.46}\\
\lim_{\beta\rightarrow\infty}\frac{\mu(P^{^{\prime}}(\beta,j))}{\mu
(P(\beta,j))}  &  =1. \tag{6.47}%
\end{align}
To prove these relations we use the following lemma.

\begin{lemma}
Let $w\equiv\beta+\tau+(j+v)\delta.$ Then the following implications:

(a) \ $w\in V_{\delta}(\rho^{\alpha_{1}})\Rightarrow j\in R_{1},\beta\in
R_{2},$

(b) \ \ $j\in S_{1},\beta\in S_{2}\Rightarrow w\in V_{\delta}(\rho^{\alpha
_{1}})\cap(R(\frac{3}{2}\rho-\rho^{\alpha_{1}-1})\backslash R(\frac{1}{2}%
\rho+\rho^{\alpha_{1}-1})),$ \ 

(c) $\ j\in S_{1},\beta\in S_{2}\Rightarrow w\in V_{\delta}^{^{\prime}}%
(\rho^{\alpha_{1}})\cap(R(\frac{3}{2}\rho-\rho^{\alpha_{1}-1})\backslash
R(\frac{1}{2}\rho+\rho^{\alpha_{1}-1}))$ hold.

The relations (6.46), (6.45) and the equality (6.44) are true.
\end{lemma}

\begin{proof}
Since $(\beta+\tau,\delta)=0$ ( see (6.42)) the inclusion $\omega\in
V_{\delta}(\rho^{\alpha_{1}})$ means that

$\mid\mid(j+v+1)\delta\mid^{2}-\mid(j+v)\delta\mid^{2}\mid<\rho^{\alpha_{1}}$
and%
\[
(\frac{1}{2}\rho)^{2}<\mid\beta+\tau\mid^{2}+\mid(j+v)\delta\mid^{2}<(\frac
{3}{2}\rho)^{2}%
\]
( see (1.10)), where$\mid v\mid<1,$ $\mid\tau\mid\leq d_{\delta}=O(1)$ ( see
(6.42)). Therefore by direct calculation we get the proof of the implications
$(a)$ and $(b).$

Now we prove $(c).$ Since $(b)$ holds and%
\[
V_{\delta}^{^{\prime}}(\rho^{\alpha_{1}})=V_{\delta}^{^{\prime}}(\rho
^{\alpha_{1}})\backslash(\cup_{a\in\Gamma(p\rho^{\alpha})\backslash
\delta\mathbb{R}}V_{a}(\rho^{\alpha_{2}})
\]
( see Definition 1.1), it is enough to show that $w\notin V_{a}(\rho
^{\alpha_{2}})$ for $a\in\Gamma(p\rho^{\alpha})\backslash\delta\mathbb{R}$.
Using $\ $the orthogonal decomposition $a_{1}+a_{2}\delta$ \ of $a\in
\Gamma(p\rho^{\alpha})$ ( see (3.2)), where $a_{1}\in\Gamma_{\delta},$
$a_{2}\in\mathbb{R}$ , $(a_{1},\delta)=0$ and $\mid a_{1}\mid<p\rho^{\alpha},$
$\mid a_{2}\delta\mid<p\rho^{\alpha},$ we obtain $\mid w+a\mid^{2}-\mid
w\mid^{2}=d_{1}+d_{2},$ where
\[
d_{1}=\mid\beta+a_{1}\mid^{2}-\mid\beta\mid^{2},d_{2}=\mid(j+a_{2}%
+v)\delta\mid^{2}-\mid(j+v)\delta\mid^{2}+2(a_{1},\tau).
\]
The requirements on $j,a_{1},a_{2}$ imply that $d_{2}=O(\rho^{2\alpha_{1}}).$
On the other hand the condition $\beta\in S_{2}$ gives $\beta\notin
V_{a}^{\delta}(\rho^{\frac{1}{2}}),$ i.e., $\mid d_{1}\mid\geq\rho^{\frac
{1}{2}}$. Since $2\alpha_{k}<\frac{1}{2}$ for $k=1,2$ ( see \ the equality in
(6.1)), we have
\[
\mid\mid w+a\mid^{2}-\mid w\mid^{2}\mid>\frac{1}{2}\rho^{\frac{1}{2}},w\notin
V_{a}(\rho^{\alpha_{2}}).
\]
Thus $(c)$ is proved.

The inclusion (6.46) follows from the implication $(a).$ If

$w\equiv\beta+\tau+(j+v)\delta$ belongs to the right-hand side of (6.45) then
using the implication $(c)$ we obtain $w\in V_{\delta}^{^{\prime}}%
(\rho^{\alpha_{1}}).$ Therefore (6.45) follows from the definitions of
$P^{^{\prime}}(\beta,j)$ and $S_{4}(\beta,j,v,\rho)$. It remains to prove the
equality (6.44). Using the definitions of $R_{1},S_{1}$ and inequalities
$\mid\delta\mid<\rho^{\alpha},\alpha_{1}>2\alpha$ we obtain that (6.44) for
$i=1$ holds.

Now we prove (6.44) for $i=2.$ If $\beta\in R_{2}$ then

$\beta+F_{\delta}\subset R_{\delta}(\frac{3}{2}\rho+2d_{\delta}+1)\backslash
R_{\delta}(\frac{1}{2}\rho-2d_{\delta}-1).$ This implies that,%
\[
\mid R_{2}\mid<(\mu(F_{\delta}))^{-1}\mu(R_{\delta}(\frac{3}{2}\rho
+2d_{\delta}+1)\backslash R_{\delta}(\frac{1}{2}\rho-2d_{\delta}-1)),
\]
since the translations $\beta+F_{\delta}$ of $F_{\delta}$ for $\beta\in
\Gamma_{\delta},$ are pairwise disjoint sets having measure $\mu(F_{\delta}).$
Suppose $\beta+\tau\in D(\rho),$ where
\[
D(\rho)=(R_{\delta}(\frac{3}{2}\rho-1)\backslash R_{\delta}(\frac{1}{2}%
\rho+1))\backslash(\bigcup_{b\in\Gamma_{\delta}(\rho^{\alpha_{d}})}%
V_{b}^{\delta}(2\rho^{\frac{1}{2}})).
\]
Then $\frac{3}{2}\rho-1<\mid\beta+\tau\mid<\frac{1}{2}\rho+1,$ $\mid\mid
\beta+\tau+b\mid^{2}-\mid\beta+\tau\mid^{2}\mid\geq2\rho^{\frac{1}{2}}$ for
$b\in\Gamma_{\delta}(\rho^{\alpha_{d}}).$ Therefore using $\mid\tau\mid\leq
d_{\delta}$ it is not hard to verify that $\beta\in S_{2}.$ Hence the sets
$\beta+F_{\delta}$ for $\beta\in S_{2}$ is cover of $D(\rho).$ Thus
\[
\mid S_{2}\mid\geq(\mu(F_{\delta}))^{-1}\mu(D(\rho).
\]
This, the estimation for $\mid R_{2}\mid,$ and the obvious relations
$\mid\Gamma_{\delta}(\rho^{\alpha_{d}})\mid=O(\rho^{(d-1)\alpha_{d}})),$%
\begin{align*}
\mu((R_{\delta}(\frac{3}{2}\rho-1)\backslash R_{\delta}(\frac{1}{2}\rho+1)))
&  =O(\rho^{d-1}),\\
\mu((R_{\delta}(\frac{3}{2}\rho-1)\backslash R_{\delta}(\frac{1}{2}%
\rho+1))\cap V_{b}^{\delta}(2\rho^{\frac{1}{2}}))  &  =O(\rho^{d-2}\rho
^{\frac{1}{2}}),
\end{align*}
$(d-1)\alpha_{d}<\frac{1}{2}$ (see the equality in (6.1)),%
\[
\lim_{\rho\rightarrow\infty}\frac{\mu((R_{\delta}(\frac{3}{2}\rho-1)\backslash
R_{\delta}(\frac{1}{2}\rho+1)))}{\mu(R(\frac{3}{2}\rho+2d_{\delta
}+1)\backslash R_{\delta}(\frac{1}{2}\rho-2d_{\delta}-1))}=1,
\]
$S_{2}(\rho)\subset R_{2}(\rho)$ imply (6.44) for $i=2$
\end{proof}

\begin{theorem}
The simple set $B_{\delta}$ has an asymptotically full measure in the
resonance set $V_{\delta}(\rho^{\alpha_{1}})$ in the sense that (6.43) holds.
\end{theorem}

\begin{proof}
The proof of the Theorem follows from (6.44)-(6.47). By Lemma 6.2 we need to
prove (6.47). Since the translations $P(\beta,j)-\beta-j\delta$ and
$P^{^{\prime}}(\beta,j)-\beta-j\delta$ of $P(\beta,j)$ and $P^{^{\prime}%
}(\beta,j)$ are $\{\tau+v\delta:v\in\lbrack0,1),\tau\in F_{\delta}\}$ and
$\{\tau+v\delta:v\in S_{3}(\beta,\rho),$ $\tau\in S_{4}(\beta,j,v,\rho)\}$
respectively, it is enough to prove
\begin{equation}
\lim_{\rho\rightarrow\infty}\mu(S_{3}(\beta,\rho))=1,\text{ }\mu(S_{4}%
(\beta,j,v,\rho))=\mu(F_{\delta})(1+O(\rho^{-\alpha})), \tag{6.48}%
\end{equation}
where $j\in S_{1},\beta\in S_{2},v\in S_{3}(\beta,\rho),$ and $O(\rho
^{-\alpha})$ does not depend on $v.$ To prove the first equality in (6.48) it
is enough to show that
\begin{equation}
\mu(A(\beta,\rho))=O(\rho^{-\alpha}), \tag{6.49}%
\end{equation}
since $W(\rho)\supset A(\varepsilon(\rho))$ and $\mu(A(\varepsilon
(\rho))\rightarrow1$ as $\rho\rightarrow\infty$ (see Lemma 3.7). Using the
definition of $\ A(\beta,\rho)$ and the obvious relation $\mid\Gamma_{\delta
}(\rho^{\alpha_{d}})\mid=O(\rho^{(d-1)\alpha_{d}})$ we see that (6.49) holds
if $\mu(A(\beta,b,\rho))=O(\rho^{-d\alpha_{d}}).$ In other word we need to
prove that
\begin{equation}
\mu\{s\in\mathbb{R}:\mid f(s)\mid<4d_{\delta}\rho^{\alpha_{d}}\}=O(\rho
^{-d\alpha_{d}}), \tag{6.50}%
\end{equation}
where $f(s)=2(\beta,b)+\mid b\mid^{2}+s^{2}\mid\delta\mid^{2},$ $\beta\in
S_{2},$ $b\in\Gamma_{\delta}(\rho^{\alpha_{d}}).$ The last inclusions yield
\[
\mid2(\beta,b)+\mid b\mid^{2}\mid\geq\rho^{\frac{1}{2}}%
\]
for $\mid b\mid<\rho^{\alpha_{d}}$. This and the inequalities $\mid
f(s)\mid<4d_{\delta}\rho^{\alpha_{d}}$ ( see (6.50)), $\alpha_{d}<\frac{1}{2}$
( see the equality in (6.1)) imply that $s^{2}\mid\delta\mid^{2}>\frac{1}%
{2}\rho^{\frac{1}{2}}$ from which we obtain $\mid f^{^{\prime}}(s)\mid
>\mid\delta\mid\rho^{\frac{1}{4}}.$ Therefore (6.50) follows from the equality
in (6.1)). Thus (6.49) and hence the first equality in (6.48) is proved.

Now we prove the second equality in (6.48). For this we consider the set
$S_{4}(\beta,j,v,\rho)$ for $j\in S_{1},$ $\beta\in S_{2}$, $v\in S_{3}%
(\beta,\rho).$ By the definitions of $S_{4}$ and $B_{\delta}$ the set
$S_{4}(\beta,j,v,\rho)$ is the set of $\tau\in F_{\delta}$ such that
$E(\lambda_{j,\beta}(v,\tau))$ satisfies the conditions (6.6), (6.7). So we
need to consider these conditions. For this we use the decompositions
$\gamma+t=\beta+\tau+(j+v)\delta,$ $\gamma^{^{\prime}}+t=\beta^{^{\prime}%
}+\tau+(j^{^{\prime}}+v(\beta^{^{\prime}},t))\delta,$ (see Remark 3.1) and the
notations%
\[
\lambda_{j,\beta}(v,\tau)=\mu_{j}(v)+\mid\beta+\tau\mid^{2},\lambda_{i}%
(\gamma^{^{\prime}}+t)=\mid\gamma^{^{\prime}}+t\mid^{2}+r_{i}(\gamma
^{^{\prime}}+t)
\]
( see Lemma 3.1(b) and Remark 2.2). Denoting by $b$ the vector $\beta
^{^{\prime}}-\beta$ we write the decomposition of $\gamma^{^{\prime}}+t$ in
the form $\gamma^{^{\prime}}+t=\beta+b+\tau+(j^{^{\prime}}+v(\beta
+b,t))\delta.$ Then to every $\gamma^{^{\prime}}\in\Gamma$ there corresponds
$b=b(\gamma^{^{\prime}})\in\Gamma_{\delta}$. For $\gamma^{^{\prime}}\in M_{1}$
denote by $B^{1}(\beta,b(\gamma^{^{\prime}}),j,v)$ the set of all $\tau$ not
satisfying (6.6). For $\gamma^{^{\prime}}\in M_{2}$ denote by $B^{2}%
(\beta,b(\gamma^{^{\prime}}),j,v)$ the set of all $\tau$ not satisfying (6.7),
where $M_{1}$ and $M_{2}$ are defined in (6.6) and (6.7). Clearly, if
\[
\tau\in F_{\delta}\backslash(\cup_{s=1,2}(\cup_{\gamma^{^{\prime}}\in M_{s}%
}(B^{s}(\beta,b(\gamma^{^{\prime}}),j,v))
\]
then the inequalities (6.6), (6.7) hold, that is, $\tau\in S_{4}%
(\beta,j,v,\rho)$. Therefore using $\mu(F_{\delta})\sim1$ and proving that
\begin{equation}
\mu(\cup_{\gamma^{^{\prime}}\in M_{s}}B^{s}(\beta,b(\gamma^{^{\prime}%
}),j,v))=O(\rho^{-\alpha}),\text{ }\forall s=1,2, \tag{6.51}%
\end{equation}
we get the proof of the second equality in (6.48). Now we prove (6.51). Using
the above notations and (6.6), (6.7) it is not hard to verify that if $\tau\in
B^{s}(\beta,b(\gamma^{^{\prime}}),j,v),$ then
\begin{equation}
\mid2(\beta,b)+\mid b\mid^{2}+\mid(j^{^{\prime}}+v(\beta+b))\delta\mid
^{2}+2(b,\tau)-\mu_{j}(v)+h_{s}(\gamma^{^{\prime}}+t)\mid<2\varepsilon_{1},
\tag{6.52}%
\end{equation}
where $h_{1}=F_{k_{1}-1}-E_{k_{2}-1},$ $h_{2}=r_{i}-E_{k_{2}-1},$
$\gamma^{^{\prime}}\in M_{s},s=1,2.$ First we prove that if $b\equiv
b(\gamma^{^{\prime}})\in\Gamma_{\delta}(\rho^{\alpha_{d}}),$ then (6.52) does
not hold. The assumption $v\in S_{3}(\beta,\rho)$ implies that $v\notin
A(\beta,\rho).$ This means that $\mid2(\beta,b)+\mid b\mid^{2}+\mid
(j^{^{\prime}}+v(\beta+b))\delta\mid^{2}\mid\geq$ $4d_{\delta}\rho^{\alpha
_{d}}.$ Therefore if
\begin{equation}
\mid2(b,\tau)-\mu_{j}(v)+h_{s}(\gamma^{^{\prime}}+t)\mid<3d_{\delta}%
\rho^{\alpha_{d}}, \tag{6.53}%
\end{equation}
then (6.52) does not hold. Thus to prove that (6.52) does not hold it is
enough to show that (6.53) holds. Now we prove (6.53). The relations
$b\in\Gamma_{\delta}(\rho^{\alpha_{d}}),\tau\in F_{\delta}$ imply that
$\mid2(b,\tau)\mid<2d_{\delta}\rho^{\alpha_{d}}.$ The inclusion $j\in S_{1}$
and (3.6) imply that $\mu_{j}(v)=O(\rho^{2\alpha_{1}}).$ By (2.8) and (3.53),
$h_{1}=O(\rho^{\alpha_{1}}).$ Since $\alpha_{d}=3^{d}\alpha=3^{d-1}\alpha
_{1},$ (6.53) for $s=1$ is proved. Now we prove that $r_{i}=O(\rho^{\alpha
_{1}})$ which implies that $\mid h_{2}\mid=O(\rho^{\alpha_{1}})$ and hence
ends the proof of (6.53). The inclusion $\tau\in B^{2}(\beta,b(\gamma
^{^{\prime}}),j,v),$ means that (6.7) does not holds , that is,%
\[
\mid E(\lambda_{j,\beta}(v,\tau))-\lambda_{i}(\gamma^{^{\prime}}%
+t)\mid<2\varepsilon_{1}.
\]
On the other hand the inclusion$\ \gamma^{^{\prime}}\in M_{2}$ implies that
$\gamma^{^{\prime}}\in M$ ( see the definitions of $M_{2},$ and $M$) and hence%
\[
\mid E(\lambda_{j,\beta}(v,\tau))-\mid\gamma^{^{\prime}}+t\mid^{2}\mid
\leq\frac{1}{3}\rho^{\alpha_{1}}%
\]
The last two inequalities imply that $r_{i}(\gamma^{^{\prime}}+t)=O(\rho
^{\alpha_{1}}).$ Thus \ (6.53) is proved. Hence (6.52) for $b\in\Gamma
_{\delta}(\rho^{\alpha_{d}})$ does not hold. It means that the sets
$B^{1}(\beta,b,j,v)$ and $B^{^{2}}(\beta,b,j,v)$ for $\mid b\mid<$
$\rho^{\alpha_{d}}$ are empty.

To estimate the measure of the set $B^{s}(\beta,b(\gamma^{^{\prime}}),j,v)$
for $\gamma^{^{\prime}}\in M_{s}$, $\mid b(\gamma^{^{\prime}})\mid\geq
\rho^{\alpha_{d}},$ $b\in\Gamma_{\delta}$ we choose the coordinate axis so
that the direction of $b$ coincides with the direction of $(1,0,0,...,0),$
i.e., $b=(b_{1},0,0,...,0),b_{1}>0$ and the direction of $\delta$ coincides
with the direction of $(0,0,...,0,1).$ Then $H_{\delta\text{ }}$ and
$B^{s}(\beta,b,j,v)$ can be considered as $\mathbb{R}^{d-1}$ and as subset of
$F_{\delta}$ respectively, where $F_{\delta}\subset\mathbb{R}^{d-1}.$ Now let
us estimate the measure of $B^{s}(\beta,b,j,v)$ by using (5.22) for
$D=B^{s}(\beta,b,j,v),m=d-1,k=1.$ For this we prove that
\begin{equation}
\mu((B^{s}(\beta,b,j,v))(\tau_{2},\tau_{3},...,\tau_{d-1}))<4\varepsilon
_{1}\mid b\mid^{-1}, \tag{6.54}%
\end{equation}
for all fixed $(\tau_{2},\tau_{3},...,\tau_{d-1}).$ Assume the converse. Then
there are two points

$\tau=(\tau_{1},\tau_{2},\tau_{3},...,\tau_{d-1})\in F_{\delta},$
$\tau^{^{\prime}}=(\tau_{1}^{^{\prime}},\tau_{2},\tau_{3},...,\tau_{d-1})\in
F_{\delta}$ of $B^{s}(\beta,b,j,v),$ such that
\begin{equation}
\mid\tau_{1}-\tau_{1}^{^{\prime}}\mid\geq4\varepsilon_{1}\mid b\mid^{-1}.
\tag{6.55}%
\end{equation}
Since (6.52) holds for $\tau^{^{\prime}}$ and $\tau$ we have
\begin{equation}
\mid2b_{1}(\tau_{1}-\tau_{1}^{^{\prime}})+g_{s}(\tau)-g_{s}(\tau^{^{\prime}%
})\mid<4\varepsilon_{1}, \tag{6.56}%
\end{equation}
where $g_{s}(\tau)=h_{s}(\beta^{^{\prime}}+\tau+(j^{^{\prime}}+v(\beta
+b))\delta).$ Using (2.34), (2.36), (3.54), and the inequality

$\mid b\mid\geq\rho^{\alpha_{d}}$, we obtain
\begin{align}
&  \mid g_{1}(\tau)-g_{1}(\tau^{^{\prime}})\mid<\rho^{-\alpha_{1}}\mid\tau
_{1}-\tau_{1}^{^{\prime}}\mid<b_{1}\mid\tau_{1}-\tau_{1}^{^{\prime}}%
\mid,\tag{6.57}\\
&  \mid g_{2}(\tau)-g_{2}(\tau^{^{\prime}})\mid<3\rho^{\frac{1}{2}\alpha_{d}%
}\mid\tau_{1}-\tau_{1}^{^{\prime}}\mid<b_{1}\mid\tau_{1}-\tau_{1}^{^{\prime}%
}\mid. \tag{6.58}%
\end{align}
These inequalities and (6.56) imply that $b_{1}\mid\tau_{1}-\tau_{1}%
^{^{\prime}}\mid<4\varepsilon_{1}$ which contradicts (6.55). Hence (6.54) is
proved. Since $B^{s}(\beta,b,j,v)\subset F_{\delta}$, $d_{\delta}=O(1)$, we
have $\mu(\Pr_{1}B^{s}(\beta,b,j,v))=O(1).$ Therefore formula (5.22), the
inequalities (6.54) and $\mid b\mid\geq\rho^{\alpha_{d}}$ yield
\[
\mu((B^{s}(\beta,b(\gamma^{^{\prime}}),j,v)=O(\varepsilon_{1}\mid
b(\gamma^{^{\prime}})\mid^{-1})=O(\rho^{-\alpha_{d}}\varepsilon_{1})
\]
for $\gamma^{^{\prime}}\in M_{s}\subset M$ and $s=1,2.$ This implies (6.51),
since $\mid M\mid=O(\rho^{d-1}),$ $\varepsilon_{1}=\rho^{-d-2\alpha}$ and
$O(\rho^{d-1-\alpha_{d}}\varepsilon_{1})=O(\rho^{-\alpha})$
\end{proof}

\end{document}